\documentclass{emulateapj}

\shorttitle{Stellar Turbulent Convection. I.} \shortauthors{Meakin
\& Arnett}

\def\nuc#1#2{\relax\ifmmode{}^{#1}{\protect\text{#2}}\else${}^{#1}$#2\fi}

\def\mcol{\multicolumn}
\def\mult{$\times$}
\def\msun{M$_{\odot}$}

\def\c12{$^{12}$C}
\def\o16{$^{16}$O}
\def\ne20{$^{20}$Ne}
\def\na23{$^{23}$Na}
\def\p31{$^{31}$P}
\def\si28{$^{28}$Si}
\def\s32{$^{32}$S}
\def\s31{$^{31}$S}
\def\gam{$\gamma$}
\def\alph{$\alpha$}
\def\havg#1{\langle #1\rangle}

\def\avg#1{\overline{\langle #1\rangle}}
\def\tnm#1{\tablenotemark{#1}}
\def\tnt#1{\tablenotetext{#1}}
\def\jats{Journal of Atmospheric Sciences}
\def\qjrm{Quart. J. Roy. Meteor. Soc.}

\begin{document}
  \title{Turbulent Convection in Stellar Interiors. I. Hydrodynamic
  Simulation}

  \author{Casey A. Meakin\altaffilmark{1,2} and David Arnett\altaffilmark{1}}
\altaffiltext{1}{Steward Observatory, University of Arizona, Tucson
AZ} \altaffiltext{2}{FLASH Center, University of Chicago, Chicago
IL}
  \email{cmeakin@as.arizona.edu, darnett@as.arizona.edu}

\begin{abstract}
We describe the results of three-dimensional (3D) numerical simulations designed to
study turbulent convection in the stellar interiors, and compare them to stellar
mixing-length theory (MLT). Simulations in 2D are significantly different from 3D,
both in terms of flow morphology and velocity amplitude. Convective mixing regions
are better predicted using a {\em dynamic boundary condition} based on the bulk
Richardson number than by purely local, static criteria like Schwarzschild or Ledoux.
MLT gives a good description of the velocity scale and temperature gradient for a
mixing length of $\sim 1.1 H_p$ for shell convection, however there are other
important effects that it does not capture, mostly related to the dynamical motion of
the boundaries between convective and nonconvective regions. There is asymmetry
between up and down flows, so the net kinetic energy flux is not zero. The motion of
convective boundaries is a source of gravity waves; this is a necessary consequence
of the deceleration of convective plumes. Convective "overshooting" is best described
as an elastic response by the convective boundary, rather than ballistic penetration
of the stable layers by turbulent eddies.  The convective boundaries are rife with
internal and interfacial wave motions, and a variety of instabilities arise which
induce mixing through in process best described as turbulent entrainment. We find
that the rate at which material entrainment proceeds at the boundaries is consistent
with analogous laboratory experiments as well as simulation and observation of
terrestrial atmospheric mixing. In particular, the normalized entrainment rate
E=$u_E/\sigma_H$, is well described by a power law dependance on the bulk Richardson
number $Ri_B = \Delta b L/\sigma_H^2$ for the conditions studied, $20\lesssim Ri_B
\lesssim 420$. We find $E = A Ri_B^{-n}$, with best fit values, $\log A = 0.027 \pm
0.38$, and $n = 1.05 \pm 0.21$. We discuss the applicability of these results to
stellar evolution calculations.
\end{abstract}


\section{INTRODUCTION}

\par We have simulated three-dimensional (3D), turbulent, thermally-relaxed,
nearly adiabatic convection (high P\'eclet number). Such flow is relevant to deep
convective regions in stars (i.e., to most stellar mass which is convective, but not
mildly sub-photospheric and surface regions). We simulate oxygen shell burning on its
natural time scale, and core hydrogen burning driven at 10 times its natural rate.
The simulations develop a robust quasi-steady behavior in a statistical sense, with
significant intermittency. We analyze this statistical behavior quantitatively, and
compare it to predictions of astrophysical mixing length theory \citep{bv58}.
Mixing-length Theory (MLT) gives a good representation of many aspects of convection,
but omits others (especially wave generation and mass entrainment) which are related
to the dynamical behavior of stably stratified layers adjacent to the convection.

\par In Section~2 we briefly summarize some results of the study of turbulent
entrainment in geophysics, to prepare the reader for its appearance in our
astrophysical simulations. This process is not included in the standard approach to
stellar evolution \citep{cg68,clay83,kw90,hansen}. In Section~3 we discuss our
numerical and theoretical tools. In Section~4 we present our simulations of oxygen
shell burning, which attain a thermal steady state (this is possible because of the
rapidity of nuclear heating and neutrino cooling). In Section~5 we discuss a less
advanced burning stage, core hydrogen burning, which we are able to examine with the
use of an artificially enhanced hydrogen burning rate (by a factor of ten). We find
that the behavior is similar to the oxygen burning shell, suggesting that our results
may have broad application for stellar evolution. In Section~6 we compare our results
to the assumptions of MLT, and in Section~7 we show that our results lead to a simple
model of turbulent entrainment, an effect not in MLT nor in standard stellar
evolutionary calculations.

\par This paper is the first in a series.  In subsequent papers, we incorporate
the "empirical" convection model developed in this paper into the TYCHO stellar
evolution code \citep{ya05} and begin to assess its influence on stellar evolution,
on nucleosynthetic yields, and on the structure of supernova progenitors.

\section{Turbulent Entrainment}\label{turb-entrain-intro-sec}
\par
\begin{figure*}
    \epsscale{0.8}
    \plotone{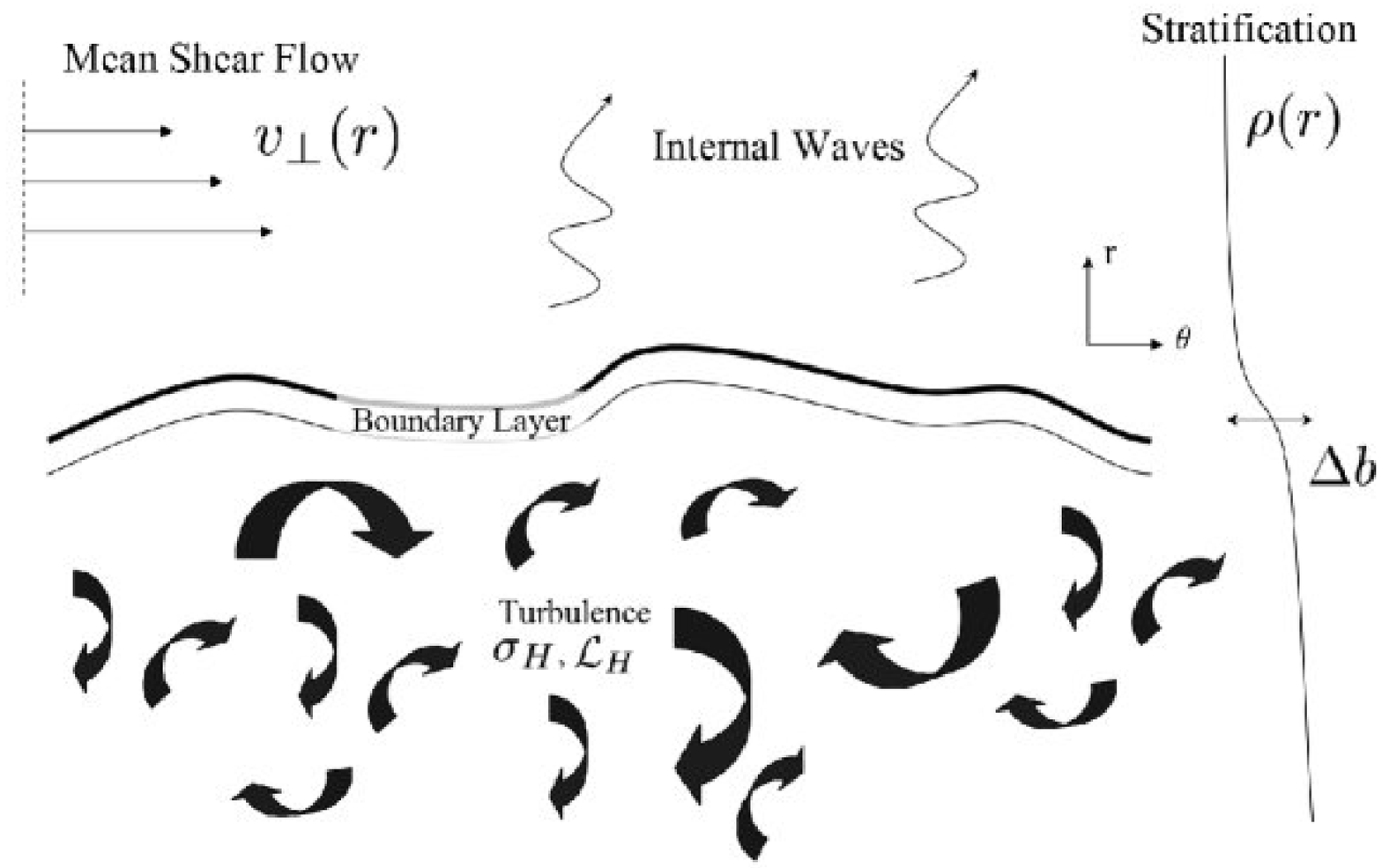}
  \caption{Diagram illustrating the salient features of the density and velocity
    field for the turbulent entrainment problem.  Three layers are present: a
    turbulent convection zone is separated from an overlying stably stratified
    region by a boundary layer of thickness $h$ and buoyancy jump $\Delta b \sim N^2 h$.
    The turblence near the interface is characterized by integral scale and RMS
    velocity $\mathcal{L}_H$ and $\sigma_H$, respectively.  The stably stratified layer
    with buoyancy frequency $N(r)$ propagates internal waves which are excited by the
    adjacent turbulence. A shear velocity field $v_{\perp}(r)$, associated with
    differential rotation, may also be present.  After \citet{strang01}.
    \label{diagram}}
\end{figure*}

 The presence of a turbulent layer contiguous with a stably stratified
layer is common in both astrophysical and geophysical flows. Turbulence in a
stratified media is often sustained by strong shear flows or thermal convection and
bound by a stabilizing density interface.  Over time, the turbulent layer
``diffuses'' into the stable layer and the density interface recedes, thus increasing
the size of the mixed, turbulent region. The basic features of this {\it turbulent
entrainment problem} are illustrated in Figure \ref{diagram}.  The rate at which the
density interface recedes into the stable layer $u_E = \partial r_i/\partial t$ is
called the entrainment rate, and its dependance on the parameters characterizing the
turbulent and the stable layers has been the subject of numerous experimental and
theoretical studies. It is generally ignored in stellar evolutionary studies.

\par Experimental studies have mostly been of ``mixing box'' type which involves a tank
of fluid with a turbulent layer and a density stratified layer. The turbulence is
generated by thermal convection or an oscillating wire mesh, and density
stratification imposed by either a solute or thermal gradient \citep{turner80}.
Complementary to these shear-free mixing box models are shear driven-models.
Shear-driven turbulence experiments involve either a recirculation track which
propels one layer of fluid above a stationary layer, or a rotating plate in contact
with the fluid that drives a circulation in the upper layer. Shear instabilities
sustain a turbulent mixed layer in the overlying fluid which then entrains fluid from
the lower, stationary layer \citep{kantha77,strang01}. In all of these laboratory
experiments, a variety of flow visualization techniques are used to study both the
overall entrainment rate $u_E$ and the physical mechanisms which underly the
entrainment process.

\par One of the primary conclusions of these studies is that the entrainment
rate depends on a Richardson number, which is a dimensionless measure of the
``stiffness'' of the boundary relative to the strength of the turbulence. In
shear-free turbulent entrainment the bulk Richardson number,

\begin{equation}
  \label{RiB-eq}
  Ri_B =  \frac{\Delta b L}{\sigma^2},
\end{equation}

\noindent is most commonly studied.  Here, $\Delta b$ is the
buoyancy jump across the interface, $\sigma$ is the r.m.s.
turbulence velocity adjacent the interface, and $L$ is a length
scale for the turbulent motions often taken to be the horizontal
integral scale of the turbulence at the interface. The relative
buoyancy is defined by the integral,

\begin{equation}
    \label{buoyancy-equation}
  b(r) = \int\limits_{r_i}^{r} N^2 dr
\end{equation}

\noindent where $N$ is the buoyancy frequency defined by,

\begin{equation}
\label{brunt-frequency}
  N^2 = -g \Big(\frac{\partial\ln\rho}{\partial r} - \frac{\partial\ln\rho}{\partial
  r}\Big|_s\Big).
\end{equation}

\noindent The entrainment coefficient $E$ is the interface migration speed $u_e$
normalized by the R.M.S. turblent velocity at the interface $E = u_E/\sigma$, and is
generally found to obey a power law dependance on $Ri_B$,

\begin{equation}
  \label{entrainment-law-equation}
  E = A Ri_B^{-n}.
\end{equation}

\noindent  The exponent is usually found to lie in the range $1 \lesssim  n \lesssim
1.75$ and has been the subject of many theoretical studies of the entrainment
process.  Dimensional analysis suggests that $Ri_B$ should be the controlling
parameter, so long as microscopic diffusion plays a minor role \citep{phillips66}.
Basic energetic arguments in which the rate of change of potential energy due to
mixing is assumed to be proportional to the turbulent kinetic energy available at the
interface leads to an exponent of $n=1$ \citep[e.g.][]{linden1975}. This same power
law exponent has also been derived for models of the growth of the planetary boundary
layer due to turbulent entrainment by penetrative convection
\citep{stull73,tennekes74,stull76,sorbjan96}.

\par The normalization of the entrainment coefficient $A$ has been found to vary
significantly between the various laboratory and field studies conducted, with recent
values found in the range $ 0.1 < A < 0.5$ \citep[e.g.][]{stevens99}. The discrepancy
among the normalization constants has been called the 'A-dilema'
\citep{bretherton99}. A review (up to 1991) of experimental measures of the
parameters in the entrainment law of equation \ref{entrainment-law-equation} are
tabulated in \citet{fernando91} and a recent review of entrainment models used in the
atmospheric sciences is discussed by \citet{stevens02}.

\par The experimental and theoretical models discussed above are generally
motivated by geophysical problems, but are directly relevant to the
conditions found in stellar interiors.  The bulk Richardson numbers
which characterize stellar convective boundaries fall within the
same parameter range ($10 < Ri_B < 500$), and the background
stratifications posses a similar buoyancy structure, so that it is
interesting to learn from the geophysical models and compare to the
stellar case.

\section{THE NUMERICAL TOOLS}

\subsection{1D Stellar Evolution}

\par The hydrodynamic simulations which we study in this paper
are of two distinct phases in the evolution of a 23\msun \ supernova progenitor: main
sequence core convection, and convective oxygen shell burning. The initial conditions
for the multi-dimensional simulations are taken from one-dimensional stellar models
evolved with the TYCHO stellar evolution code. TYCHO \citep{ya05} is an an open
source code\footnote{http://chandra.as.arizona.edu/\~ dave/tycho-intro.html}. A
choice of standard 1D stellar evolution procedures are used. The mixing length theory
as described in \cite{kw90} is used with instantaneous mixing of composition in the
convectively unstable regions. The limits of the convection zones are determined
using the Ledoux criterion, which incorporates the stabilizing effects of composition
gradients. Semiconvective mixing has been turned off. Nuclear evolution is followed
with a 177 element network using the rates of \citep{rt00}. Opacities are from
\citep{ir96} and \citep{af94} for high and low temperature regimes, respectively. The
solar abundance of \cite{gs98} are used. Although more recent abundance
determinations have been made \citep{asplund2005} the impact on the stellar structure
of the models presented here is small, and minor variations in the abundances have a
negligible influence on the development of the hydrodynamic flow.

\subsection{Multi-Dimensional Reactive Hydrodynamics with PROMPI\label{multid-method}}

\par The core of our multi-dimensional hydrodynamics code is
the solver written by \cite{fma89} which is based on the direct
Eulerian implementation of PPM \citep{cw84} with generalization to
non-ideal gas equation of state \citep{cg85}. This code solves the
Euler equations, to which we add nuclear reactions and radiative
diffusion through an operator-split formulation.  The complete set
of combustive Euler equations, including diffusive radiative
transfer, can be written in state-vector form,

\begin{equation}
  \frac{\partial {\mathbf Q}}{\partial t} + \nabla{\cdot \mathbf \Phi} = {\mathbf S},
\end{equation}

with the {\em state vector}

\begin{equation}
  {\mathbf Q} \equiv \left[
    \begin{array}{c}
      \rho \\
      \rho u\\
      \rho E\\
      \rho X_l\\
    \end{array}
    \right],
\end{equation}

the {\em flux vector}

\begin{equation}
  {\mathbf \Phi} \equiv  \left[
    \begin{array}{c}
      \rho {\mathbf u}\\
      \rho {\mathbf u}{\mathbf u} + p\\
      (\rho E + p){\mathbf u} + {\mathbf F}_r\\
      \rho X_l \mathbf u\\
    \end{array}
    \right],
\end{equation}

and the {\em source vector}

\begin{equation}
  {\mathbf S} \equiv \left[
    \begin{array}{c}
      0\\
      \rho {\mathbf g}\\
      \rho{\mathbf {u \cdot g} + \rho\epsilon_{net}}\\
      R_l\\
    \end{array}
    \right],
\end{equation}

\noindent where $E = E_I + E_K$ is the total energy per gram
consisting of internal and kinetic energy components, and $\rho$,
$p$, ${\mathbf u}$, ${\mathbf g}$, and $T$ are the density,
pressure, velocity, gravitational force field and temperature. The
net energy source term due to nuclear reactions and neutrino cooling
is $\epsilon_{net} = \epsilon_{burn} + \epsilon_{cool}$, and the
time rate of change of composition $X_l$ due to nuclear reactions is
denoted $R_l$. The radiative flux is ${\mathbf F}_{r}=-k_{r} \nabla
T$, with radiative ``conductivity'' $k_{r} = 4acT^3/(3\kappa_R
\rho)$ and Rosseland mean opacity, $\kappa_R$. Self gravity is
implemented assuming the interior mass at each radius is distributed
with spherical symmetry. The mass interior to the inner boundary of
the hydrodynamics grid is adopted from the TYCHO stellar model.

\par The stellar models, which are calculated on a
finely meshed Lagrangian grid, are linearly interpolated onto the
Eulerian hydrodynamics grid taking into account the sub-grid
representation of mass used in the PPM scheme. Mapping the models
leads to small discrepancies in hydrostatic equilibrium. An
equilibration to hydrostatic balance occurs through the excitation
and then damping of low amplitude, standing, predominantly radial
pressure waves within the computational domain. These low amplitude
waves, which are well described by the linearized wave equation,
have a negligible affect on the convective flow.

\par To save computational resources, we simulate carefully chosen subregions
of the star. Thus, these calculations are local models of convection in the {\em box
in a star} tradition.  The advantage of local convection models is that higher
effective resolution can be used than is currently possible in global circulation
models. This approach, however, precludes investigation of the lowest order modes of
flow, and we do not yet include rotation or magnetic fields which are best studied
using global domains. The boundary conditions used are periodic in angular
directions, and stress-free reflecting in the radial direction.

\par Our simulation code, dubbed PROMPI, has been adapted to parallel
computing platforms using domain decomposition and the sharing of a
three zone layer of boundary values and uses the MPI message passing
library to manage interprocess communication.

\section{OXYGEN SHELL BURNING}

\par We have evolved a 23\msun \ stellar model with the TYCHO code to
a point where oxygen is burning in a shell which overlies a silicon-sulfur
rich core.  Approximately 60\% of the oxygen fuel available for fusion has
been depleted at the time we begin the hydrodynamic simulation, when the
star is $\sim 2\times10^7$ yrs from the zero age main sequence.  Carbon,
helium, and hydrogen burning shells are also present contemporaneously at
larger radii in the classic "onion skin" structure \citep{hoyle46}.  In one
of the models presented here (ob.2d.e), which is also discussed in
\citet{ma06a}, we adopt an outer radius that encompasses both the oxygen
and carbon burning shells. In this paper, however, we focus our analysis on
the oxygen shell burning convection zone and the stable layers which bound
it.

\par The oxygen shell burning model affords us the opportunity to
study a thermally relaxed model because the thermal balance is
determined by the very large neutrino cooling rates rather than the
much lower radiative diffusion timescale \citep[][ch.11]{da1996}.
Neutrinos dominate the
energy balance in the stable layers so that the stellar structure
and the nature of convection are determined by the interplay between
nuclear burning and neutrino emission
\citep{aufderheide1993,da72}. The effects of radiative diffusion
are both unresolved and energetically unimportant during these
evolutionary phases, and has not been included in the oxygen
shell calculations for computational efficiency.

\par
\begin{figure*}
    \epsscale{0.8}
    \plotone{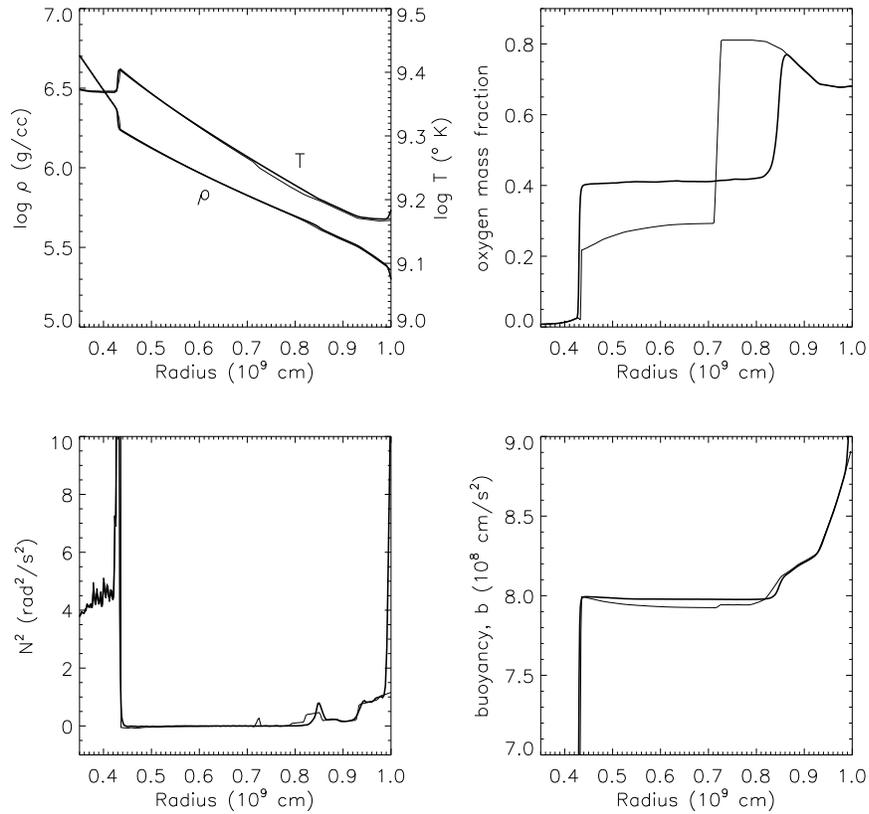}
  \caption{Radial profile of the simulated region for the oxygen shell burning
models.  The thin lines indicate the initial conditions and the thick lines indicate
the 3D model at t = 400 s. (top left) Temperature and density. (top right) Mass
fraction of $\rm ^{16}O$. (bottom left) Squared buoyancy frequency. (bottom right)
Buoyancy.
    \label{ob-profile}}
\end{figure*}

 The radial profile of the simulated region is presented in
Figure \ref{ob-profile}.  The temperature and density profiles betray the complex
structure of the model, including the narrow burning shell that resides at the very
base of the convection zone which is coincident with the temperature peak.  The
initial extent of the convection zone can be identified by the plateau in oxygen mass
fraction at $0.43<r_9<0.72$ (where $r_9 = r/10^9$cm). Characteristic of shell burning
regions, the entropy gradient is quite steep at the boundaries of the convection zone
and gives rise to peaks in the buoyancy frequency at those locations. The initial
location of the upper convective boundary is coincident with a small stable layer at
$r\sim0.72\times 10^9$cm, which is overwhelmed by the convective flow that develops
in the simulation (see \S\ref{transient-section}). A new boundary forms where the
buoyancy frequency again becomes stabilizing at $r \gtrsim 0.8\times 10^9$cm. This
mixing is shown in the change in $\rm^{16}$O abundance (Figure~\ref{ob-profile}, top
right) after 400s.

\par In Table \ref{network-table} we list the 25 nuclei used in our
network. This network reproduces to within 1\% the energy generation of the full 177
element network used to evolve the one-dimensional TYCHO model for the simulated
conditions, including oxygen and carbon burning shells. During carbon burning the
dominant reactions are \c12(\c12, \alph)\ne20 and \c12(\c12, p)\na23, leaving an ash
of \ne20, \na23, protons and alpha-particles. \ne20 is photodisintegrated through the
\ne20(\gam, \alph)\o16 reaction. The dominant reactions during oxygen burning are
\o16(\o16, \alph)\si28, \o16(\o16, p)\p31, and \o16(\o16, n)\s31, leaving an ash of
predominantly \si28 and $^{32}$S. Neglecting the non-alpha chain species \na23, \p31
and \s31 can affect the net energy generation rate during carbon and oxygen burning
by a factor of a few under the conditions studied here.  The reaction rates,
including \c12(\alph, \gam)\o16, are from \cite{rt00}.

\par Nuclear evolution is time advanced using the same reaction network subroutines
as the TYCHO code and uses implicit differencing \citep{da1996}.  We include cooling
by neutrino-antineutrino pair emission, denoted $\epsilon_{cool}$, which result from
photo, pair, plasma, bremstrahlung, and recombination processes
\citep{beaudet1967,itoh1996}.

\par The Helmholtz equation of state code of \cite{ts00} is used to represent the ion
and electron pressure with an arbitrary degree of electron degeneracy. With our 25
nuclei network, the initial conditions are thermodynamically consistent with the
initial TYCHO model to better than a few percent at all radii after mapping to the
hydrodynamics grid.

\par  We calculate oxygen shell burning models in two and three dimensions. Our baseline
model, labeled ob.2d.c, is a $90^{\circ}$ wedge embedded in the equatorial plane with
radii encompassing the oxygen burning convective shell and two stable bounding
layers.  The effects of dimensionality on the oxygen burning convective shell are
explored with a three-dimensional model ob.3d.B which has an angular extents of
(27$^{\circ}$)$^2$. The influence of the upper boundary was studied with model
ob.2d.e, which includes the overlying carbon burning convective shell as well
(additional details concerning this model are presented in \citet{ma06a}). A
preliminary resolution study is undertaken with model ob.2d.C which uses the same
domain limits but twice the linear resolution of the baseline model.   Properties of
the oxygen shell burning models presented in this paper are summarized in Table
\ref{ob-table}.

\subsection{The Correct Mixing Boundary\label{transient-section}}

\par
\begin{figure*}
    \plotone{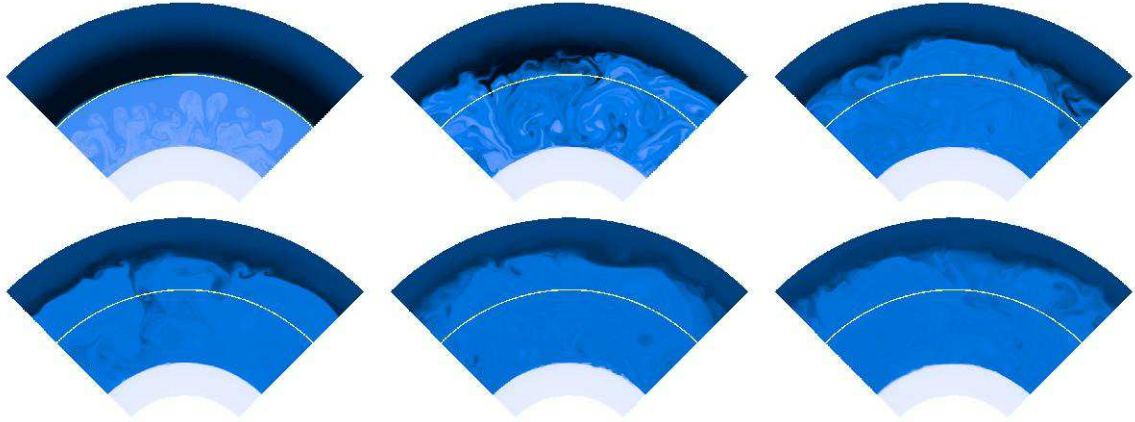}
  \caption{This time sequence shows the onset of convection in the oxygen shell
burning model.  The first 200 s of the 2D model (ob.2d.c) is shown, including the
initial transient and the settling down to a new quasi-steady state. The light yellow
line indicates the location of the convective boundary as defined in the 1D TYCHO
stellar evolution model (Ledoux criterion), which was used as initial conditions for
the simulation.
    \label{ob-initial-transient}}
\end{figure*}

 Convection is initiated through random low-amplitude (0.1\%)
perturbations in density and temperature applied to a region in the center of the
convectively unstable layer on a zone by zone basis. (Two additional simulation
models with the same characteristics as ob.2d.c were calculated which used
perturbations with a larger amplitudes and (1\%), and a low order mode distribution.
The development of the convective flow was found to be insensitive to these
differences.) The role played by the perturbations is to break the angular symmetry
of the initial model, and seed rising and sinking plumes whose growth is driven by
nuclear burning, neutrino cooling, and the slightly superadiabatic background
gradient imprinted in the initial TYCHO model. As the plumes rise they penetrate the
original convective boundary which was determined in the TYCHO code using the Ledoux
criterion. The initial evolution of the flow is presented in a time series of
snapshots in Figure \ref{ob-initial-transient}; the light yellow contour shows the
initial outer convective boundary.

\par The location of the initial outer boundary can be seen as a small bump in the
initial profile of the buoyancy frequency presented in Figure \ref{ob-profile} at
radius $r \sim .72\times 10^9$cm. The reason the boundary is stable in the 1D model
but did not survive in the multi-D simulation is because of the local nature of the
Ledoux criterion used. This can be appreciated by the fact that although the buoyancy
frequency at this location is positive, and hence locally stable to convective
turnover, the {\em buoyancy jump} across this region is very small $\Delta b \sim
3\times 10^6$ cm/s$^2$ compared to the turbulent kinetic energy in the adjacent flow,
by which it is easily overwhelmed. This type of inconsistency can be relatively
easily removed from 1D simulations by using a parameter akin to the bulk Richardson
number (eq. [\ref{RiB-eq}]) to characterize convective boundaries in place of the
Ledoux or Schwarzchild criteria.  For the original outer boundary $Ri_B \lesssim 1$,
a condition under which a boundary is expected to mix on an advection timescale, akin
to the expansion of turbulence into a homogenous medium.

\par The relationship between $Ri_B$ and the traditional Schwarzschild and Ledoux criteria
can be appreciated by writing the buoyancy frequency in terms of the well known
"nablas" used in stellar evolution,

\begin{equation}
  N^2 = \frac{g\delta}{H_P}\Big(\nabla_{ad}-\nabla_{s}+\frac{\varphi}{\delta}\nabla_{\mu}
  \Big)
\end{equation}

\noindent where $\nabla = (d\ln T/d\ln p)$, $\nabla_{s}$ is the gradient of the
stellar background, $\nabla_{ad}$ is the gradient due to an adiabatic displacement,
$\nabla_{\mu} = (d\ln\mu/d\ln p)$ is the mean molecular weight gradient, and the
thermodynamic derivatives are $\delta = -(d\ln \rho/d\ln T)$ and
$\varphi=(d\ln\rho/d\ln\mu)$. Therefore, the Ledoux criteria is simply,

\begin{equation}
N^2 > 0.
\end{equation}

\noindent The Schwarzschild criteria is the same, but with the stabilizing effect of
the mean molecular weight gradient $\nabla_{\mu}$ neglected. For comparison, the bulk
Richardson number can be written $Ri_B \sim N^2 h L/\sigma^2$, where $h$ is some
measure of the boundary width. A convective boundary will start to become stabilizing
when,

\begin{equation}
N^2 \ga \sigma^2/(hL).
\end{equation}

\noindent This latter criteria is based on a finite threshold for stability which
takes into account the strength of the convective turbulence. Additionally, the bulk
Richardson number is more than a simple stability criteria; it is also an indicator
of the rate at which boundary erosion will proceed. We conclude that the correct
criterion for determining the extent of a convective zone is neither the Ledoux nor
the Schwarzschild criterion, which are both static, linear, and local criteria, but a
{\em dynamic boundary condition}, based on the bulk Richardson number, which we will
discuss in more detail in \S\ref{entrainment-section}.

\subsection{Time Evolution}

\par
\begin{figure*}
  \epsscale{0.90}
    \plotone{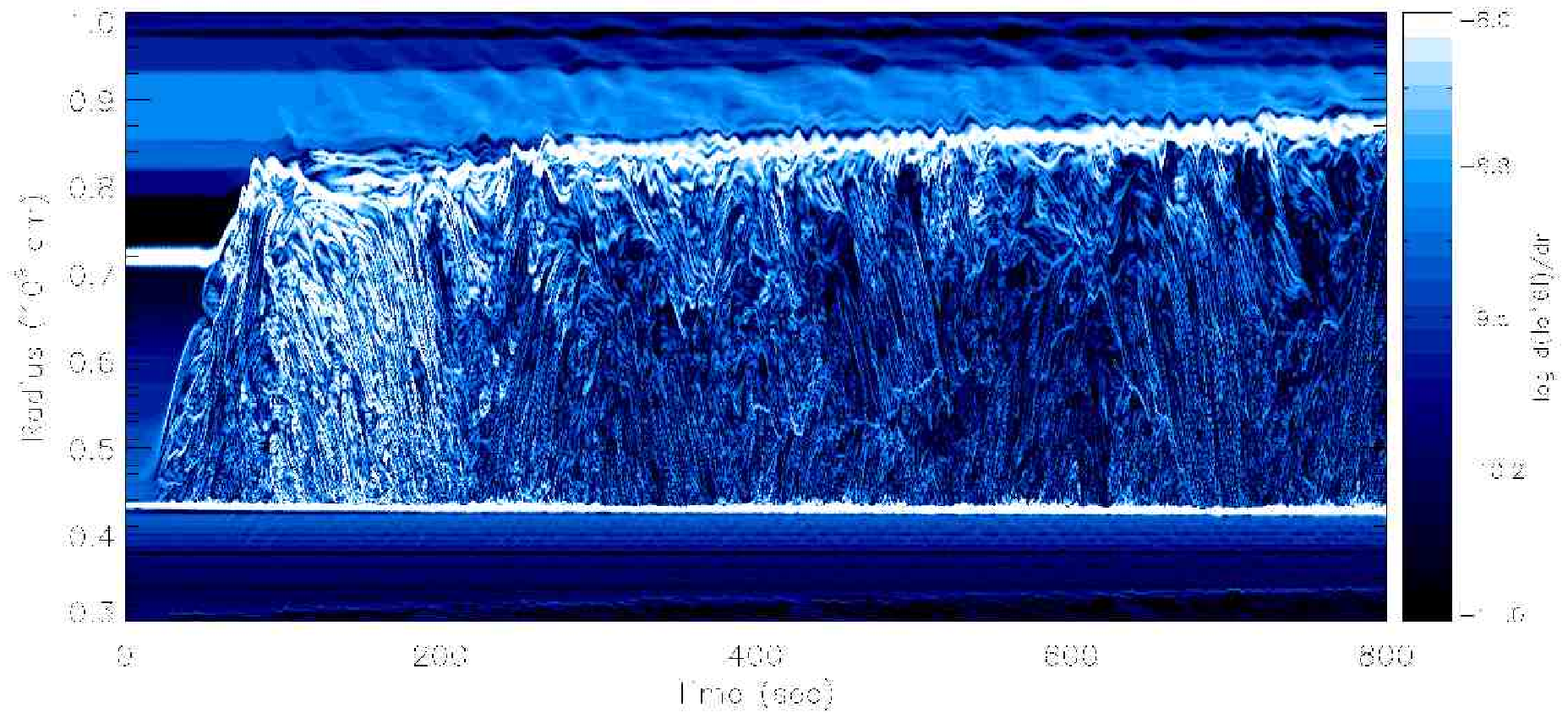}
    \plotone{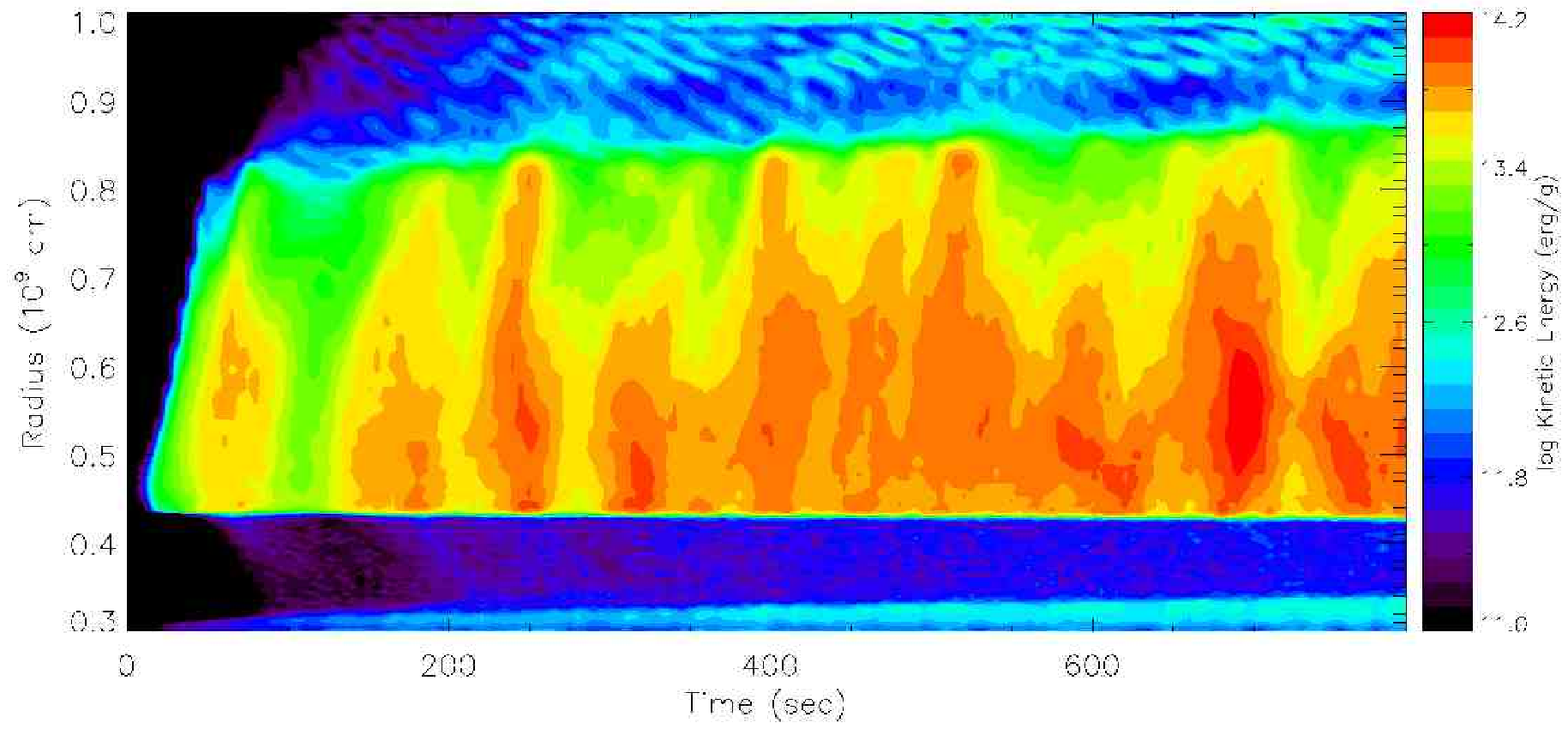}
  \caption{The time evolution of the 3D oxygen shell burning model.
  (top) The magnitude of the oxygen abundance gradient is shown and
  illustrates the migration of the convective
boundaries into the surrounding stable layers.  Interfacial oscillations are also
apparent in the upper convective boundary layer ($r\sim 0.85\times 10^9$ cm), and
internal wave motions can be seen quite clearly in the upper stable layer. (bottom)
The kinetic energy density is shown, and illustrates the intermittent nature of the
convective motions.  The upwelling chimney-like features in the convective region are
seen to excite internal wave trains in the stable layers, which propagate away from
the boundaries of the convection zones. See also Fig.
\ref{qflux-fig}.\label{ob3d-evol}}
\end{figure*}

 The rich dynamics taking place at the convective boundary are apparent in the time
evolution of the 3D flow presented in Figure \ref{ob3d-evol}, which provides a global
view of the evolution. The upper panel shows the evolution in time and radius of the
oxygen abundance gradient, represented by a colormap in which light is large and dark
is small. At the beginning of the simulation (far left) the colors are smooth as the
turbulence has not yet developed. The light line near the bottom of the panel is the
lower boundary of the convective shell, where oxygen is separated from the
silicon-sulfur core below. The short horizontal band at $r\sim 0.72\times 10^9$cm is
the initial weakly stable convective boundary discussed above; it is overwhelmed in
the first 100 seconds by convection. After $\sim$300 seconds the abundance
distribution has approached a quasi-steady state, with slow growth of the convective
region. The bottom of the convection zone moves downward, but at a much slower rate
than the upper boundary moves outward. The mottled apearance in the convection zone
is due to the ingestion of new oxygen entrained from above, followed by turbulent
mixing. At the top boundary of the convection zone an oscillatory behavior can be
seen, and in the overlying stable region wave motions are apparent.

\par The lower panel in Figure \ref{ob3d-evol} shows the radial profile of the kinetic
energy, which illustrates a major feature of the convection: intermittency. While
these simulations are well described by a statistical steady state over a few
convective turnover times, at any instant the fluctuations are significant. The flow
is episodic, with bursts of activity followed by lulls.  The bursts in kinetic energy
in the convection zone are seen to induce wave trains in both the upper and lower
stable layers. Characteristic of g-modes, the phase velocity (orientation of the wave
crests) is orthogonal to the group velocity (direction of energy transport) in these
wave trains, which can be seen by comparing the composition and kinetic energy
profiles.

\subsection{Quasi-Steady Oxygen Shell Burning Convection}

\par
\begin{figure*}
  \epsscale{0.5}
    \plotone{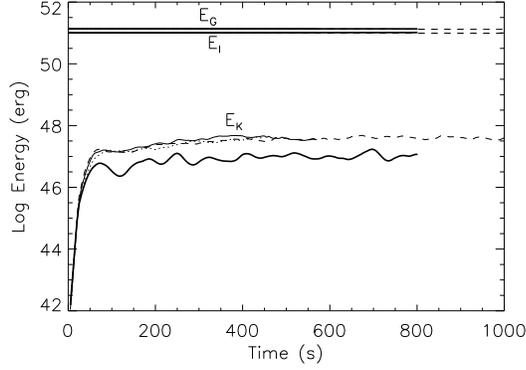}
  \caption{The time evolution of the energy budgets for the oxygen shell
burning models: the (thick line) 3D model, and (thin lines) the three 2D models are
shown, including: (thin-solid) ob.2d.c; (thin-dashed) ob.2d.e; and (thin dotted)
ob.2d.C. The energy budget includes the internal energy $E_I$, the gravitational
energy $E_G$, and the kinetic energy $E_K$. Note that the energy scale is
logarithmic, so that the 3D kinetic energy is much smaller than the 2D values.
    \label{teint-ob}}
\end{figure*}

 Following the transient readjustment of the outer convective
boundary, the oxygen burning convective shell attains a quasi-steady character. In
Figure \ref{teint-ob} we present the time evolution of the integrated internal,
gravitational, and kinetic energy. The energy is calculated by forming horizontal
averages of the flow properties and then assuming a full spherical geometry. The
gravitational energy contribution from material on the computational grid is
calculated according to,

\begin{equation}
  E_G \equiv \int \frac{G M(r)dM}{r} dr,
\end{equation}

\noindent where the mass increment is $dM = 4\pi r^2\langle\rho\rangle$, and the
integral is taken over the radial limits of the grid.

\par The total kinetic energy levels off in all of the models by $t\sim 300$s.  The 2D
models are characterized by a much larger overall kinetic energy. The total kinetic
energy settles down to a slow increase as the oxygen shell evolves; this is true for
both 2D and 3D.

\par
\begin{figure*}
    \epsscale{1.0}
    \plottwo{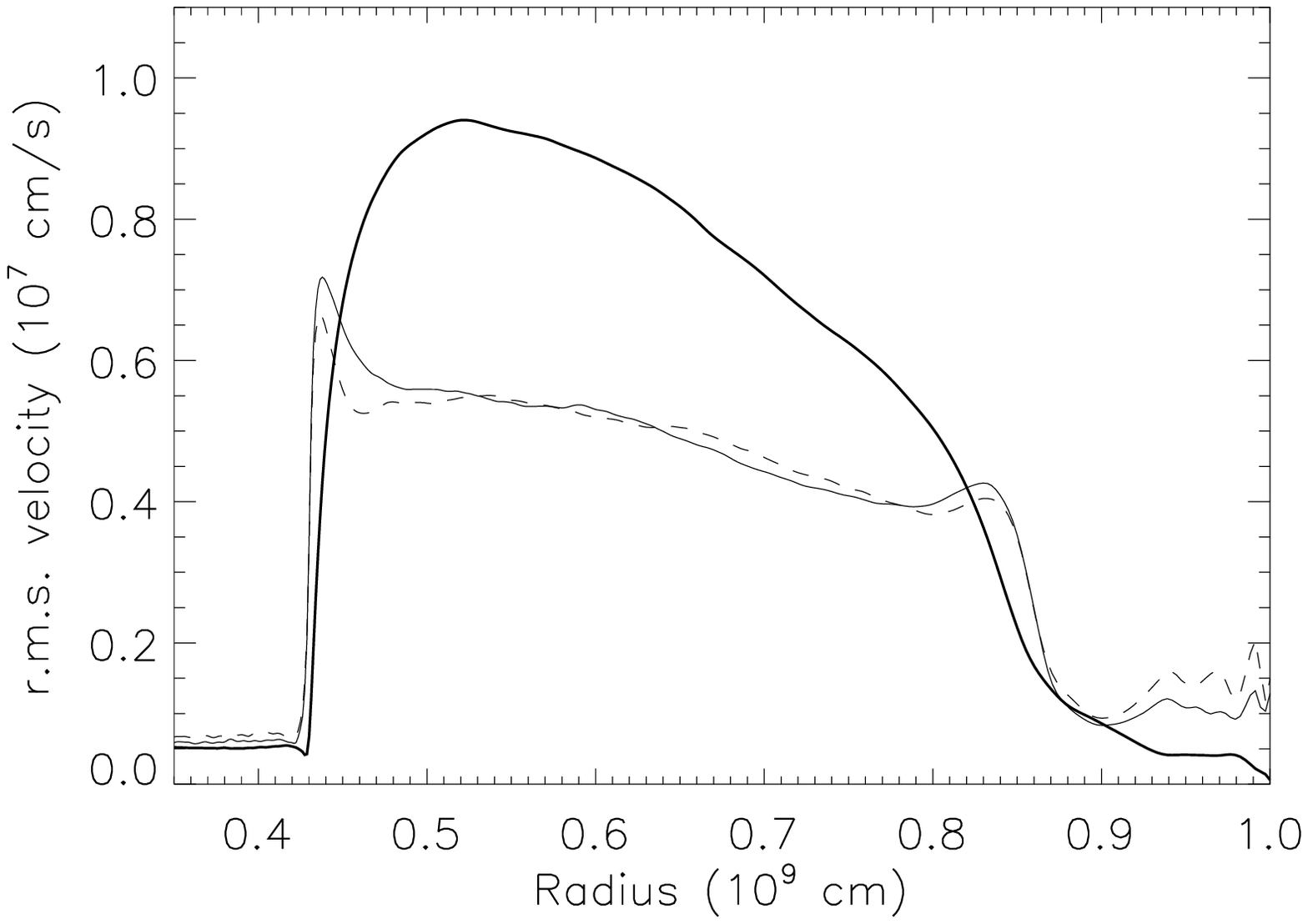}{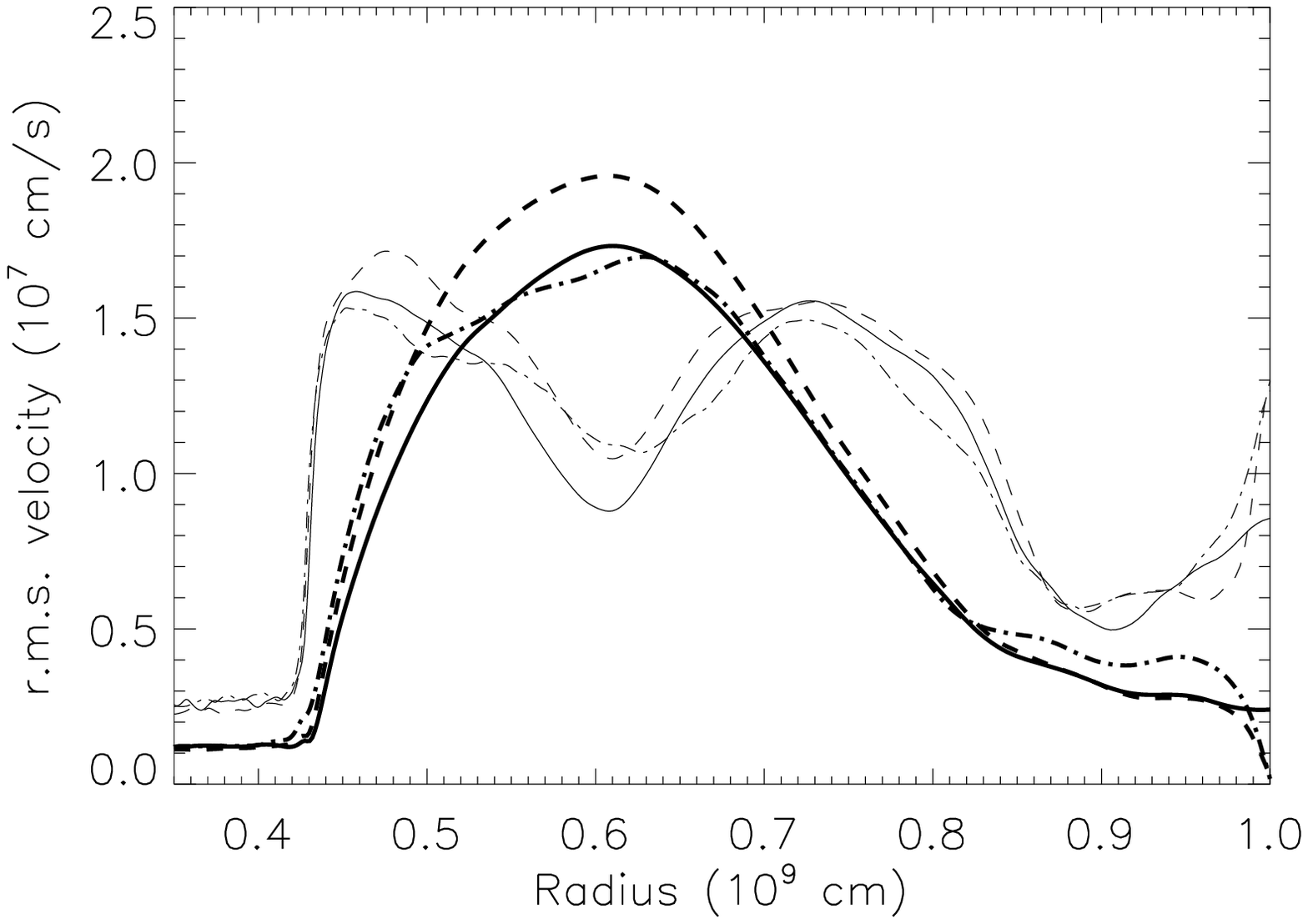}
  \caption{The r.m.s. velocity fluctuations for oxygen shell burning:
  (left) 3D model, with velocity components (thick-solid) $v_r$, (thin-solid) $v_{\theta}$, and
  (thin-dashed) $v_{\phi}$.
  (right) The 2D models, with velocity components (thick) $v_r$ and
  (thin) $v_{\phi}$ for simulations  (solid) ob.2d.e, (dashed) ob.2d.c, and (dash-dot)
  ob.2d.C.
    \label{ob-2d-3d-vrms}}
\end{figure*}

 The radial profile of the r.m.s. velocity fluctuations are presented in
Figure \ref{ob-2d-3d-vrms} for the 2D and 3D models. The velocity fluctuation
amplitudes in all of the 2D models are higher than the 3D model by a factor of
$\sim$2. The 2D models also assume a significantly different radial profile than the
3D model, with a flow structure that is dominated by large convective vortices which
span the depth of the convection zone.  The signature of these large eddies is
apparent in the horizontal velocity components, as well as the fairly symmetric shape
of the radial velocity profile within the convection zone.  The velocity components
in the 3D model reveal an up and down flowing circulation with horizontal deflection
taking place in a fairly narrow layer at the convective boundaries.

\par Although significant differences exist between 2D and 3D models, the 2D models
are found to be in good agreement with each other to the extent that the statistics
have converged, which are calculated over the time period $t\in[300,450]$s. The time
period for calculating statistics was limited by the model ob.2d.C, which was only
run as far as $t\sim 450$s. The agreement among the 2D models shows that the outer
boundary condition (tested by model ob.2d.e) and the grid resolution (tested by model
ob.2d.C) are not playing a decisive role in determining the overall structure of the
flow, at least in these preliminary tests.  The agreement in overall velocity
amplitude in the upper stable layer in model ob.2d.e indicates that the stable layer
velocity amplitudes are not strongly affected by the details of the modes that are
excited in that region.  This gives credence to the analysis in \citep{ma06b} which
assumes that the stable layer velocity amplitudes are determined by the dynamical
balance between the convective ram pressure and the wave induced fluctuations.

\par The convective turnover times $t_c = 2\Delta R/v_{conv}$ for the 2D models are
all of order $t_c \sim 40$s, and they span between 10 and 55 convective turnovers.
The turnover time for the 3D model is $t_c \sim 103$s, and the model spans
approximately 8 convective turnovers.

\subsection{Stable Layer Dynamics During Shell Burning}

\par In both the 2D and 3D models, the stably stratified layers are characterized by
velocity fluctuations throughout their extents (Figure \ref{ob-2d-3d-vrms}).  These
fluctuations are the signature of g-modes which are excited by the convective
motions.  In the 2D model, the amplitudes of the stable layer velocity fluctuations
are higher. In the lower stable layer, the 2D models also have a much smaller ratio
of horizontal to radial velocity amplitude. The velocity amplitude ratio is roughly
proportional to the ratio of the mode frequency and buoyancy frequency,
$v_r/v_{\perp} \approx \omega/N$ \citep{press81}, so that the waves excited in the 2D
model are of lower frequency. The velocity ratios in the upper stable layer are
comparable between the 2D and 3D models, though the 2D amplitudes are higher by a
factor of $\sim 2$.

\par During late burning stages, multiple concentric convective shells form which are
separated by stably stratified layers. These intervening stable layers act as
resonating cavities for g-modes that are excited by the turbulent convection. In
\citet{ma06a} it was shown that the stable layer motions in model ob.2d.e can be
decomposed into individual g-modes that are well described by the linearized
non-radial wave equation \citep{unno89}. \citet{ma06b} showed that a good estimate
for the amplitudes of the wave motions (and the associated thermodynamic
fluctuations) in both the 2D and 3D models can be made by assuming that the pressure
fluctuations associated with the g-modes balance the ram pressure of the turbulent
convection.  In the latter paper, a single mode (frequency and horizontal scale) was
assumed, based on integral properties of the turbulence (convective turnover time,
and mixing length scale). In this section we present the spectrum of motions present
in the stable layers and turbulent regions for the more realistic 3D model.

\par
\begin{figure*}
  \epsscale{0.32}
  \plotone{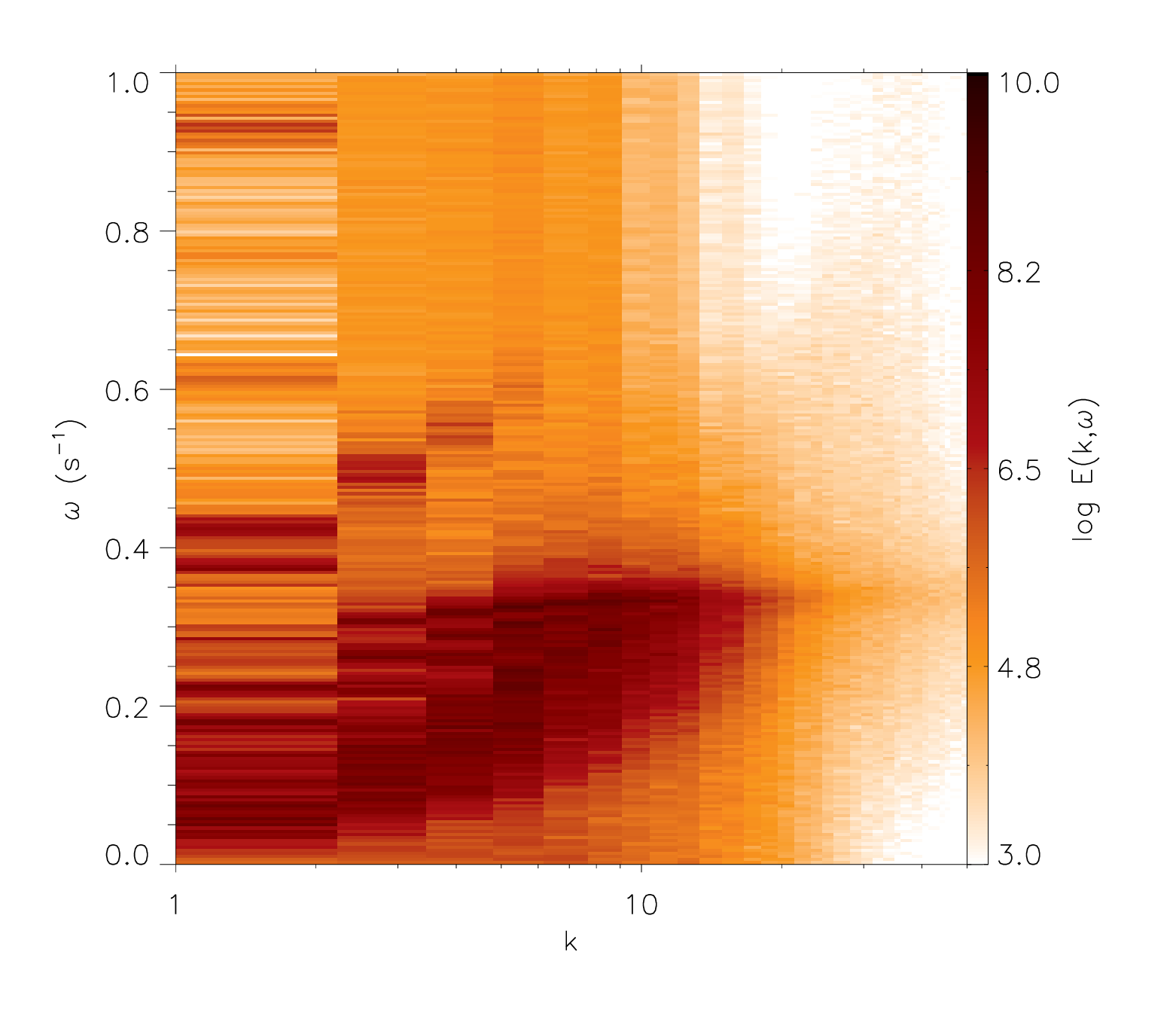}
  \plotone{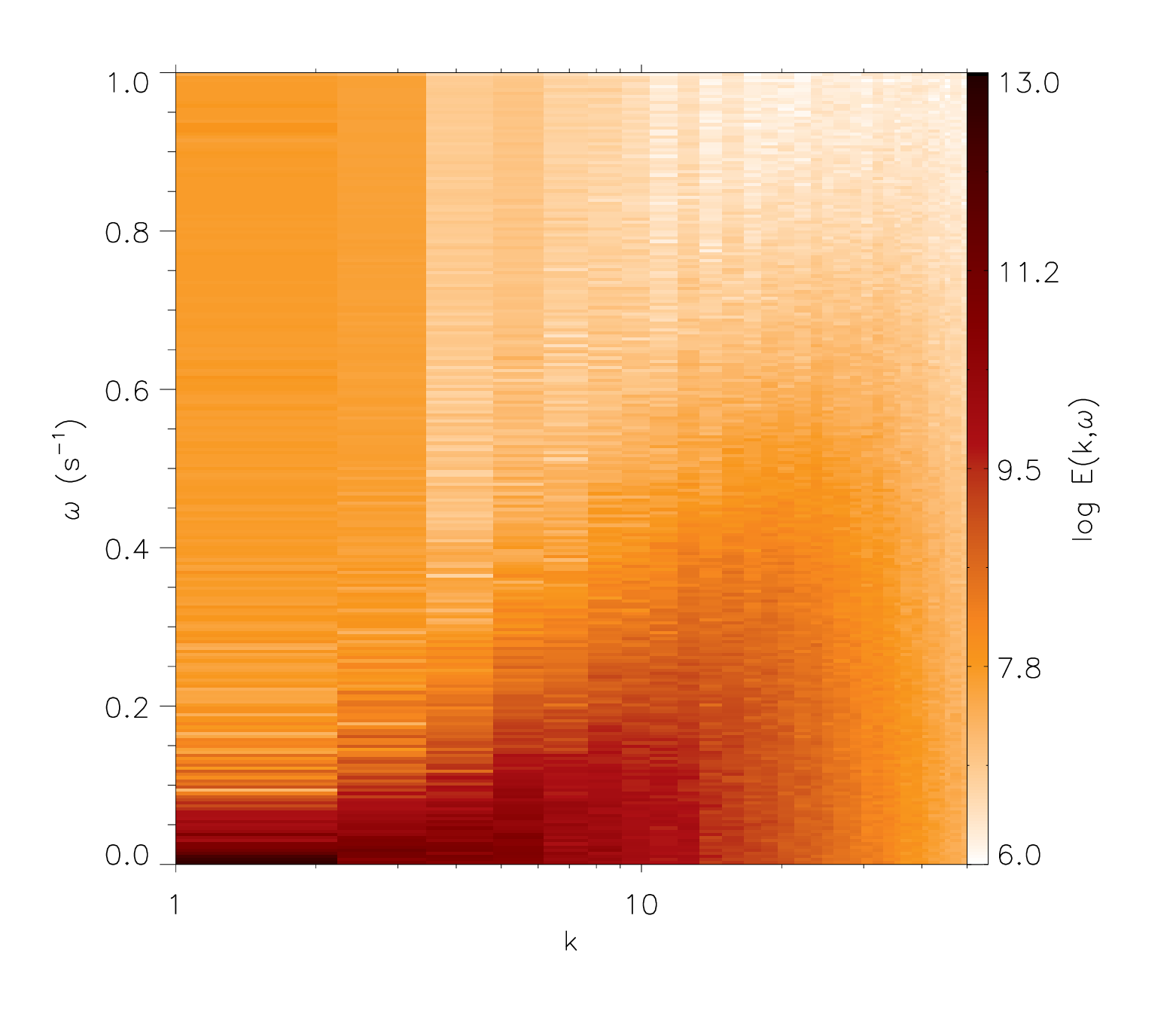}
  \plotone{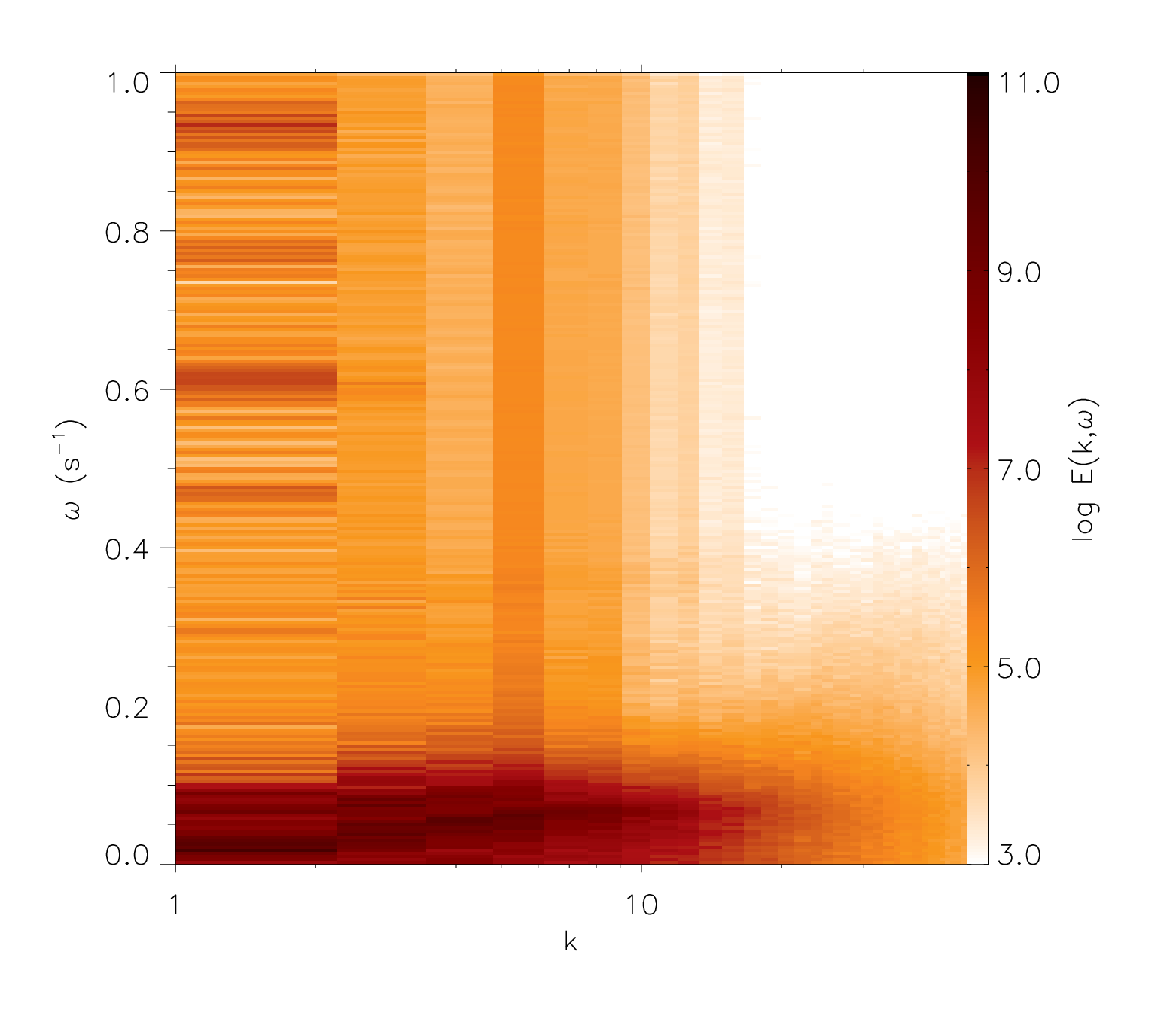}
  \caption{Mode diagrams for several radial positions in the oxygen shell burning
    model show the dominant spatial and time scales on which motions occur.
    The abscissa measures $k$ which is related to the wavenumber index $l$ of the mode
    by $l = 12\times k$.  The three locations shown here include:
    (left) Lower stable layer, just beneath the convective shell $r=0.4\times 10^9$ cm.
    (middle) Middle of convective shell, $r=0.6\times 10^9$.
    (right) Upper stable layer, just above the convective shell $r=0.9\times 10^9$ cm.
    \label{ob3d-komega}}
\end{figure*}

 For a given background structure, a spectrum of eigenmodes exist which are
solutions to the non-radial wave equation and boundary conditions. Individual modes
can be uniquely identified by a horizontal wavenumber index $l$ and oscillation
frequency $\omega$. In Figure \ref{ob3d-komega}, $l-\omega$ diagrams are presented
for the convection zone, and the two bounding stable layers. The individual
$l-\omega$ components have been isolated through Fourier transforms of a time
sequence of the simulation data.

\par Several modal components or "branches" can be identified in the stable layer
diagrams (left and right panels). These include: (1) p-modes, seen as a series of
points at the lowest $l$ values that extend to high frequencies; (2) g-modes, which
appear as ridges that are bound above by the buoyancy frequency; and (3) f-modes,
which appears as a ridge separating the g- and p-modes. The $f$-modes are interfacial
waves, and are most prominently seen in the lower-boundary diagram at a radius $r =
0.4\times 10^9$ cm.  The f-mode signature is due to interfacial waves running along
the convective boundary at $r\sim 0.43\times 10^9$cm, where there is a spike in
buoyancy frequency.

\par In the convection zone, the spectrum is dominated by power at low temporal and
spatial frequencies. This strong non-modal convection signature is also present,
though at lower amplitude, in the stable layers. This "turbulence" spectrum can be
seen extending from the lower left corner of the diagrams. This same feature was also
present in the simulations of He-shell burning by \cite{herwig2006}.

\section{CORE CONVECTION}

\par Are the hydrodynamic features of oxygen shell burning of more general applicability?
To investigate this, we examined core convection during hydrogen burning. Because of
the long thermal time scale for radiative diffusion in such stars, we focus on the
hydrodynamic behavior of a model in which the inner boundary provides a driving
luminosity about ten times larger than natural. This allows us to simulate the
convection with our compressible hydrodynamics code; an anelastic method (if
multi-fluid) would allow this to be done in the star's natural time scale. While our
calculation is not thermally relaxed on a Helmholtz-Kelvin time scale, it does relax
dynamically, and provides some clue as to the convective behavior.

\par We have previously evolved a 23 \msun \ star onto the main sequence with TYCHO, to
an age of 2.4\mult10$^5$ yrs, at which point hydrogen is burning in a convective
core. The model is then mapped onto the PROMPI hydrodynamics grid for simulation.
This model represents an early point in main sequence evolution, in which the core
hydrogen content has been depleted by only 1.7\% ($X_{core}$ = 0.689, $X_{init}$ =
0.701, $\Delta X$ = 0.012). The inner radius of the simulation was chosen such that
the convectively unstable region covers $\sim$1 pressure scale height (convective
cores are usually only of order a pressure scale height because of the divergence of
the scale height towards the stellar center). The entire domain covers $\sim$5
pressure scale heights and $\sim$3.3 density scale heights. The density contrast
across the computational domain is $\sim$30 with a contrast of $\sim$2 across the
convectively unstable region.

\par
\begin{figure*}
  \epsscale{0.8}
    \plotone{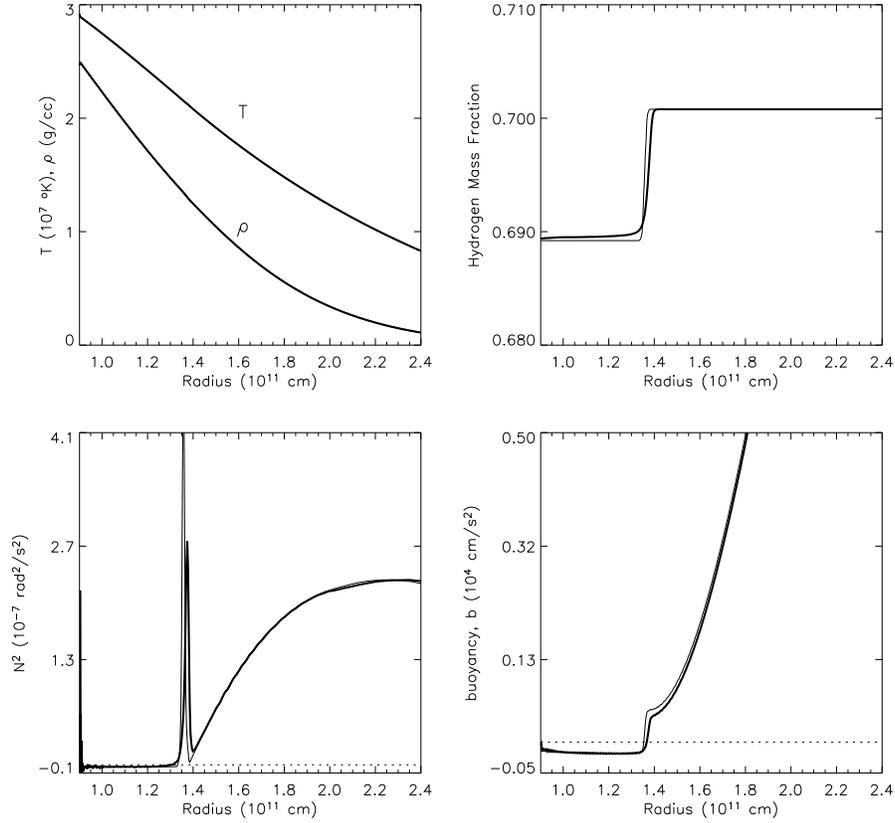}
  \caption{Radial profile of the simulated region for the main sequence core convection model. The thin
    lines show the initial conditions and the thick lines show the state of the 3D model at t = 10$^6$ s.
    (top left) Temperature and density.
    (top right) Hydrogen abundance.
    (bottom left) Squared buoyancy frequency.
    (bottom right) Buoyancy.\label{msc-profile}}
\end{figure*}

 The radial profile of the simulated region is presented in Figure \ref{msc-profile}
including the run of temperature, density, composition, buoyancy frequency, and
relative buoyancy. The entropy jump at the edge of the convective core, due to the
fuel-ash separation, gives rise to a buoyancy jump (spike in buoyancy frequency).

\par The equation of state for the main sequence model is well described by an
ideal gas with radiation pressure component.  The ratio of gas to total pressure lies
in the range $0.85 < \beta < 0.95$, with an increasing contribution from radiation
pressure as the stellar center is approach. A single composition representing
hydrogen has been evolved to keep track of nuclear transmutation and the mean
molecular weight of the plasma. A metalliicity of $Z=0.01879$ has been used to
represent the additional 175 species in the initial TYCHO model and helium is
calculated according to $Y=1 - (X+Z)$, where $X$ is the self consistently evolved
hydrogen mass fraction.

\par The luminosity due to nuclear burning in the computational domain is a small
fraction of the total stellar luminosity (2.4\%) which is dominated by burning in the
inner regions of the core and $L_{tot} = 7.8\times10^4 L_{\odot}$. Core burning is
incorporated into the simulation as an input luminosity at the inner boundary of the
computational domain.

\par The Kelvin-Helmholtz timescale for this model is $t_{KH}\sim10^5$ years,
which is many orders of magnitude longer than the dynamical timescales that are
feasible to simulate.  Additionally, the small luminosity of the star produces a
convective velocity scale that is very subsonic ($M\sim 10^{-3}$).  Since we are not
interested in the thermal relaxation of the model, we have boosted the input
luminosity by a factor of 10 to increase the velocity scale of the flow.  This was
necessary because our fully compressible code is limited by the sound crossing time.
(An anelastic or low-mach number method would be ideal for simulating this core
convection flow at the natural velocity scale.)

\par Radiation transport is treated in the diffusion limit. Opacities approximated
by Thompson scattering, which agrees well with the OPAL opacities \citep{ir96} used
in the 1D TYCHO model for the region simulated. The effects of radiative diffusion,
however, are found to be unresolved in the current simulation (with the diffusion
time across a single zone $\tau_{rad}=\Delta^2/k_{rad} \gg t_{conv}$, with grid zone
size $\Delta$) and therefore energy transport in the convection zone occurs primarily
on the subgrid level due to numerical diffusion. This is a high P\'eclet number
simulation.

\par A 2D and a 3D model have been calculated. The simulated wedges have angular
extents of 30$^{\circ}$ in both the polar and azimuthal directions and are centered
on the equator to avoid zone convergence problems near the poles. This minimal
angular domain size was chosen by calculating models of increasing angular size in 2D
domains until the flow pattern converged. The angular domain size used in the present
simulations encompasses a large convective roll in 2D. Smaller 2D domains were found
to distort the convective roll while domains larger by integer multiples contained
proportionally more rolls of the same flow amplitude and morphology. The boundary
conditions in the radial direction are reflecting and stress free, and periodic
conditions are used in both angular directions. The grid zoning, domain limits, and
simulation run times are summarized in Table \ref{msc-table} for the 2D and 3D
models.

\subsection{Quasi-Steady Core Convection \label{msc-qss}}

\par
\begin{figure*}
\plotone{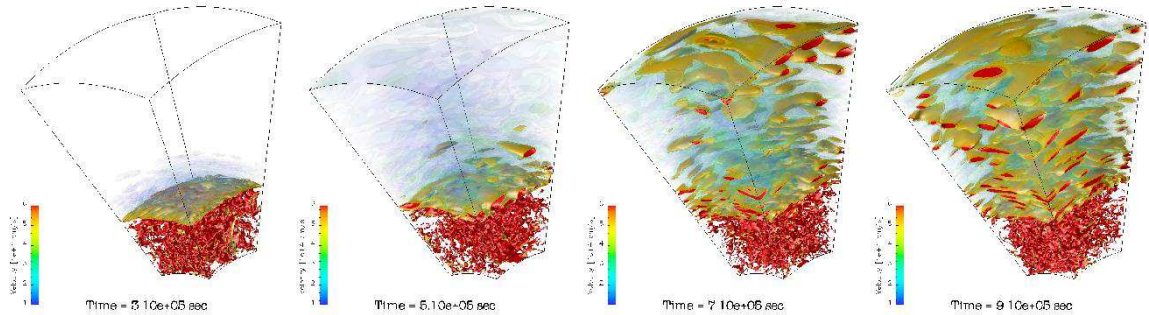}
  \caption{Velocity isocontours show the development of the flow in the 3D core convection
    model. The turbulent convective flow excites internal waves which radiate into the
    overlying stably stratified layer.  By the end of the time sequence shown
    the stable layer cavity is filled with resonant modes.
    \label{msc-3d-isov-init}}
\end{figure*}

\begin{figure*}
    \plottwo{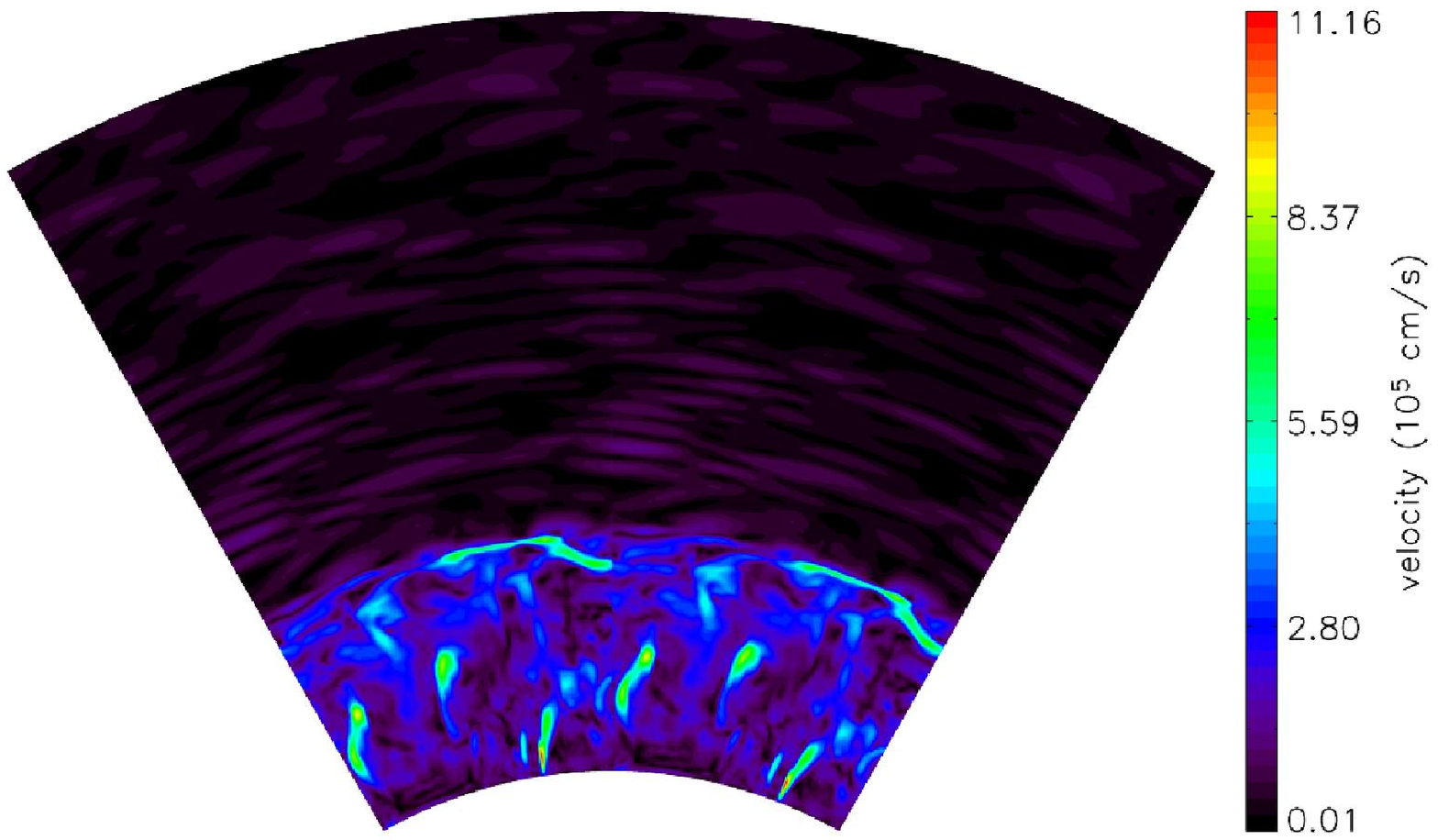}{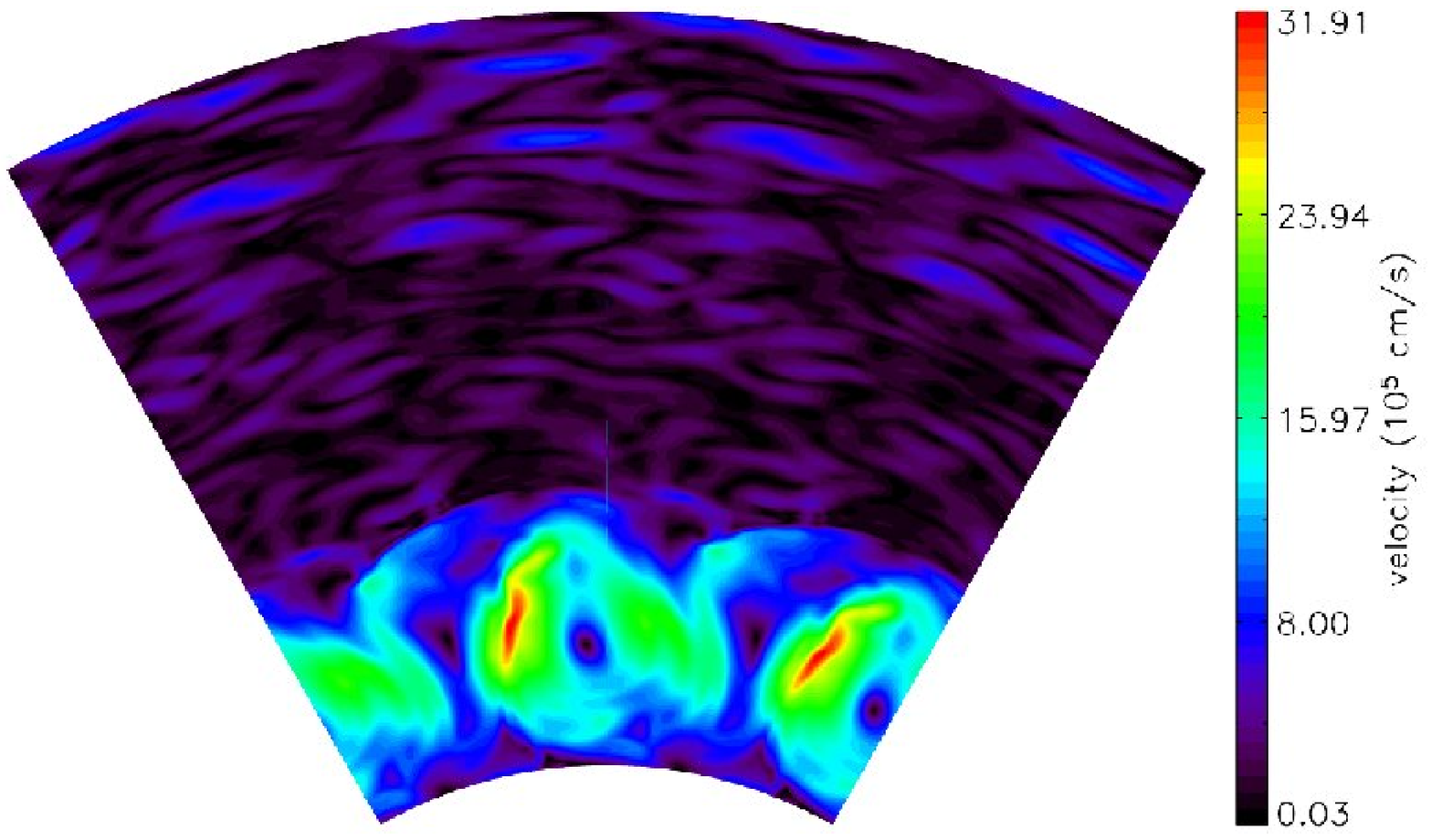}
  \caption{The velocity magnitude for the core convection model at t=10$^6$ s:
    (left) a slice through the 3D model; and (right) the 2D model.
    The topology of the convective flow is significantly different
    between 2D and 3D models: the 3D convective flow is dominated by small plumes and eddies
    while the 2D flow is much more laminar, and dominated by a large vortical eddies which span
    the depth of the layer.  The wave motions in the stable layer have similar morphology in 2D
    and 3D, but the velocity amplitudes are much larger in 2D.
    The computational domains have been tiled once in angle for presentation.
    \label{2d-3d-vtot}}
\end{figure*}

 Convection is initiated through random low-amplitude ($0.1\%$) perturbations
in density and temperature applied as in the oxygen shell simulation. In order to
save computing time, the 3D model was initiated on a domain one quarter as large in
azimuthal angle which was then tiled four times in angle once convective plumes began
to form. The initial development of the flow in the 3D model is presented in Figure
\ref{msc-3d-isov-init} as a time sequence of velocity isosurfaces. The turbulent
structure of the convective flow, as well as the excitation of internal waves which
radiate into the overlying stably stratified layer, are clearly illustrated. A
comparison of the flow morphology between the 2D and 3D models is presented in Figure
\ref{2d-3d-vtot}. The 2D convective flow is much more organized and laminar, and is
dominated by a single large convective cell while the 3D convection is composed of
many smaller scale plume-like structures and is more obviously turbulent. In both
models the stably stratified regions are rife with internal waves excited by the
convection.

\par
\begin{figure*}
  \epsscale{0.5}
    \plotone{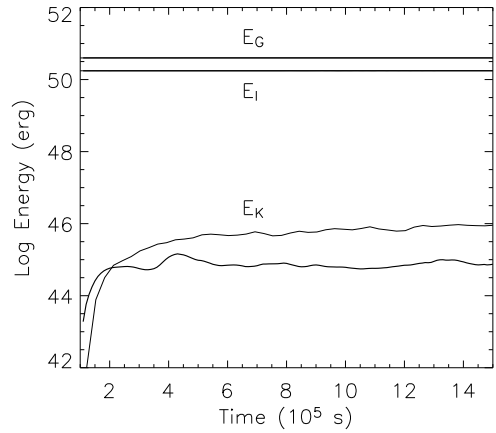}
  \caption{The time evolution of the energy budget for the main sequence core convection
    models: the (thick line) 3D model; and (thin line) the 2D
    model are shown. The energy budget includes the total internal energy $E_I$, gravitational
    energy $E_G$, and kinetic energy $E_K$ on the computational grid.
    \label{teint-msc}}
\end{figure*}

 The 3D convective flow attains a quasi-steady character after approximately
6\mult10$^5$ s, or approximately two convective turnovers. The evolution of the
internal, gravitational, and total kinetic energy components on the computational
grid for the 2D and 3D models, are presented in Figure \ref{teint-msc} and are
calculated in the same way as for the oxygen burning model.

\par In both the 2D and 3D models, the total kinetic energy fluctuates in times
with excursions from the mean as larger $\delta E_K/\overline{E_K} \sim 0.4$ in 3D
and $\delta E_K/\overline{E_K}\sim 0.6$ in 2D. The kinetic energy in the 2D model
grows on a slightly longer timescale, and achieves a steady character after $\sim
10^6$ s, at which time the kinetic energy growth rate tapers off. The total energy is
conserved to better than $\sim$0.2\% for both the 2D and 3D flows by the end of the
calculation.

\par
\begin{figure*}
\epsscale{1.0}
    \plottwo{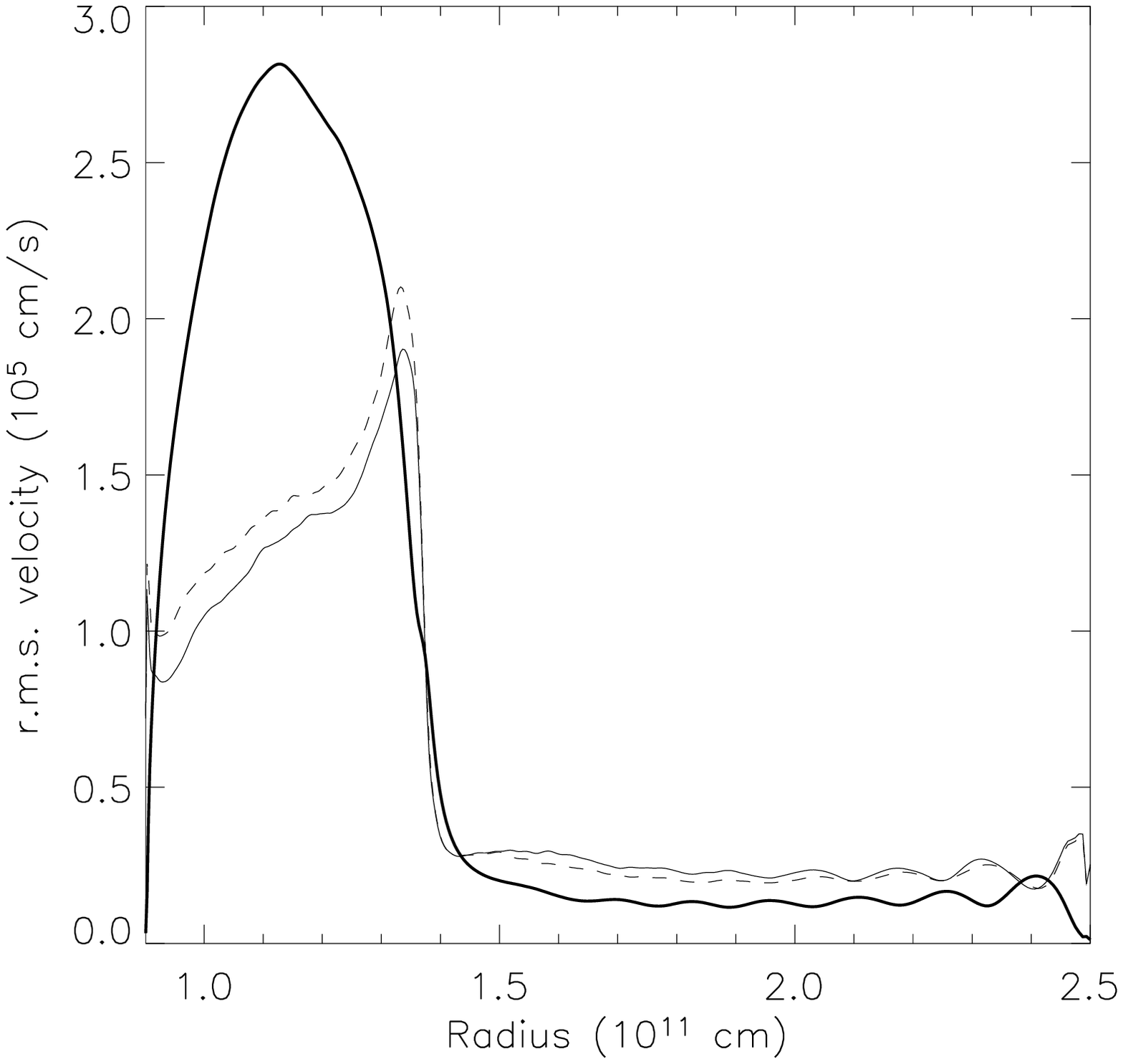}{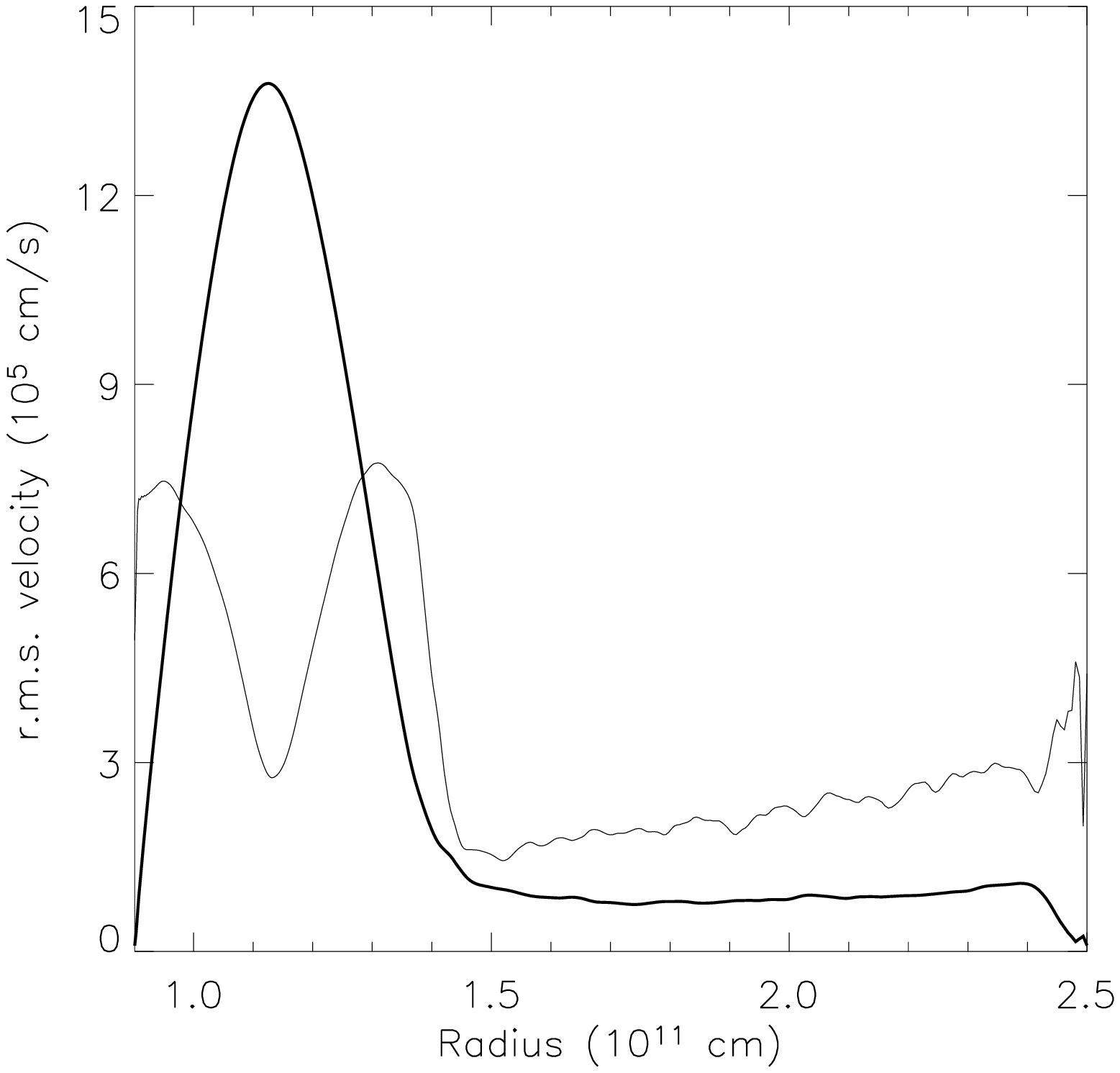}
  \caption{The r.m.s. velocity fluctuations for the core convection model:
    (left) the 3D model, and (right) the 2D model.
    In each plot, the thick line indicates the radial velocity component
    and thin line is used to indicate horizontal velocity components, with the dashed
    line used to show the polar angle component in the 3D model.
    \label{msc-2d-3d-vrms}}
\end{figure*}

 The r.m.s. velocity fluctuations are presented in Figure \ref{msc-2d-3d-vrms}
for the 2D and 3D models.  The resultant flows in both the 2D and 3D models are
similar to that found for the oxygen shell burning model. The velocity amplitudes are
higher in 2D by a factor of $\sim$5 (see axis scale in Figure \ref{msc-2d-3d-vrms}),
and the flows are dominated by large eddies spanning the depth of the convective
region.  The horizontal deflection of matter is also found to occur in a much
narrower region in the 3D model.  The hard-wall lower boundary of the 3D model is
characterized by an even narrow horizontal flow, probably due to the absence of a
stable layer which is host to g-modes.

\par The time averaged convective flow velocity for the 3D model is
$v_c\approx 2.8\times 10^{5}$ cm/s. The turnover time is $t_c = 2\Delta R/v_c \approx
3.2\times 10^5$ s, and the simulation spans approximately 5 convective turnovers. The
peak velocity fluctuation is $v_{peak}\sim 2\times 10^6$ cm/s, corresponding to a
peak Mach number of $M\sim$ 0.03, and the maximum density fluctuations within the
convective flow are $\sim$ 0.02\%, which is of order $M^2$ as expected for low Mach
number thermal convection \citep{gough69}.  The time averaged convective flow
velocity in the 2D model is $v_c\approx 1.3\times 10^{6}$ cm/s, and the convective
turnover time for this model is $t_c \approx 7\times 10^4$ s. The simulation spans
$1.5\times 10^6$ s which is $\sim$21 convective turnovers. The peak velocity
fluctuation in the 2D model is comparable to that in the 3D simulation, with
$v_{peak}\sim 2\times 10^6$ cm/s and the peak density fluctuations is a little more
than twice that found in the 3D model, $\sim$0.05\%. The turnover times and
convective velocity scales are summarized in Table \ref{msc-table}.

\subsection{The Stable Layer Dynamics Overlying the Convective Core}
\label{msc-stable-section}

\par As in the oxygen shell burning model, the stably stratified layers in the core
convection models are characterized by velocity fluctuations throughout. Similar to
shell burning, the 2D stable layer velocity amplitudes are larger and have a smaller
radial to horizontal component ratio $v_r/v_{\perp}\approx \omega/N$ compared to the
3D flow.

\par The stable layer motions in the core convection simulation are
predominantly resonant modes, which compare well to the analytic eigenmodes of the
linearized wave equation, and are analogous to those discussed for the oxygen shell
burning model.  The region outside the convective core will act as a resonant cavity,
with the outer boundary at the location where the buoyancy frequency and Lamb
frequency cross.

\par
\begin{figure*}
\epsscale{0.5}
    \plotone{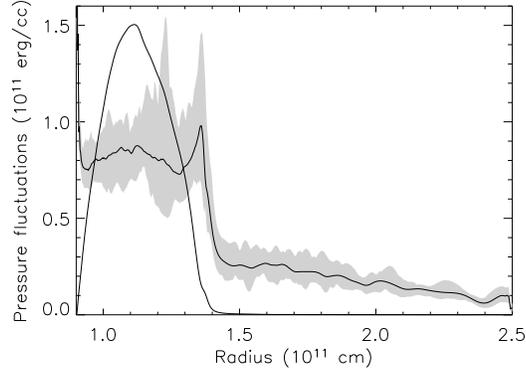}
  \caption{Pressure fluctuations in core convection model:
    The time averaged horizontal r.m.s. pressure fluctuations
    are shown as the thick line, with extreme values over two convective turnovers indicated
    by the shaded region.  The thin line shows the radial component of the turbulent ram pressure
    $\rho v_r^2$ averaged over a convective turnover.  At the upper boundary, the
    curves cross at a point where the turbulent pressure is balanced by the wave induced
    pressure fluctuations in the stable layer.  This crossing point is coincident with the location
    of the convective boundary.  The pressure perturbations at the lower boundary are due to
    the input luminosity which drives the convective flow.
    \label{msc-ppert}}
\end{figure*}

  The amplitudes of the internal waves excited will

 be
determined by the ram pressure of the turbulence at the convective boundary. In
Figure \ref{msc-ppert} the ram pressure and horizontal r.m.s. pressure fluctuations
are presented for the 3D model, and can be seen to balance at the interface between
the convective core and the stably stratified layer. Using this condition of pressure
balance, \citet{ma06b} estimate the amplitudes of the excited internal wave
velocities and the induced thermodynamic fluctuations and find this to be in good
agreement with the oxygen shell burning simulations. The relationship between the
density fluctuations, the convective velocity scale, and the stellar structure (i.e.,
$N$ and $g$) was found to be,

\begin{equation}\label{rhopert-equation}
\frac{\rho'}{\havg{\rho}} \sim M_c^2 + \frac{v_c N}{g}.
\end{equation}

\begin{figure*}
\epsscale{1.0}
  \plottwo{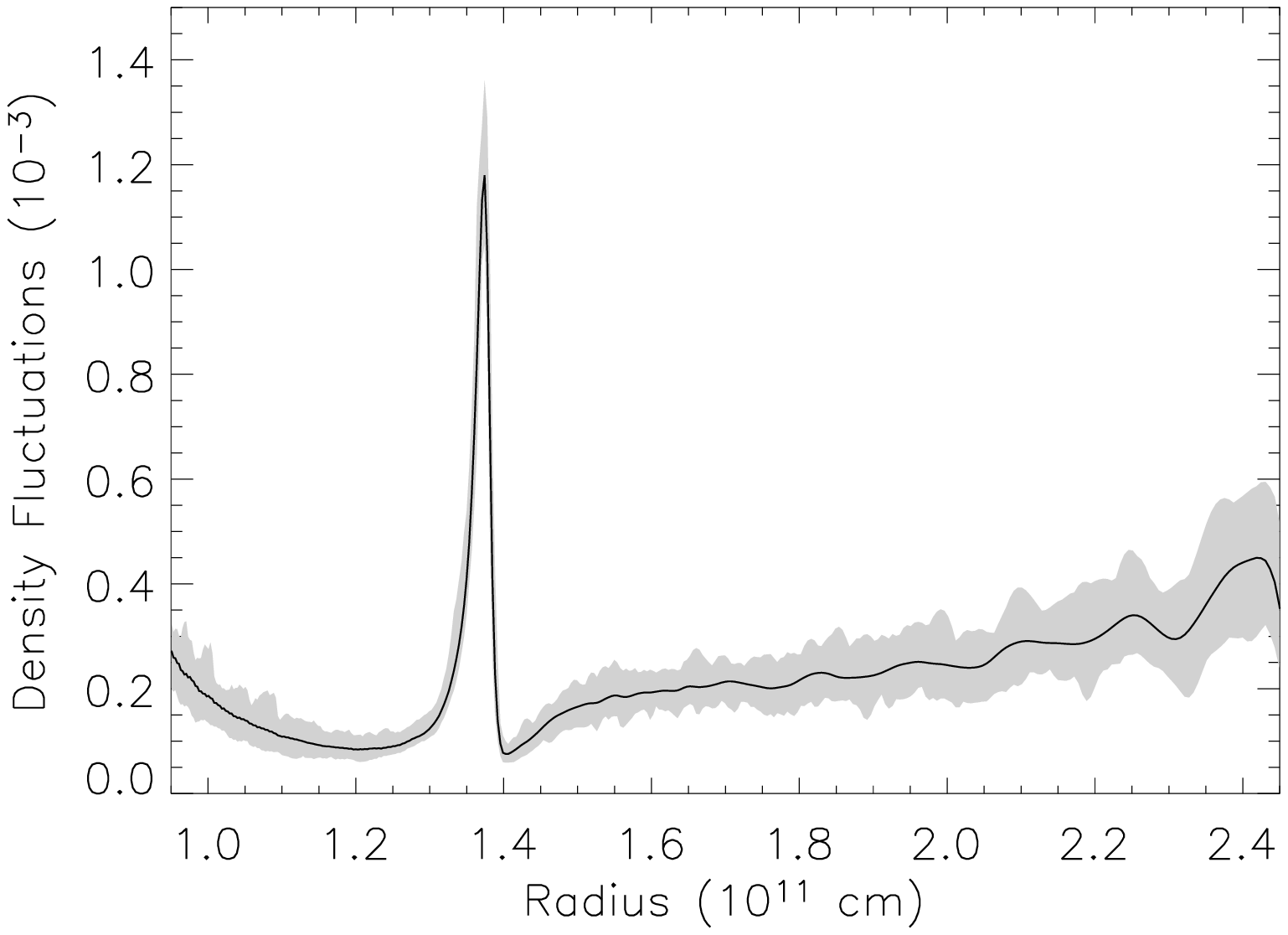}{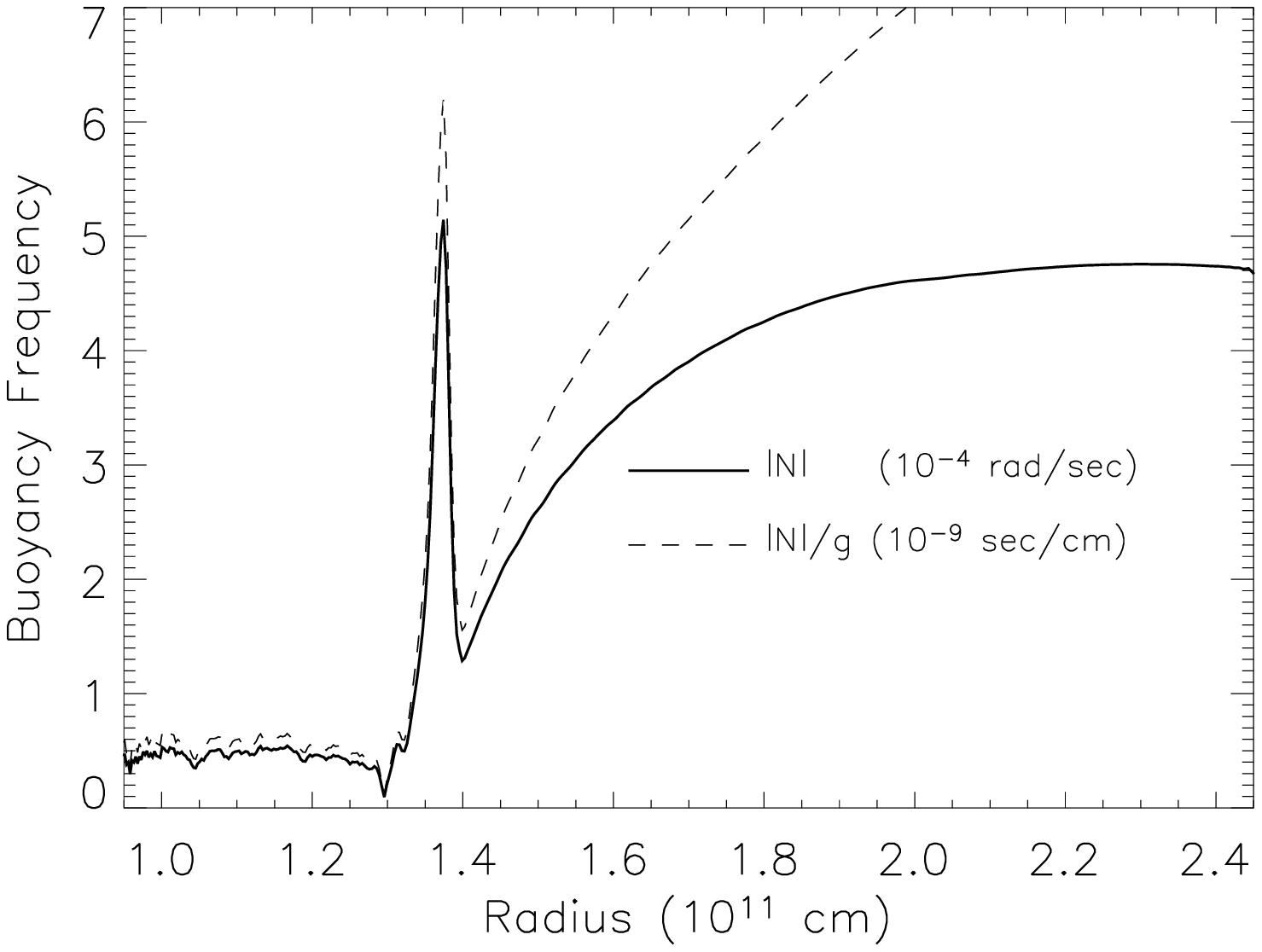}
  \caption{(left) Density fluctuations in the 3D core convection model:
    The time averaged maximum density fluctuation is shown as
    the thick line, with extreme values for the averaging period
    (two convective turnovers) shown by the shaded region.  The
    largest fluctuations occur at the interface between the turbulent
    convective region and the stably stratified layer. The maximum
    fluctuation at the interface is $\rho'/\havg{\rho}\sim$0.12\%.
    (right) The buoyancy frequency is shown in units of ($10^{-4}$ rad/sec).
    Also shown by the dashed line is the buoyancy
    frequency normalized by the gravity which sets the scale of the
    density fluctuations at the convective boundary through
    equation \ref{rhopert-equation}. The expected density fluctuation is
      $\rho'/\havg{\rho}   \sim v_c |N|/g \sim 0.12$\%, where a velocity
      scale of $v_c \sim 2\times 10^5$ cm/s has been used
      (see Figure \ref{msc-2d-3d-vrms}).\label{msc-rhopert}}
\end{figure*}

\noindent That this proportionality holds in the core convection model as well, where
fluctuation amplitudes are lower than those in the oxygen shell burning model by an
order of magnitude, is illustrated in Figure \ref{msc-rhopert}, which presents the
buoyancy frequency and density fluctuation profiles for the boundary region. The
measured density fluctuation and the value calculated according to equation
\ref{rhopert-equation} compare remarkably well, with $\rho'/\havg{\rho}\sim$ 0.12\%.

\section{Simulations and Mixing Length Theory}\label{ob-mlt}

\par In this section we compare our 3D oxygen shell burning simulation
results to the mixing length theory of convection. We choose to compare this
particular simulation since it represents the most physically complete model in our
suite of calculations, both in terms of dimensionality and thermal evolution. Unless
otherwise specified, the time period over which averaging is performed on the
simulation data is $t\in[400,800]s$, which is approximately 4 convective turnovers.
We find that this period is sufficiently long compared to the time evolution of the
flow that average values are not affected appreciably by increases in the averaging
time period.

\subsection{Mixing Length Theory Picture}
\par The basic picture underlying the mixing length theory, which is
the standard treatment of convection used in one-dimensional stellar evolution
modeling \citep[see][]{cg68,clay83,kw90,hansen}, is one in which large eddies are
accelerated by an unstable temperature gradient and advect across a certain distance
until they suddenly lose their identity by turbulently mixing with the {\em
background}. Energy is transported through this process because the envisioned
turbulent {\em blobs} which are moving radially outward have a higher entropy at
their formation location than the location in which they {\em dissolve}. The vertical
extent over which large eddies retain their identity as they advect through a
convection zone is a fundamental parameter in the mixing length theory. This {\em
mixing length} $\Lambda$ is generally taken to be a multiple of the local pressure
scale height $\Lambda = \alpha_{\Lambda} H_p$.

\par Within this physical picture, the mixing length theory develops a
relationship between the convective flux, the temperature gradient, the velocity
scale, and the geometrical factors which describe the large scale eddies.
The starting point in mixing length theory is the radial enthalpy flux, which is
written in terms of fluctuations in the flow properties, and is taken to be (assuming
a horizontally isobaric flow),

\begin{equation}
F_c = v_c\rho c_P T'.
\end{equation}

\noindent The temperature fluctuations in mixing length theory are related to the
temperature gradient and the distance traveled by,

\begin{equation}
\label{mlt-tpert-eq} T'/T  = \Big(\frac{\partial\ln T_e}{\partial r}
-\frac{d\ln T_0}{d r}\Big)\frac{\Lambda}{2}
 =(\Delta\nabla)\frac{1}{H_p}\frac{\Lambda}{2}
\end{equation}

\noindent where the subscript "e" indicates properties of the large convective eddies
and the dimensionless temperature gradient $\nabla$ is used (see
\S\ref{transient-section}), and the difference between the gradient in the eddy as it
moves and the averaged stellar background is written,

\begin{eqnarray*}
\Delta \nabla = (\nabla - \nabla_{e}).
\end{eqnarray*}

\noindent The factor of 1/2 in equation \ref{mlt-tpert-eq} represents the idea that
on average the large convective eddies have traversed about half a mixing length
before reaching the current position. The velocity obtained by the convective eddy is
computed by calculating the work done by the buoyancy force over a mixing length,

\begin{equation}\label{mlt-v-eq}
v_c^2 = g\beta(\Delta\nabla)\frac{\Lambda^2}{8 H_p}.
\end{equation}

\noindent Here again, the eddy is assumed to have been accelerated over half of a
mixing length and an additional factor of 1/2 is incorporated on the right hand side
to account for energy lost driving other flows, such as small scale turbulence and
horizontal motions (e.g., note that the r.m.s. horizontal velocity is of the same
order as the r.m.s. radial velocity in the simulation). The average convective flux
can then be written,

\begin{equation}
\label{mlt-fenth-eq} F_c = \rho c_p T
\sqrt{g\beta}\frac{\Lambda^2}{4\sqrt{2}H_p^{3/2}}(\Delta\nabla)^{3/2}.
\end{equation}

\par The temperature gradient for the convecting material is found
by assuming that eddies follow isentropic trajectories $\nabla_e = \nabla_{ad}$.
Deviations from isentropic motion have been considered in the mixing length theory.
In the case of strong radiative diffusion losses, the eddy geometry (in terms of the
surface area to volume ratio) is an important additional parameter since the eddies
are envisioned to {\em leak} a fraction of their thermal energy over a mixing length
distance. When local cooling dominates, either through radiative losses (in optically
thin regions) or neutrino losses (such as in the present model), the geometry of the
eddies is not important since energy escapes everywhere from the large eddies, not
just at eddy "surfaces".

\par During the oxygen shell burning simulations being considered here,
non-adiabatic losses are small over a convective turnover time and the convection is
expected to be "efficient". A quantitative measure of convective efficiency is the
P\'eclet number, which is the ratio of the energy loss timescale to the convective
turnover timescale for the large eddies. In the current model, energy losses are
dominated by neutrino cooling $\epsilon_{\nu}$. Therefore, we calculate an effective
P\'eclet number using the following convective and neutrino-cooling timescales:

\begin{equation}
\tau_c \sim \frac{H_p}{v_c}
\end{equation}

\begin{equation}
\tau_{\nu} \sim \frac{c_p T'}{T'\partial\epsilon_{\nu}/\partial
T}=\frac{c_p T}{\epsilon_{\nu}}\Big(\frac{\partial\ln\epsilon_{\nu}}{\partial\ln T}\Big)^{-1}
\end{equation}

\begin{equation}
Pe =\frac{\tau_{\nu}}{\tau_c} \sim\frac{v_c c_p
T}{H_p\dot{\epsilon_{\nu}}}\Big(\frac{\partial\ln\epsilon_{\nu}}{\partial\ln
T}\Big)^{-1} \\ \sim
10^4\Big(\frac{\partial\ln\epsilon_{\nu}}{\partial\ln
T}\Big)^{-1}\label{peclet-eq}
\end{equation}

\noindent where characteristic values from the simulation have been used in equation
\ref{peclet-eq}, and the temperature dependence of the neutrino loss rates is
$(\partial\ln\epsilon_{\nu}/\partial\ln T)\lesssim 9$. Therefore, the P\'eclet number
for the convection is P\'e $\gtrsim 10^3$, and we should expect the convection zone
to be very nearly isentropic.

\subsection{The Enthalpy Flux, Background Stratification, and Temperature
and  Velocity Fluctuations}\label{alphas-section}

\par
\begin{figure*}
\epsscale{0.9}
\plottwo{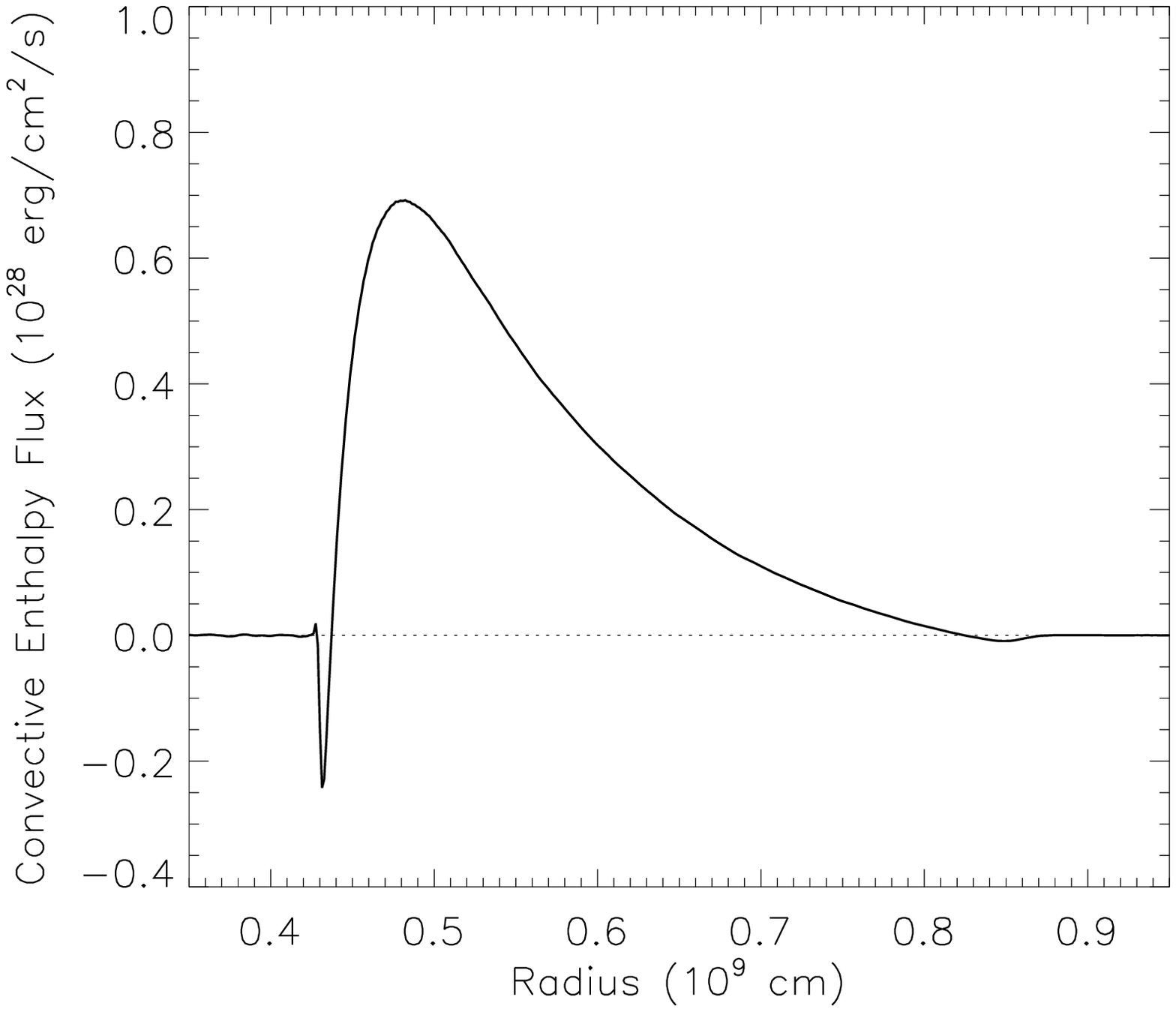}{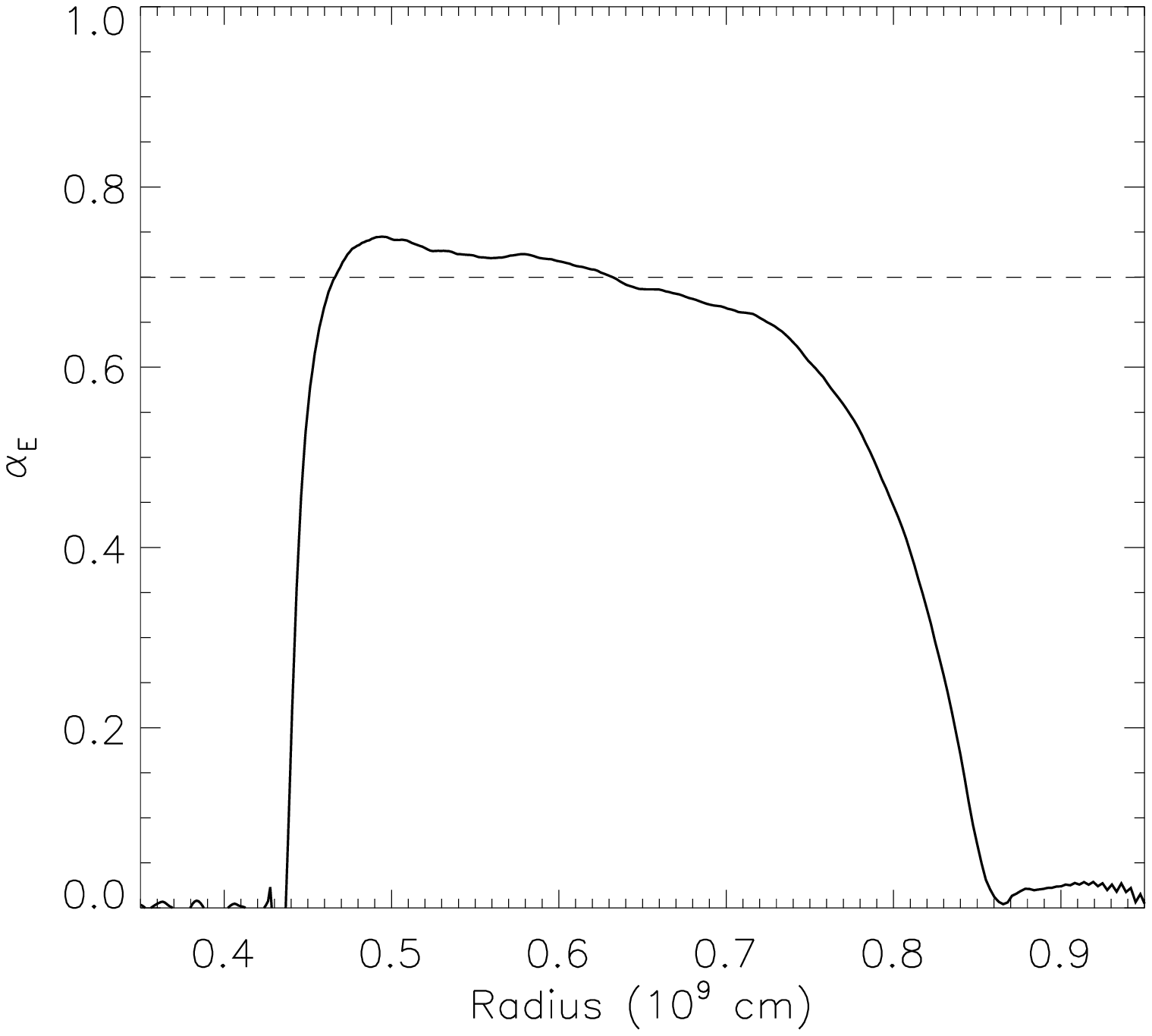}
 \caption{(left) Convective enthalpy flux, $F_c = \avg{\rho c_p v_r T'}$.
    (right) Temperature-velocity correlation function $\alpha_E$ calculated according to
    equation \ref{alphae-eq}, with mean value $\havg{\alpha_E} = 0.7$ shown by the dashed line.
    \label{fenth-fig}}
\end{figure*}

 The convective enthalpy flux measured in the simulation is
presented in Figure \ref{fenth-fig}. The spike at the bottom of the convection region
and the slight dip at the top reflect the braking of convective motion at these
boundaries. The enthalpy flux is calculated by performing time and horizontal
averages on the flow.  Mixing length theory, however, makes the assumption that the
velocity and temperature fluctuations are perfectly correlated, so that horizontal
averaging of fluctuations is comparable to products of the averages.  This is not
neccesarily the case.  To test the degree to which the velocity and temperature
fluctuations are correlated we can calculate the correlation coefficient, $\alpha_E$,
defined in the following way:

\begin{equation}
\label{alphae-eq} F_c = \havg{\rho c_p T' v'_c} = \alpha_E\havg{\rho
c_p}\havg{T'}\havg{v'_c}.
\end{equation}

\noindent The fluctuations, $T'$ and $v'_c$ in equation \ref{alphae-eq} are taken to
the be the r.m.s.\ fluctuations in the simulation. The radial profile of $\alpha_E$
is shown in Figure \ref{fenth-fig}.  We find $\havg{\alpha_E} = 0.7\pm 0.03$ averaged
over the radial interval $r\in [0.5,0.75]\times 10^9$ cm within the convection zone.
A value smaller than unity indicates that the horizontal distribution of temperature
and velocity fluctuations in the flow are not perfectly correlated. The degree of
correlation, however, is found to be fairly uniform throughout the convection zone.

\par
\begin{figure*}

\epsscale{0.9}
\plottwo{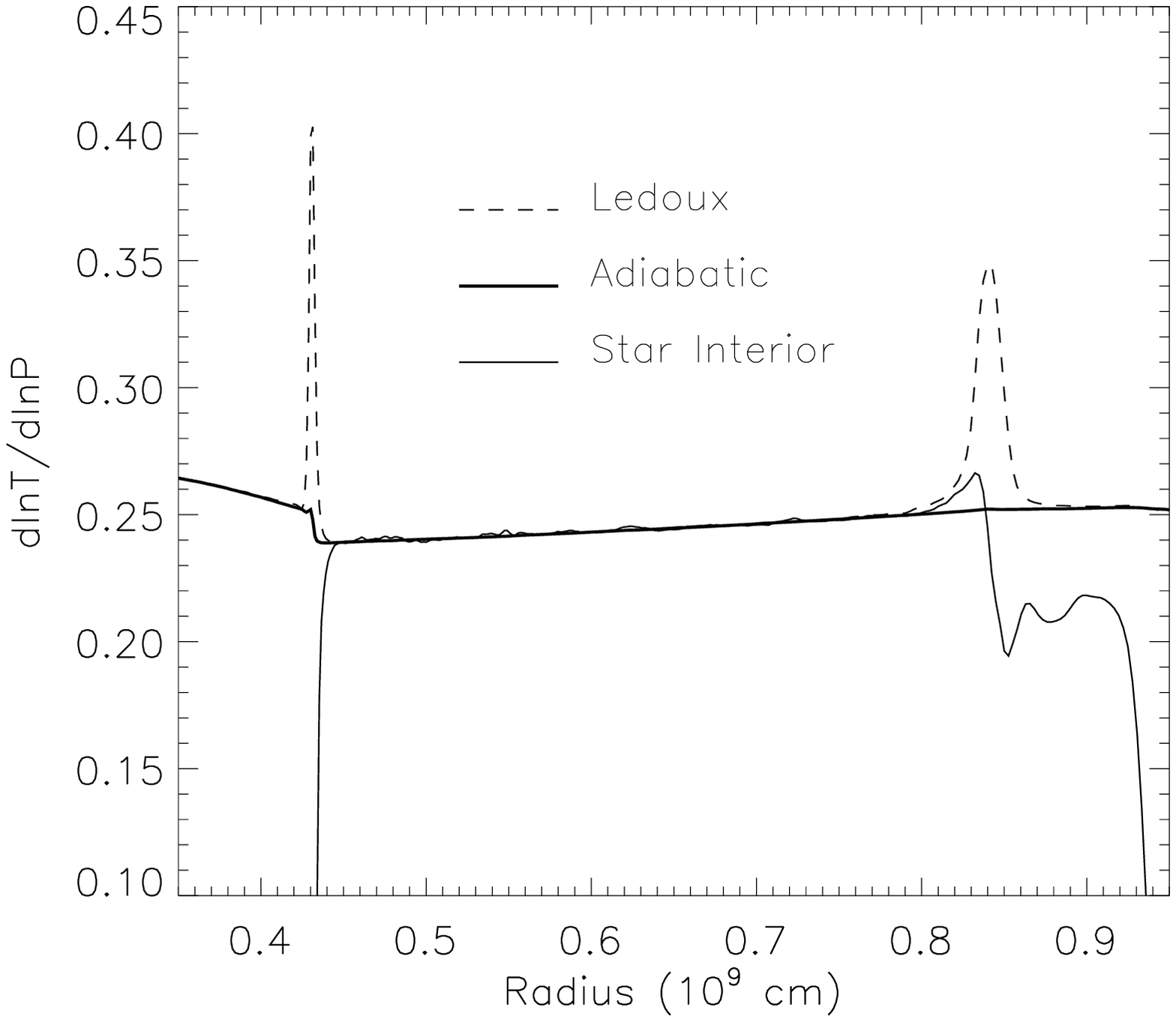}{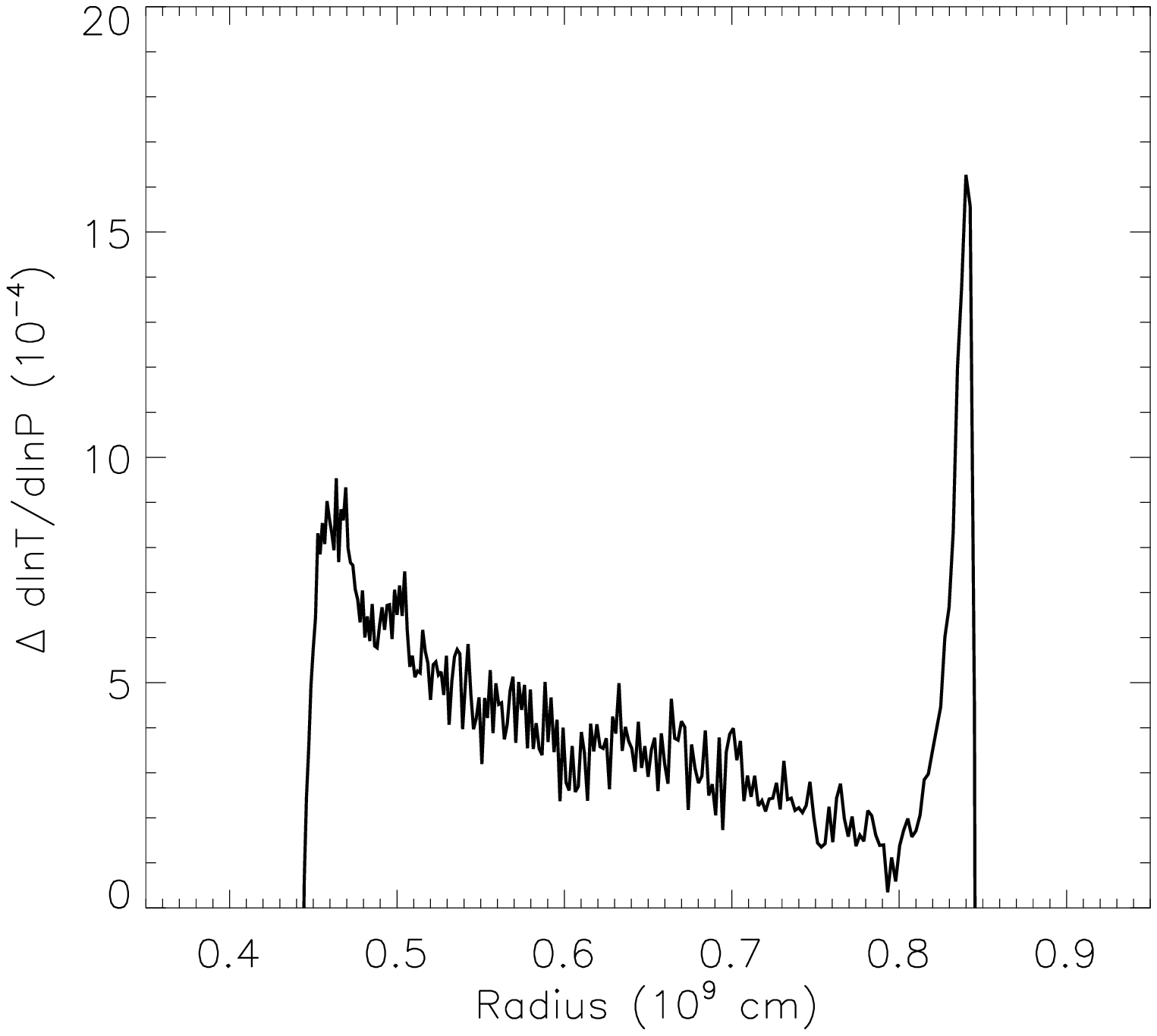}
 \caption{(left) Dimensionless temperature gradients:
the stellar interior $\nabla_s$; adiabatic $\nabla_{ad}$; and Ledoux $\nabla_{led}$
gradients are shown. (right) Super-adiabatic temperature gradient horizontally and
time averaged. \label{nablas-fig}}
\end{figure*}

 We next consider how well the velocity and temperature
fluctuations are correlated with the local temperature gradient. In Figure
\ref{nablas-fig} the temperature gradient of the horizontally averaged hydrodynamic
model profile $\nabla_s$ as well as the adiabatic $\nabla_{ad}$ and the
composition-corrected (Ledoux) gradient $\nabla_{Led} = \nabla_{ad} +
\varphi/\beta\times\nabla_{\mu}$, are presented.
The super-adiabatic temperature profile of the stellar background $\Delta\nabla =
\nabla_s - \nabla_{ad}$ is presented in the right panel of Figure \ref{nablas-fig}.
While the convection zone is found to have a super-adiabatic profile throughout, it
is very small ($\Delta\nabla \lesssim 10^{-3}$). This confirms the efficiency of the
convection, in accord with our estimate for P\'e. Stability is maintained in the
upper boundary layer by the composition gradient, where we have
$\nabla_{ad}<\nabla_s<\nabla_{Led}$.

\par In order to assess the validity of the mixing length theory temperature and velocity
fluctuation amplitudes given by equations \ref{mlt-tpert-eq} and \ref{mlt-v-eq} we
calculate the correlation coefficients $\alpha_T$ and $\alpha_v$ which are defined
by,

\begin{equation}\label{alphat-eq}
T'/T %
=(\Delta\nabla){\alpha_T}
\end{equation}

\noindent and

\begin{equation}\label{alphav-eq}
v_c = \frac{\alpha_v}{2} \sqrt{g\beta(\Delta\nabla)H_p}.
\end{equation}

\par An important question concerns how to interpret and measure the temperature and
velocity fluctuations $T'$ and $v_c$ in the simulations for comparison to mixing
length theory. In the mixing length theory, these fluctuations are identified with
the properties of large eddies. Therefore, a direct comparison would entail isolating
the large eddies from the rest of the flow, and measuring their properties.  In lieu
of this complicated procedure we identify the fluctuations in the large eddies with
two distinct quantities for comparison: (1) the r.m.s. fluctuations in the flow; (2)
the difference between the horizontally averaged background value and the mean values
in the up and down flowing material.

\par
\begin{figure*}

\epsscale{0.9}
\plottwo{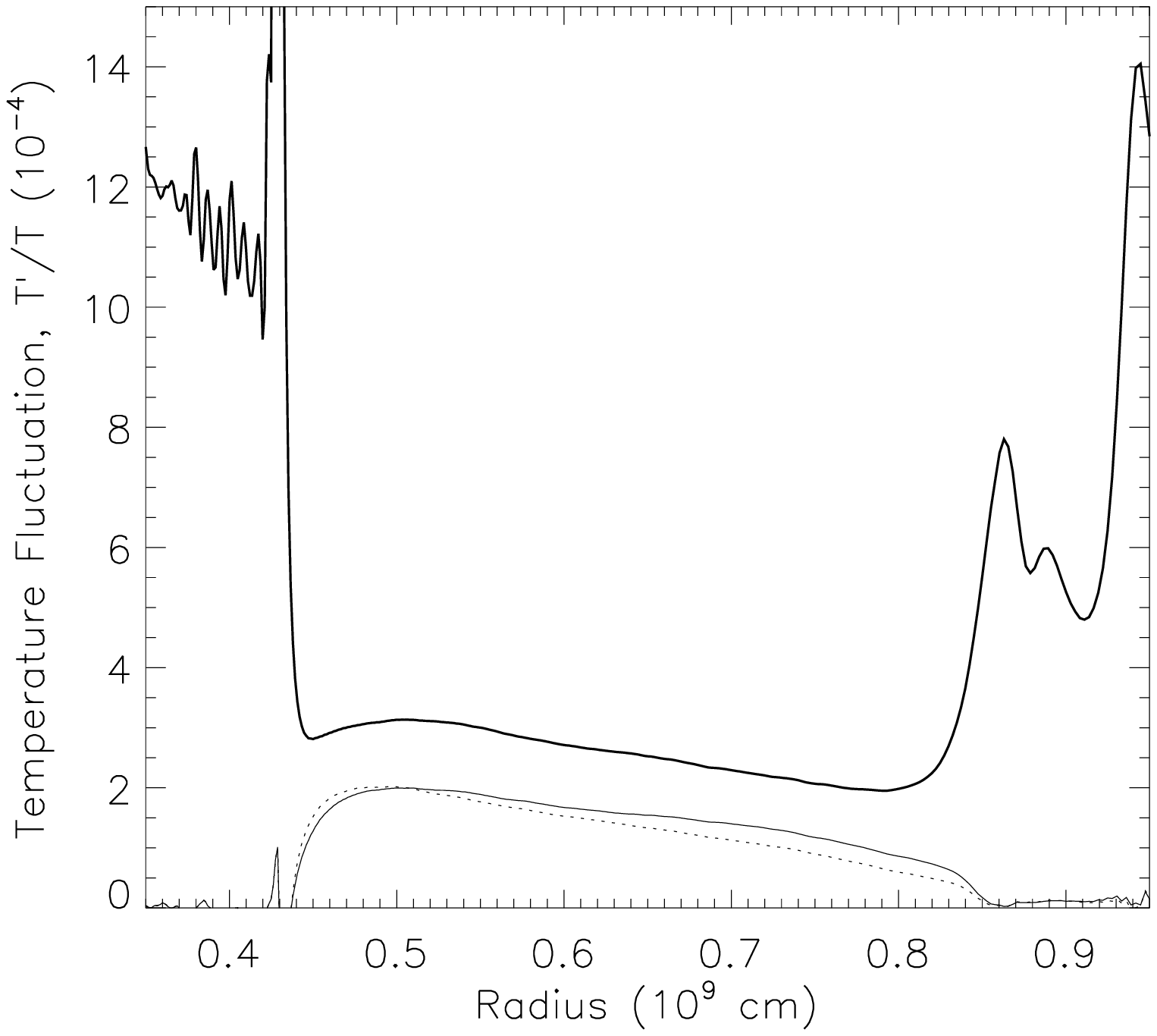}{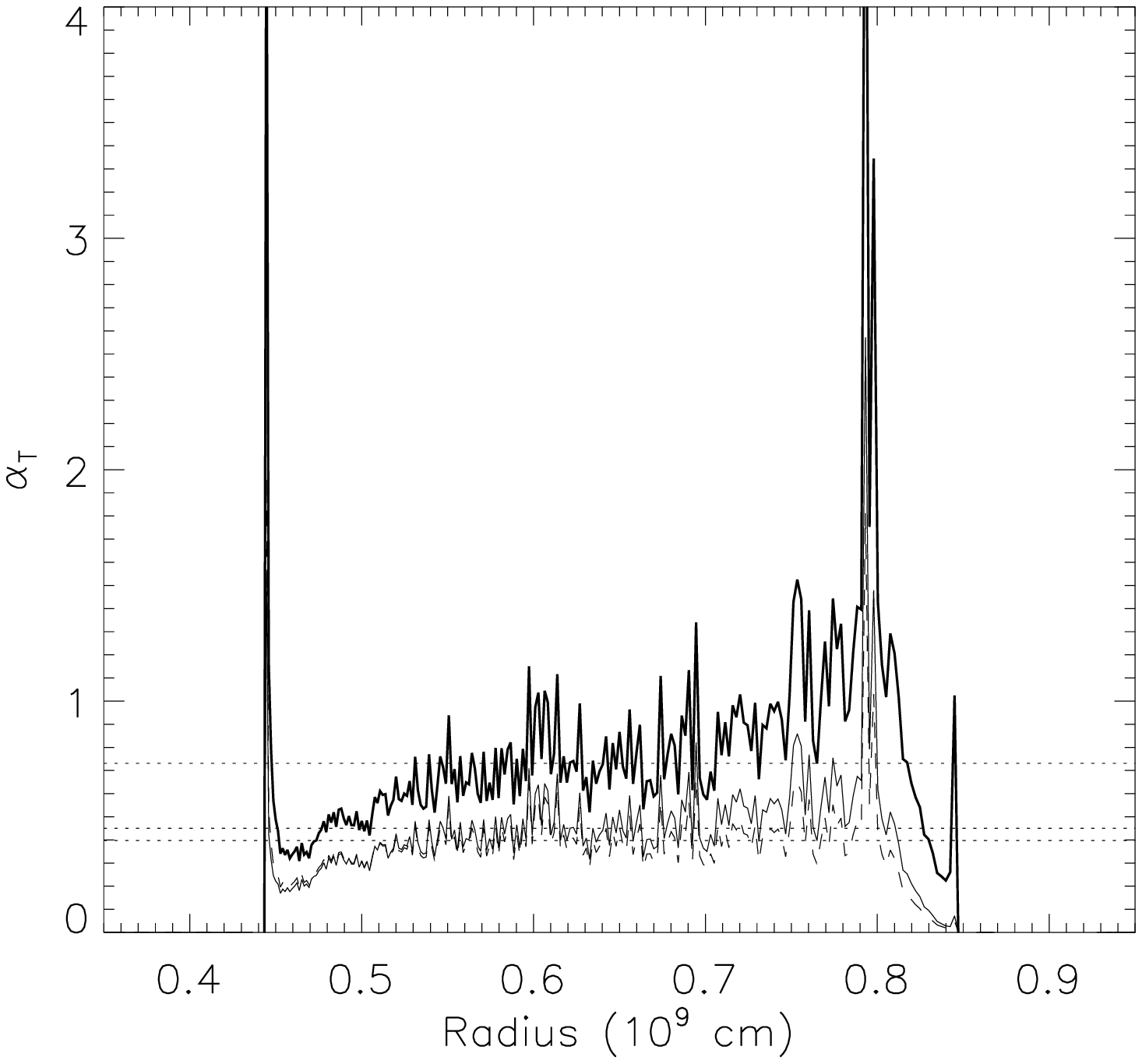}
 \caption{(left) Time averaged r.m.s. temperature fluctuations: (thick solid)
    line shows the r.m.s. fluctuations; the (thin solid) and (thin dotted) lines show
    the fluctuations in the upward and downward directed flow components, respectively.
    (right) The radial dependence of the "thermal mixing length" parameters $\alpha_T$
    defined by equation \ref{alphat-eq} are shown the temperature fluctuations presented in
    the left panel, using the same line types. The mean values, averaged over
    $r\in[0.5,0.75]\times 10^9$cm are shown by the thin dotted lines.\label{tpert-fig}}
\end{figure*}

 The temperature fluctuations calculated using these two methods are
presented in Figure \ref{tpert-fig}.  The temperature fluctuations in the convection
zone follow a trend similar to the super-adiabatic gradient, i.e., decreasing with
increasing radius. In the right panel of Figure \ref{tpert-fig} the radial profile of
$\alpha_T$ is shown using both definitions of the fluctuations. The nonzero
temperature fluctuations outside the convective region are due to distortions in
stable layers due to convective buoyancy braking \cite{ma06b}; the use of separate up
and down flows is cleaner, eliminating these. The slope in the temperature
fluctuation profiles are slightly overcompensated for by the super-adiabatic gradient
when forming the ratio $\alpha_T$.  Within the scatter, however, $\alpha_T$ is fairly
well represented by a constant value. The mean value within the body of the
convection zone (taken to be $r\in[0.5,0.75]\times 10^9$ cm) is larger for the r.m.s.
fluctuations $\havg{\alpha_T(\hbox{rms})} =0.73$ compared to
$\havg{\alpha_T(\hbox{up})}= 0.45$ and $\havg{\alpha_T(\hbox{down})} = 0.40$.  The
largest departures from the mean, within the convective region, occur at the base of
the convection zone, $r\la0.52\times 10^9$ cm, where the nuclear flame is driving the
convective flow. The departures are also large at the top, in the region of buoyancy
braking.

\par
\begin{figure*}
\plottwo{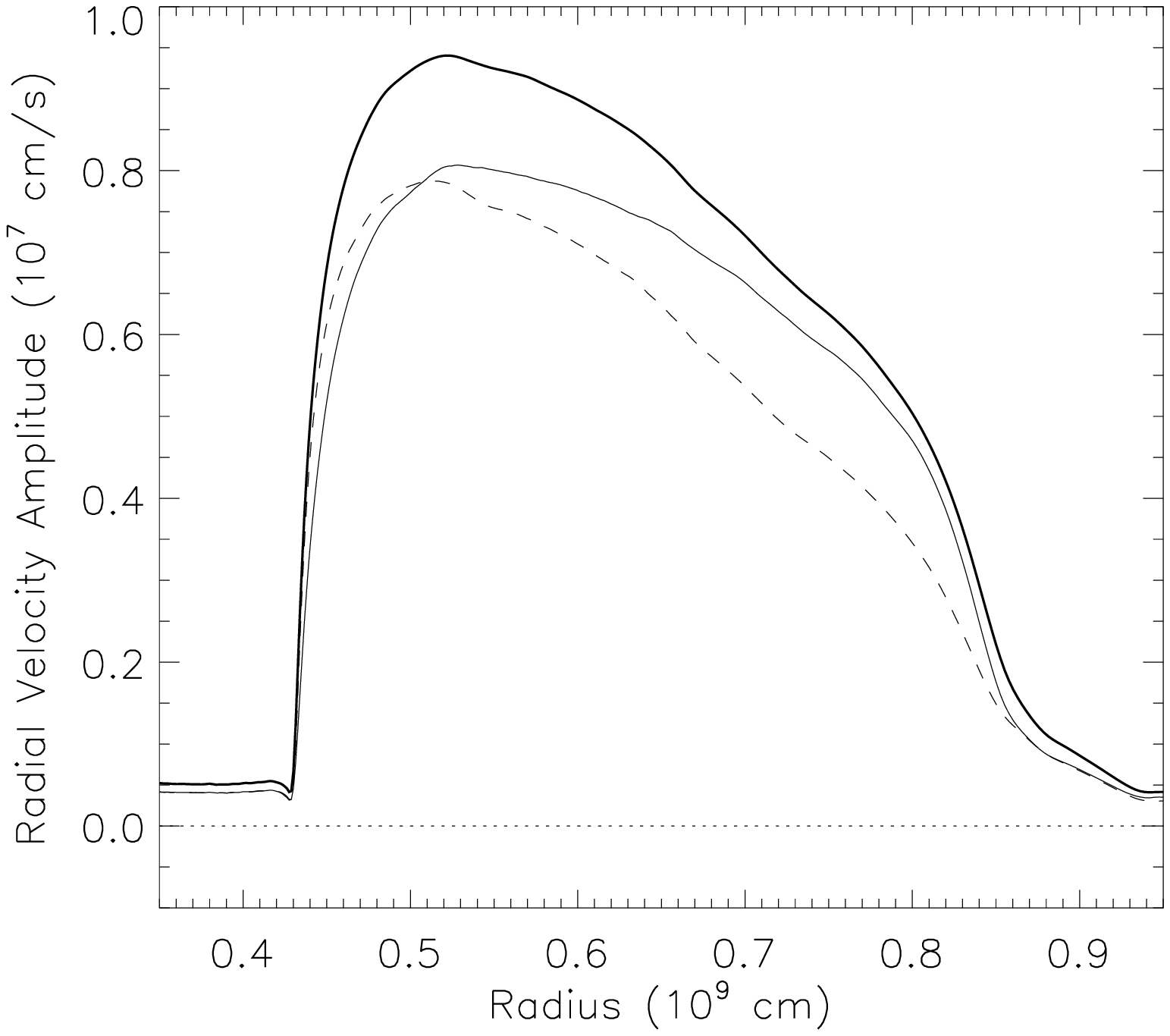}{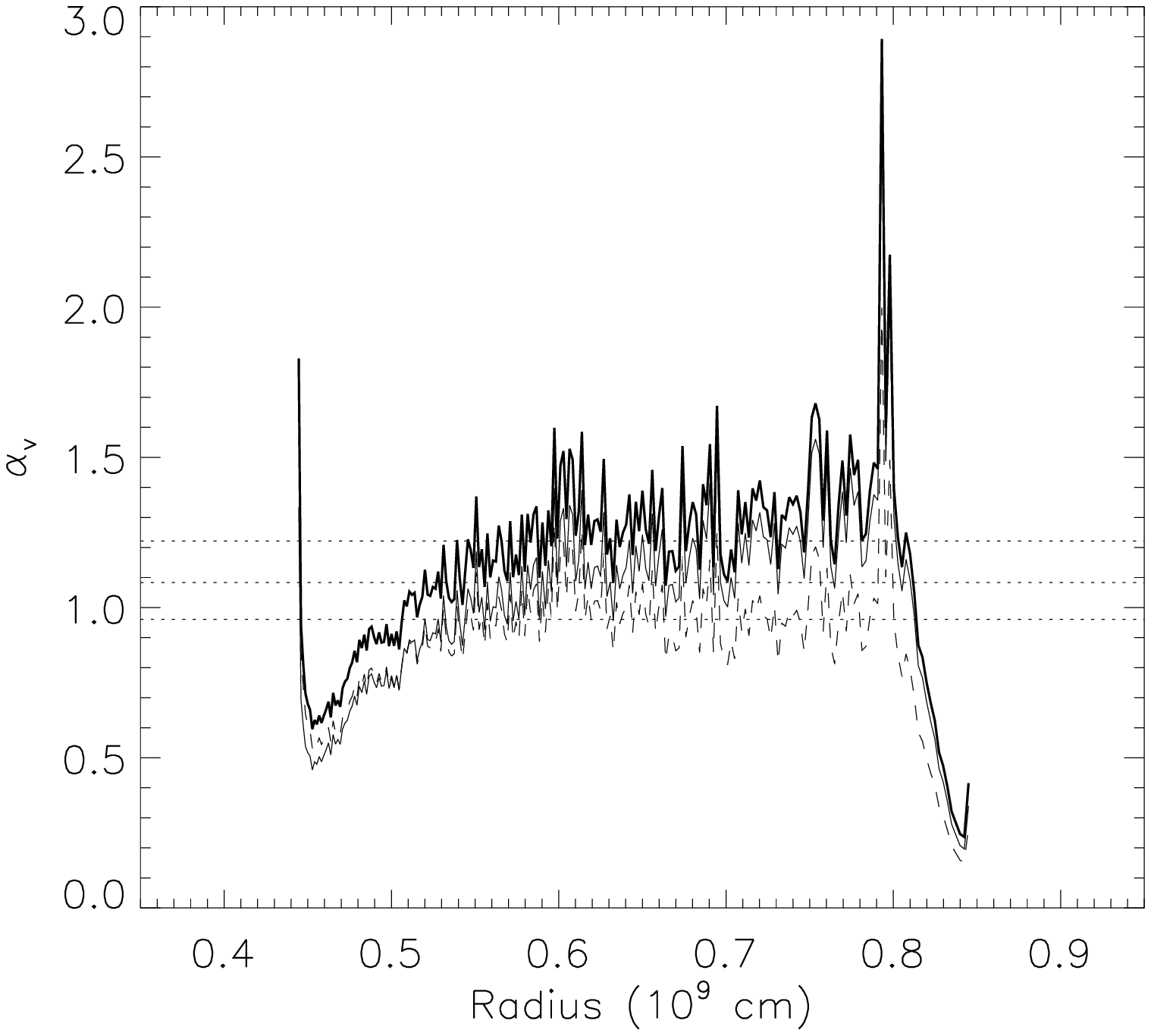}
 \caption{(left) Radial velocity amplitudes: (thick solid) r.m.s. value;
 the (thin solid) and (thin dashed) show the mean up- and down-flow velocities,
 respectively.  (right) The radial dependance of the "velocity mixing length"
 parameters $\alpha_v$ defined by equation \ref{alphav-eq} are shown for the velocity amplitudes
 presented in the left panel, using the same line types. The mean values, averaged over
 $r\in[0.5,0.75]\times 10^9$cm are shown by the thin dotted lines.\label{vuvd-fig}}
\end{figure*}

 The corresponding analysis for the velocity fluctuations is presented
in Figure \ref{vuvd-fig}.  The overall trends are similar for $\alpha_T$ and
$\alpha_v$. The mean values of $\alpha_v$ within the body of the convection zone are
$\havg{\alpha_v(\hbox{rms})} =1.22$,$\havg{\alpha_v(\hbox{up})}= 1.08$, and
$\havg{\alpha_v(\hbox{down})} = 0.96$.  The largest departure from constancy is again
found to be at the base of the convection zone (the flame region).

\par The sharp decrease in the effective mixing length near the lower boundary is
not entirely surprising. The distance to the convective boundary provides an upper
limit to the mixing length, while further away from the boundaries the mixing length
is limited by the distance over which eddies can maintain their coherence. This
effect is possibly more exaggerated at the lower boundary because of the steep
gradient in velocity which is needed to move the energy out of the burning zone. In
contrast, the upper boundary is characterized by a more gentle deceleration of
material and a "softer" boundary (i.e., lower $N^2$). Ignoring this boundary effect
and using the same mixing length parameter throughout the convection zone would
result in a shallower temperature gradient near the boundary.  The stiff temperature
dependance of the nuclear reaction rates may therefore be affected.

\par The absolute calibration of $\alpha_T$ and $\alpha_v$ are somewhat
arbitrary, and are scaled by factors of order unity for a particular implementation
of the mixing length theory based on the heuristic arguments discussed above.
According to equations \ref{mlt-tpert-eq}, \ref{mlt-v-eq}, \ref{alphat-eq}, and
\ref{alphav-eq} the equivalencies are $\alpha_{\Lambda,T} = 2\times\alpha_T$ and
$\alpha_{\Lambda,v} = \sqrt{2}\times\alpha_v$ where the values subscripted by
$\Lambda$ indicate the mixing length theory values defined by \citet{kw90}.  The
corresponding values measured in the simulation are
$\havg{\alpha_{\Lambda,v}(\hbox{rms})}=1.73$,
$\havg{\alpha_{\Lambda,v}(\hbox{up})}=1.53$, and
$\havg{\alpha_{\Lambda,v}(\hbox{down})}=1.35$, for velocity fluctuations; and
$\havg{\alpha_{\Lambda,T}(\hbox{rms})}=1.46$,
$\havg{\alpha_{\Lambda,T}(\hbox{up})}=0.9$, and
$\havg{\alpha_{\Lambda,T}(\hbox{down})}= 0.8$ for temperature fluctuations.

\par The ratios $\alpha_{\Lambda,T}/\alpha_{\Lambda,v}$ are 0.84, 0.59, and 0.60 for
the r.m.s., up-flow, and down-flow values, respectively. In relation to the present
simulation, a higher degree of consistency (i.e., $\alpha_{\Lambda,v} =
\alpha_{\Lambda,T}$) can be brought to this implementation of the mixing length
theory by scaling the velocity fluctuation in equation \ref{mlt-v-eq} by the inverse
of the ratio $\alpha_{\Lambda,T}/\alpha_{\Lambda,v}$. Physically, this translates
into a higher efficiency (by a factor $\sim$1.2 - 1.7) for the buoyancy work to
accelerate the large eddies over the value of 1/2 adopted above, which is reasonable
considering the heuristic argument used. Alternatively, agreement can be made by
scaling the temperature fluctuations in equation \ref{mlt-tpert-eq} by the same
ratio, which amounts to decreasing the distance over which eddies remain coherent and
adiabatic as they move across the convection zone. Both of the these effects are
plausible, as well as a combination of the two so long as the ratio is maintained.
Which is operating in the present simulation? Unfortunately, the degeneracy between
these two parameters cannot be broken because they combine linearly when calculating
the enthalpy flux, which therefore does not provide a further constraint.  Finally,
{\em it is possible that the effective mixing lengths for temperature and velocity
fluctuations are different}, a notion that is supported by the correlation lengths
which we discuss next.

\subsection{Correlation Length Scales}\label{correlation-lengths-section}

\par
\begin{figure*}
\plottwo{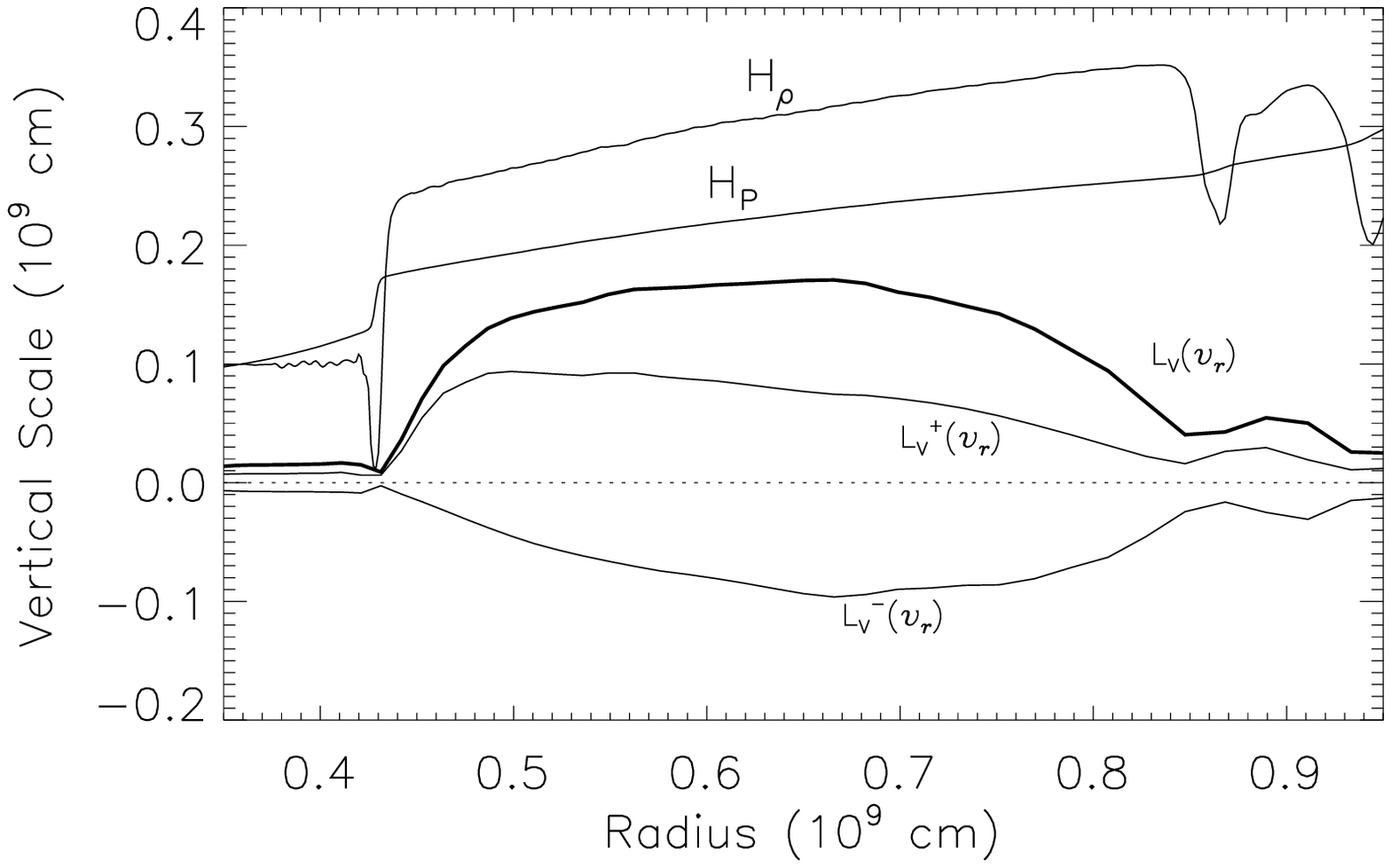}{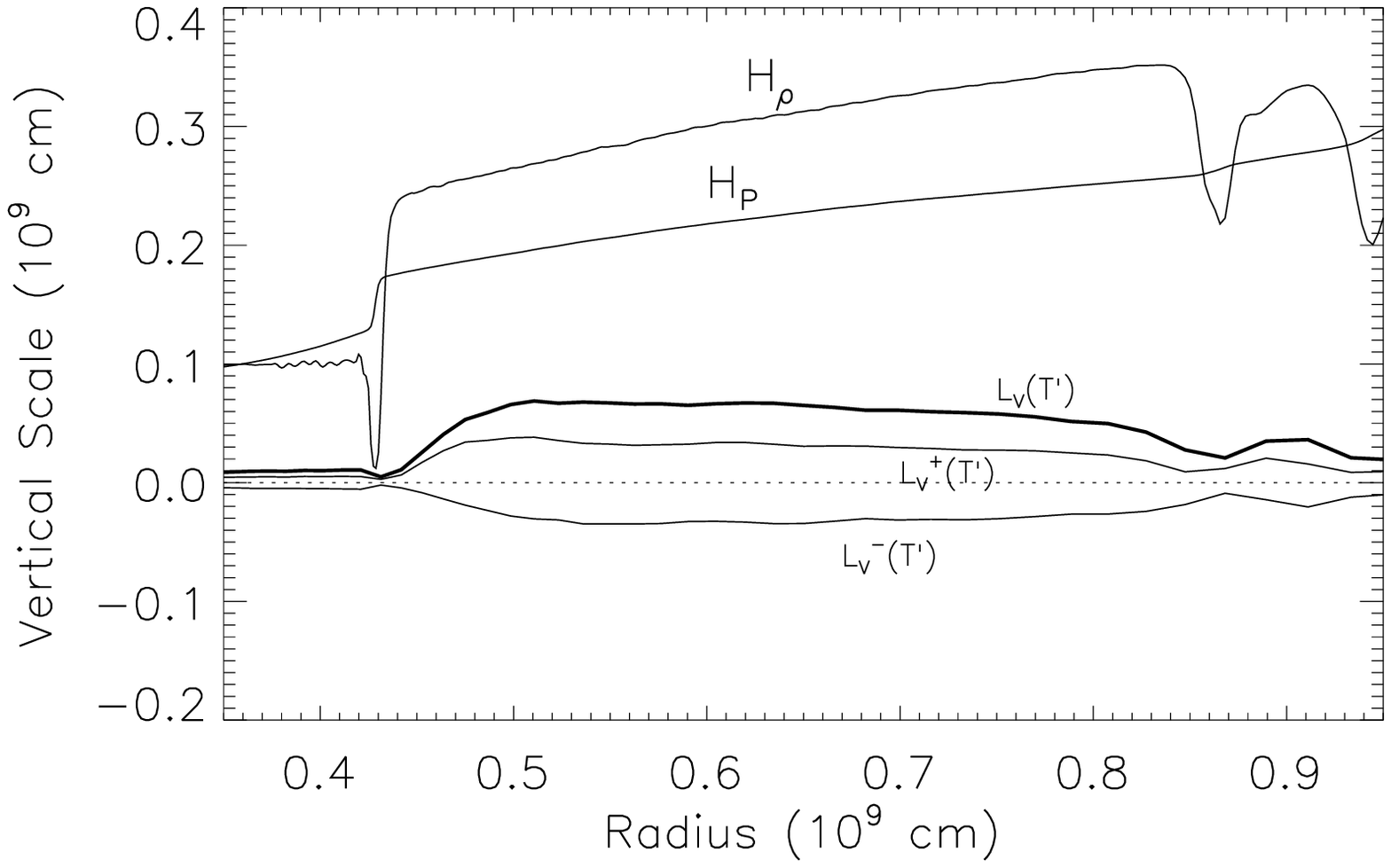} \plottwo{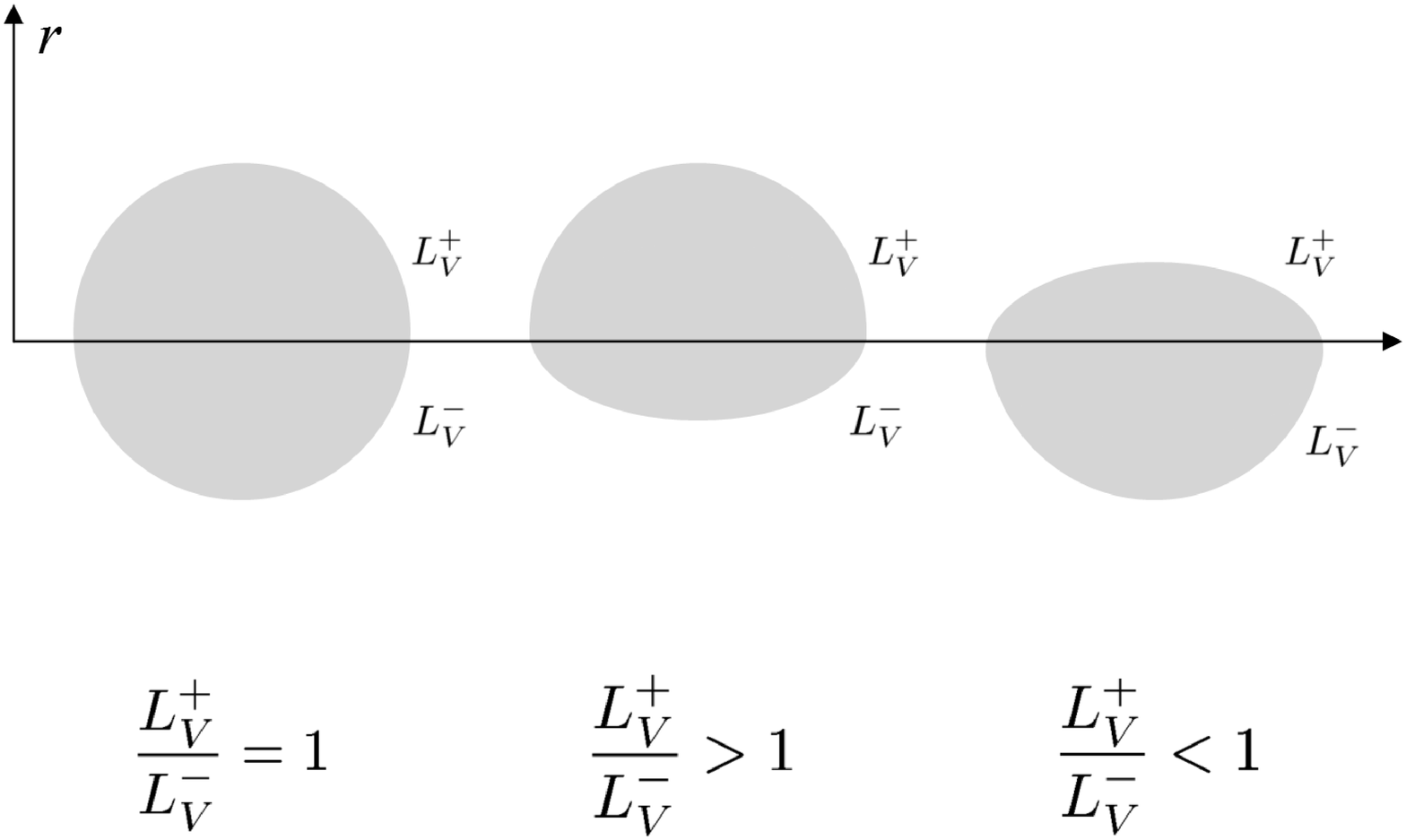}{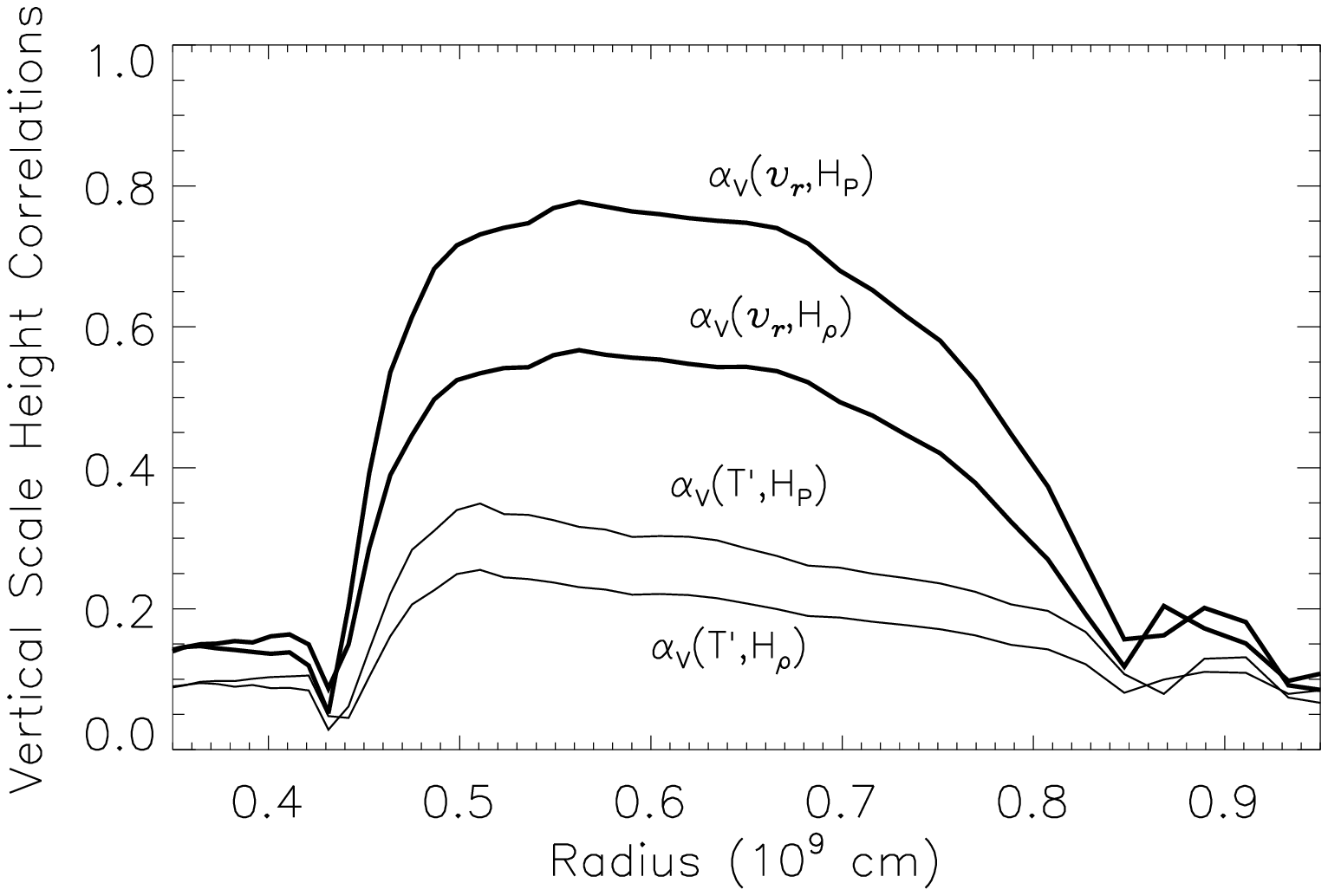}
  \caption{The vertical correlation length scales $L_V$ as defined in
  \S\ref{correlation-appendix}. (top left) $L_V$ for velocity fluctuations, $v_r'$;
  (top right) $L_V$ for temperature fluctuations,
  $T'$.  The pressure scale height $H_p$ and density scale height $H_{\rho}$ are
  shown for comparison. (bottom left) Illustration of the relationship between eddy shape and the
  correlation length scales, $L_V^+$ and $L_V^-$. The grey patches represent
  the shapes of the eddies and the $L_V^{+/-}$ values are measured in the radial
  direction, away from the horizontal line. (bottom right) Correlations
  lengths $L_V$ scaled to pressure and density scale heights, e.g., $\alpha_V(v_r,H_p) = L_V(v_r)/H_p$
  \label{vscale-fig}}
\end{figure*}

 In the top two panels of Figure \ref{vscale-fig} the vertical
correlation length scales, calculated according to equation \ref{vscale-equation},
are presented for the velocity and temperature fluctuations.  The vertical scale
height is defined as the full width at half maximum of the correlation function, and
can be written in terms of the correlation length in the positive and negative
directions, $L_V = L_V^+ - L_V^-$.  The relative values of $L_V^+$ and $L_V^-$ give
an indication of asymmetries in the eddies (Figure \ref{vscale-fig}, lower-left):
$L_V^+/L_V^-=1$ is a symmetric eddy; $L_V^+/L_V^->1$ is an eddy flattened on the
bottom; and $L_V^+/L_V^-<1$ is an eddy flattened at the top.  Based on this simple
diagnostic both the temperature and velocity correlations indicate that the eddies
near the lower boundary are flattened on the bottom, and those at the upper boundary
are flattened on the top. The "overshooting" distance ($h \sim 10^7$ cm at the upper
boundary and $h\la 10^6$ cm at the lower boundary), which is best described as an
elastic response to the incoming turbulent elements, is very small compared to the
correlation lengths measured here. Therefore, these eddies are effectively hitting a
"hard wall" upon reaching the boundaries.

\par The signature of this eddy "flattening" is also present in the radial profile of
the full width length scale, $L_V$. In the case of velocity, which has larger
correlation length scales, significant asymmetries are present throughout the
convection zone. The smaller length scales associated with the temperature
fluctuations permit a broad region throughout the convection zone where the eddies
are roughly symmetric ($L_V^+/L_V^-\approx1$) and appear to be uninfluenced by the
boundaries. In this intermediate region, away from the boundaries, the temperature
fluctuation length scales are relatively constant in size, even decreasing with
radius, in contrast to the pressure and density scale heights which are increasing
with radius.

\par  In the standard mixing length theory, the sizes of convective eddies are
assumed to be comparable to the size of the mixing length. How do the correlation
length scales compare to the mixing length parameters found above?  The lower left
panel of Figure \ref{vscale-fig} shows the ratios of $L_V$ to the pressure and
density scale heights. None of these curves are particularly constant within the
convection zone, and boundary effects are particularly strong throughout the
convection zone in the case of the velocity correlations. Interestingly, the velocity
correlation parameter $\alpha_v(v_r,H_p)=L_V(v_r)/H_p$ is larger than the temperature
correlation parameter $\alpha_v(T',H_p)=L_V(T')/H_p$. This is in accord with the
ratio $\alpha_{\Lambda,T}/\alpha_{\Lambda,v} < 1$ found in the mixing length analysis
above.  Concerning the absolute calibration, however, {\em the correlation length
scales are smaller than the mixing length values by as much as a factor of a few.} In
an analogous comparison by \citet{rob04} for subgiant atmosphere models, the vertical
correlation lengths were also found to be smaller than the mixing length used to
construct the initial model, and that the ratio varied significantly throughout the
convection zone.

\par
\begin{figure*}
\epsscale{0.5}
\plotone{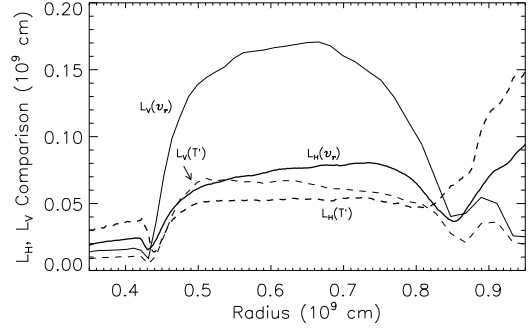}
  \caption{The horizontal and vertical correlation length scales, $L_H$ (thick line) and $L_V$ (thin line) are
    shown for temperature (dashed) and velocity (solid) fluctuations.\label{hvscale-fig}}
\end{figure*}

 The horizontal correlation lengths $L_H$ are shown in Figure \ref{hvscale-fig}
together with the vertical scales for comparison. For the velocity, the horizontal
scale is much smaller than the vertical, indicative of eddies which are significantly
elongated in the vertical direction.  The temperature fluctuations appear to be much
more symmetric, with only a small degree of elongation in the vertical direction
which is slightly more pronounced near the top of the convection zone. In the stable
layers, the horizontal scales are larger than the vertical,  which is a
characteristic of the horizontal "sloshing" motions associated with g-modes.

\subsection{The Kinetic Energy Flux, Flow Asymmetry, and
                Moving Beyond the Mixing Length Theory}

\par The kinetic energy flux associated with convection is
ignored in the mixing length theory since it arises from the {\em
asymmetries} in the flow and MLT assumes that the flow is symmetric.
An order of magnitude estimate for the kinetic energy flux, however,
can be made:

\begin{equation}
\frac{F_K}{F_c} \sim \frac{\rho v_c^2/2}{\rho c_p T'}\frac{v_c}{v_c}
\sim \frac{\alpha_{\Lambda}}{8}\frac{\beta P}{T\rho c_p} =
\frac{\alpha_{\Lambda}}{8}\nabla_{ad} \sim 0.03
\end{equation}

\noindent where mixing length relationships have been used to calculate $v_c$ and
$T'$, $\alpha_{\Lambda}$ is assumed to be of order unity, and $\nabla_{ad}\sim0.25$
has been adopted from the simulation. This result tells us that the kinetic energy
flux will be a few percent of the convective enthalpy flux. This estimate is an upper
limit since up-flows and down-flows will cancel to some degree
\citep[][\S6.1]{b-v92}. In the simulation, the ratio of kinetic to enthalpy flux is
found to be $F_{K,max}/F_{c,max}\sim 0.01$, which is of order the simple MLT scaling,
but down by a factor of a few as expected.

\par
\begin{figure*}

\epsscale{0.5}
\plotone{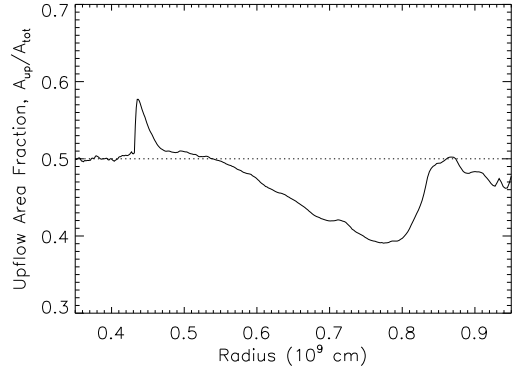}
 \caption{The fractional area occupied by the upward
flowing material $f_u$ is shown as a function of radial position. The downward
flowing area is $f_d = (1-f_u)$ and the dashed line at 1/2 indicates up-down
symmetry. \label{fup-fig}}
\end{figure*}

\begin{figure*}
\epsscale{1.0} \plottwo{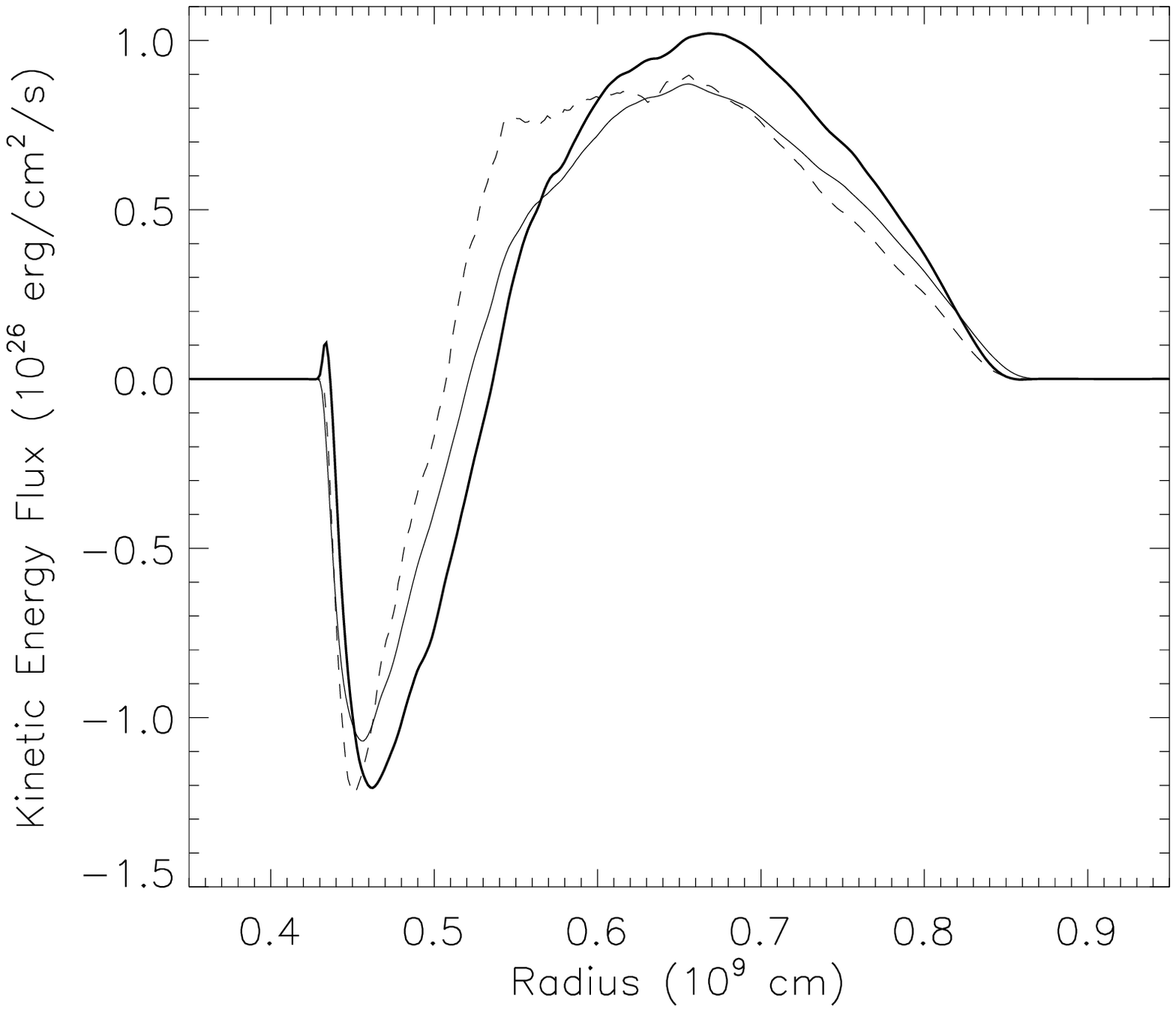}{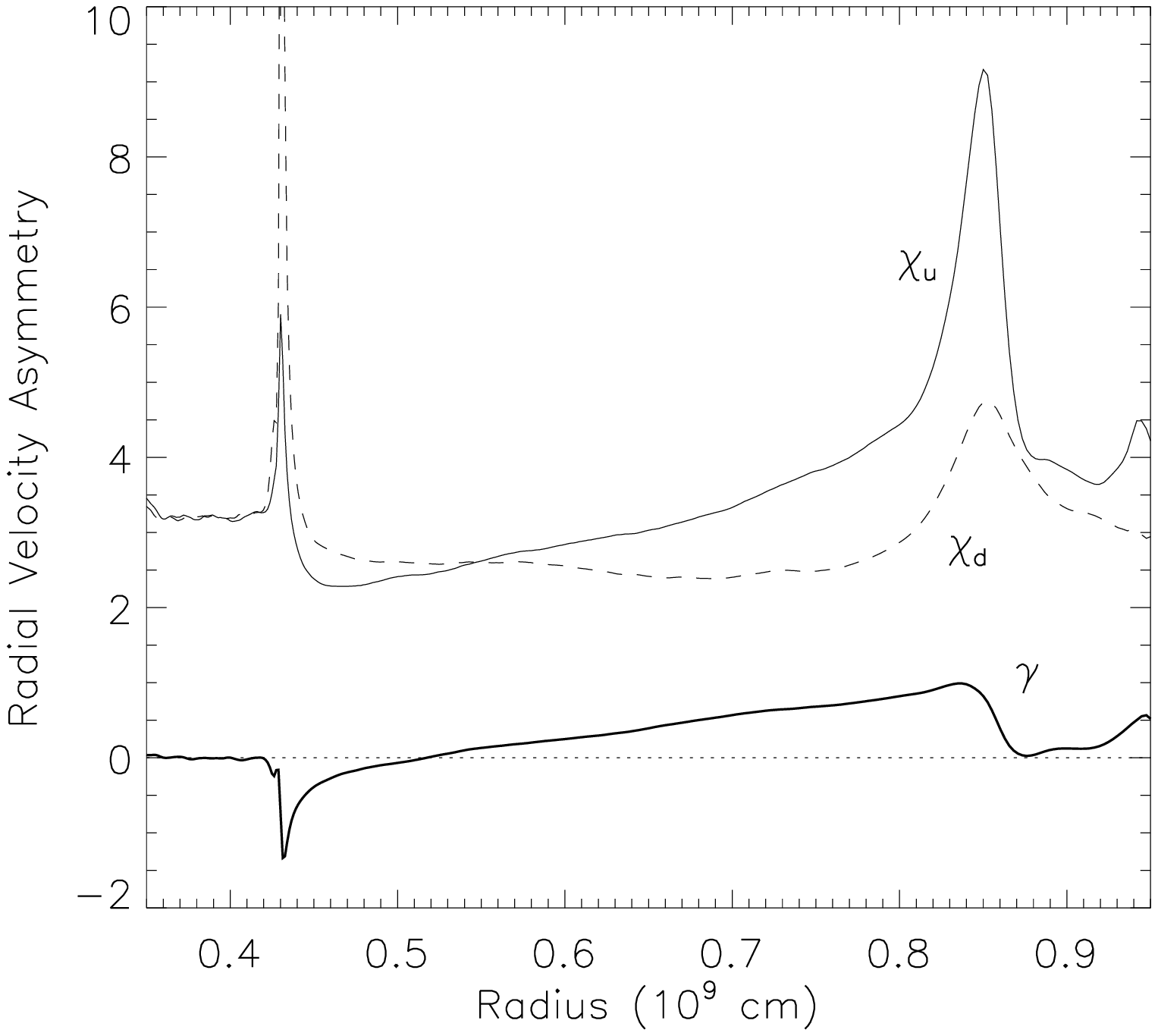}
 \caption{(left) Kinetic energy flux: (thick) line shows the value measured in the
 simulation averaged over two convective turnovers; the (thin solid) line shows  $F_K$
 calculated according to equation \ref{fke-eq};  the (thin dashed) line shows $F_K$
 calculated according to equation \ref{fke-eq} but uses $c\havg{v}^3$ in place of $\havg{v^3}$,
 and a correlation constant of $c=5$.
 (right) Asymmetries in radial velocity: the (thick) line show the skewness in the velocity
    field, $\gamma = \havg{v^3}/\sigma_v^3$; the (thin solid) and (thin dashed) lines show the
    correlations $\chi = \havg{v^3}/\havg{v}^3$ where the subscripts $u$ and $d$ indicate up- and
    down-flows,respectively.\label{fke-fig}}
\end{figure*}

  We can directly relate the kinetic energy flux to the flow asymmetry in the
following way. The upflow area covering fraction $f_u = A_{up}/A_{tot}$ is shown in
Figure \ref{fup-fig}.  We can then write an estimate for the kinetic energy flux as,

\begin{equation}
F_{K,net} = \frac{1}{2}\rho_0 (f_u v_u^3 - f_d v_d^3)
\end{equation}

\noindent which can be written in terms of just the flow velocities,

\begin{equation}
\label{fke-eq} F_{K,net} = \frac{1}{2}\rho_0 \Big[\frac{v_u^3 +
v_d^3}{v_u/v_d+1}- v_d^3\Big]
\end{equation}

\noindent where we have used the mass conservation equation, $f_u v_u + f_d v_d = 0$
assuming $\rho_u \approx \rho_d$ which is a good approximation in these simulations.
The kinetic energy flux in the simulation is shown in Figure \ref{fke-fig}. Shown by
the thin line is $F_K$ calculated according to equation \ref{fke-eq}, which is in
good agreement. Here, we have used the horizontal and time averaged values for
$\havg{v}$ and $\havg{v^3}$. The mixing length theory, however, does not provide
information about $\havg{v^3}$, but only $\havg{v}$. We overplot with the dashed line
$F_K$ calculated using $\havg{v}^3$ in place of $\havg{v^3}$; it has to be scaled by
a factor of 5 to fit the simulation data.

\par The scaling factor needed to calculate the kinetic energy flux is required to
account for the skewness in the radial velocity field. More precisely, the
correlation coefficient $\chi = \avg{v_u^3}/\avg{v_u}^3$ is needed, which is related
to the skewness $\gamma = \avg{v^3}/\sigma_v^3$. Both $\chi$ and $\gamma$ are
presented in Figure \ref{fke-fig}. Note, that the skewness is a good proxy for the
down-flow covering fraction ($f_d = 1-f_u$; see Figure \ref{fup-fig}), and therefore
its sign is indicative of the direction of the kinetic energy flux.

\par Convective regions which are spanned by several pressure scale heights are
found to have kinetic energy to enthalpy flux ratios larger than a few percent as
found in this study. For instance, the simulations of \citet{cattaneo91} and
\citet{chan89} which each span $\sim$5 pressure scale heights achieve $|F_K/F_c|\sim$
35\%, and the domain in \citet{chan96} spanning $\sim 7$ pressure scale heights
achieves $|F_K/F_c|\sim$ 50\%. A key result in the analysis of \citet{cattaneo91} is
that the kinetic energy flux is dominated by coherent, downward-directed flows which
are correlated over distances comparable to the simulation domain. Additionally, the
enthalpy flux and kinetic energy fluxes associated with these downflows essentially
cancel with $c_p T' \sim v_c^2$, which was shown to follow if the downflows can be
described as Bernoulli streamlines.

\par The long range correlations just described, together with the boundary effects which
dominate our shell burning model, undermine the basic mixing length theory picture of
convection. The large coherence of the flows seen in these simulations, however, and
present even in turbulent parameter regimes, suggest that modeling these {\em
coherent structures} is a viable approach. Already, models incorporating multiple
streams or "plumes" as closure models \citep[e.g.][]{rempel04,lesaffre05,belkacem06}
are providing enticing alternatives to the mixing length theory.

\subsection{Related Studies}

\par Although the mixing length parameters calculated above deviate from
constancy near the convective boundaries, a mean value is a good approximation for
most of the convection zone.  It would be interesting if these parameters $\alpha_E$,
$\alpha_T$, and $\alpha_v$ were universal, as assumed by mixing-length theory. If we
restrict consideration to 3D compressible convection simulations for simplicity and
homogeneity, there are several previous studies which have confronted mixing length
theory to which we can compare our results. These studies investigate convection
under diverse conditions, including slab convection
\citep{chan87,chan89,chan96,pw00}, a red giant envelope \citep{pwj00}, and solar and
sub-giant surface layers \citep{kim95,kim96,rob03,rob04,rob05}.
The number of zones used range from $1.9 \times 10^4$ \citep{chan89} to $6.7 \times
10^7$ \citep{pw00}. The equations of state used include a gamma-law
\citep{chan89,pw00}, ionized gas \citep{kim96,rob04}, and a combined relativistic
electron plus ion gas \citep{ts00} in this paper. Subgrid scale physics was treated
by a Smagorinsky model \citep{smag63} or by ignoring it. We note that \cite{styne00}
have shown that PPM methods solving the Euler equations converge to the same limit as
solutions to the Navier-Stokes equations, as resolution is increased and viscosity
reduced. Additionally, the subgrid scale turbulence "model" implicit in the numerical
algorithm of PPM is known to be well behaved (Fernando Grinstein, personal
communication). Given this already inhomogeneous set of simulations, determining
consistent convection parameters is difficult. Our attempt is given in Table
\ref{conv-table}, in which we summarize the convection parameters found in these
studies for comparison to our own.

\par How well do these compare? In some respects the agreement is
striking. The parameter $\alpha_E$ is in the range $\sim$0.7 - 0.8 for all groups.
Further, all agree that for their case, the mixing-length theory gives a fairly
reasonable representation of the simulations in the sense that the alphas are roughly
constant throughout the body of the convection zone. The difficulty is that the
specific values of these alphas depend upon the case considered. The two
best-resolved simulations, ours and \citep{pw00}, use the same solution method, PPM,
yet have the most differing alphas. This suggests to us that the differences are due
to the physical parameters of the respective convection zones.  \cite{pwj00} have
already shown that slab and spherical geometry give qualitatively different behavior
for the alphas. Our shell is only two pressure scale heights in depth, and is
relatively slab-like; \cite{pw00} have a convection zone which is more than twice as
deep by this measure. There is a suggestion in Table~\ref{conv-table} that the alphas
increase with the depth of the convection zone. This would be reasonable if a
convective plume were accelerated through the whole convection region before it is
decelerated at the nonconvective boundary. However, the other differences mentioned
above probably contribute to the scatter in the alpha parameters in \ref{conv-table}.

Further efforts on this issue are needed. If convection does depend upon the
nonlocal, physical structure of the star, calibration of the mixing length to fit the
sun, as is traditionally done, is not wise. Furthermore, it is well known that the
mixing length theory is particularly prone to problems in the surface layers where
convection becomes inefficient. Therefore, the empirical agreement of mixing length
calibration to the sun and to Population II giants \citep{ferr06} may be a fortuitous
coincidence.

\section{MIXING AT CONVECTIVE BOUNDARIES} \label{entrainment-section}

\par
\begin{figure*}
   \plotone{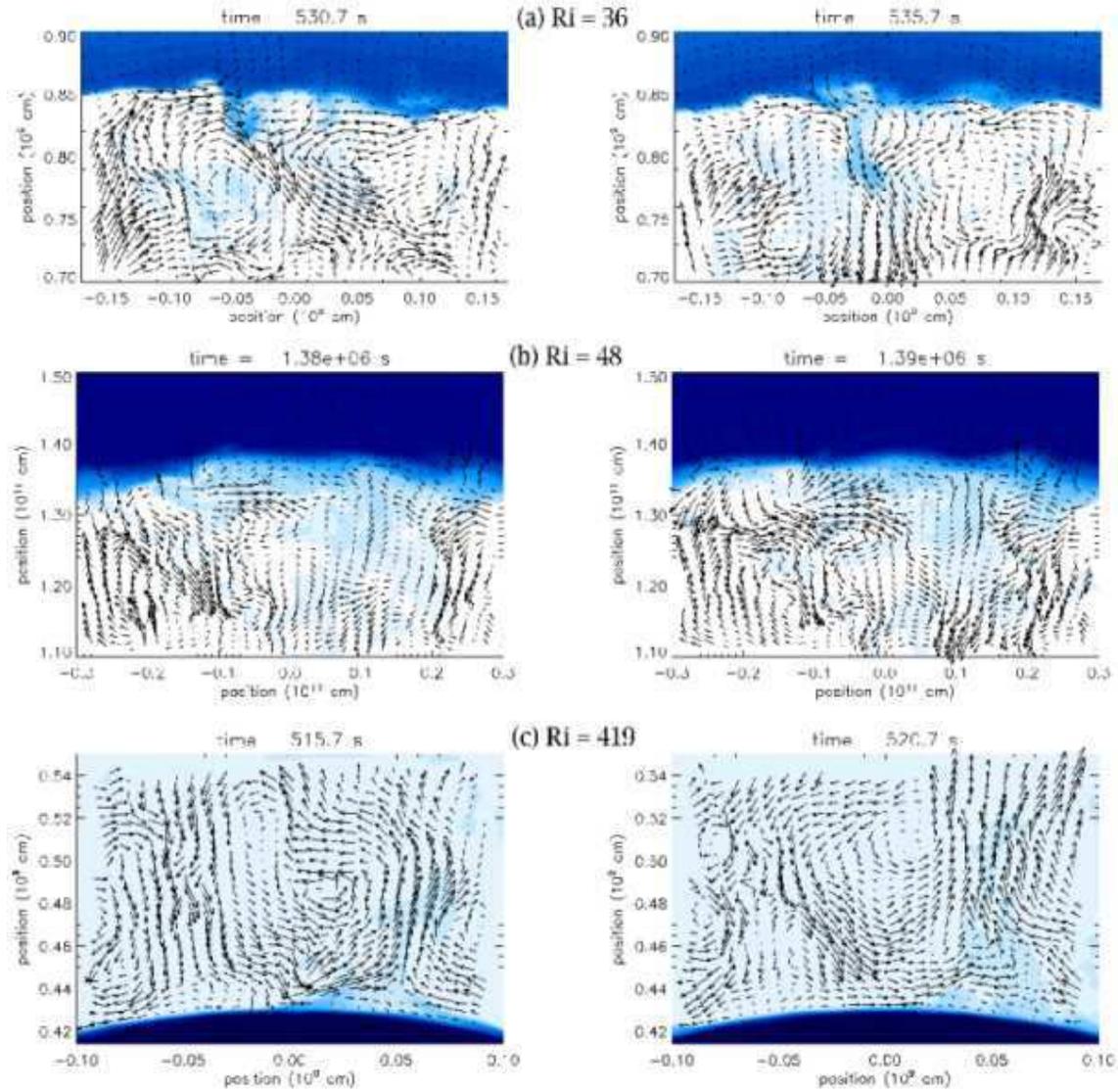}
  \caption{Equatorial slices showing the flow in the vicinity of the convective
  boundaries in the 3D simulations, ordered by relative stability:
  (row a) upper shell convection boundary, $Ri_B\sim$ 36;
  (row b) core convection boundary, $Ri_B \sim 48$;
  (row c) lower shell convection boundary, $Ri_B \sim 419$.
   The colormap indicates composition abundance, where the darker tones
   trace stable layer material entrained across the interface.
  The velocity vectors have been sampled every third zone in each dimension.
  \label{boundary-snapshots-fig}}
\end{figure*}

 The boundaries which separate the convective regions from the stably
stratified layers in our 3D simulations span a range of relative stability, with $1
\lesssim Ri_B \lesssim 420$.  At the lowest values of $Ri_B$, the boundary is quickly
overwhelmed by turbulence, as described in \S\ref{transient-section}.  Once $Ri_B$
becomes large enough, the boundary becomes stabilizing and evolves over a much longer
timescale. Snapshots of the quasi-steady shell burning and the core convection
boundaries are presented in Figure \ref{boundary-snapshots-fig}, ordered by $Ri_B$
with spans the range $36 \lesssim Ri_B \lesssim 420$. The anatomy of the convective
interfaces includes the turbulent convection zone, the distorted boundary layer of
thickness $h$, and the stably stratified layer with internal wave motions (compare to
Figure \ref{diagram}).

\par The boundary becomes more resilient to thickening, and distortion by the turbulence
as $Ri_B$ increases. A region of partial mixing exists primarily on the turbulent
side of the interface, where material is being drawn into the convection zone. The
"ballistic" picture of penetrative overshooting \citep{zahn91} in which convective
eddies are envisioned to pierce the stable layer does not obtain.  Instead, material
mixing proceeds through instabilities at the interface, including shear instabilities
and "wave breaking" events, which break the boundary up into wisps of material that
are then drawn into the turbulent region and mixed. The convective interface remains
fairly sharp in all cases, and the effective width is well described by the elastic
response of the boundary layer to incoming eddies, $h\sim v_c/N$. The convective
interfaces seen in our simulations bear a striking resemblance to those observed in
laboratory studies of turbulent entrainment of comparable $Ri_B$ \citep[see
e.g.][Figs. 2-5]{mcgrath97}.

\par The mixing that occurs due to the instabilities and eddy scouring
events at the interface leads to a steady increase in the size of
the convection zone. In this section we quantify the entrainment
rates at the convective boundaries, we discuss these results in
terms of the the buoyancy evolution of the interface, and we
describe how the "turbulent entrainment" process can be incorporated
into a stellar evolution code as a dynamic boundary condition to be
used in addition to the traditional static Ledoux and Schwarzschild
criteria. We conclude the section with some comments on numerical
resolution.

\subsection{Quantifying the Boundary Layer Mixing Rates\label{quantify-entrainment-section}}

\par As evident in Figure \ref{boundary-snapshots-fig}, the
convective boundary layers are significantly distorted from
spherical shells.  To estimate the radial location of the interface
we first map out its shape in angle $r_i = r_i(\theta,\phi)$. At
each angular position the surface is taken to be coincident with the
radial position where the composition gradient is the steepest (this
is comparable to the location of minimum density scale height
$H_{\rho} = [\partial \ln\rho/\partial r]^{-1}$). The interface
thickness $h$ is taken to be the r.m.s. variation of the surface
$r_i$ with angle, $h = \sigma [r_i(\theta,\phi)]$, which provides a
quantitative measure of the amplitudes of the distortions imparted
to the interface. The mass interior to the interface is calculated
according to,

\begin{equation}
  M_i = \int\limits_{r_0}^{\havg{r_i}} 4\pi r^2 \langle\rho\rangle dr
\end{equation}

\noindent where $r_0$ is the inner boundary of the computational
domain, $\langle\rho\rangle$ is the horizontally averaged density,
and the mean interface radius is used for the upper limit on the
integral.  The time derivative $\dot{M_i}$ is the rate at which mass
is entrained into the convection zone.

\par
\begin{figure*}
 \epsscale{0.3}
\plotone{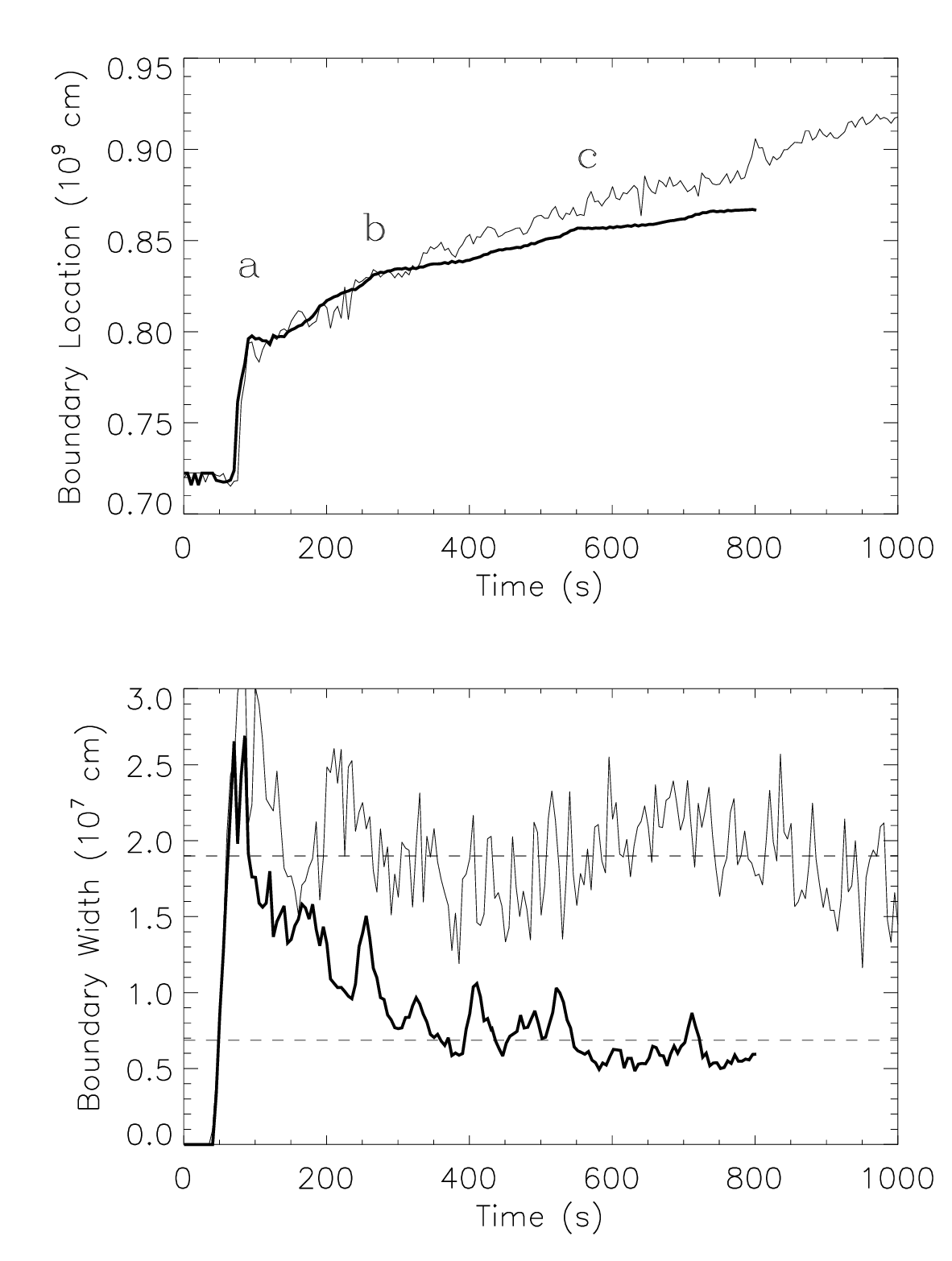} \plotone{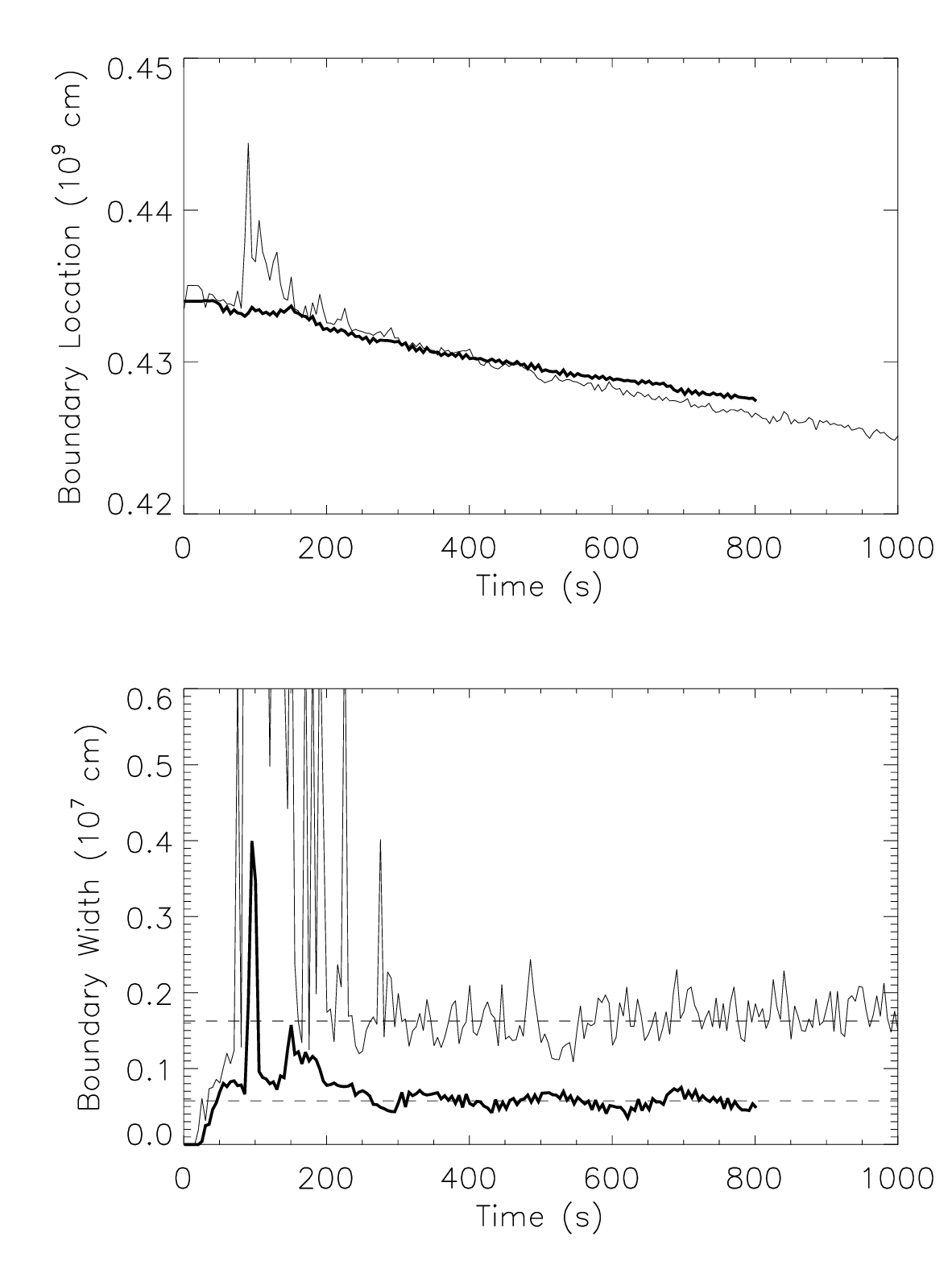} \plotone{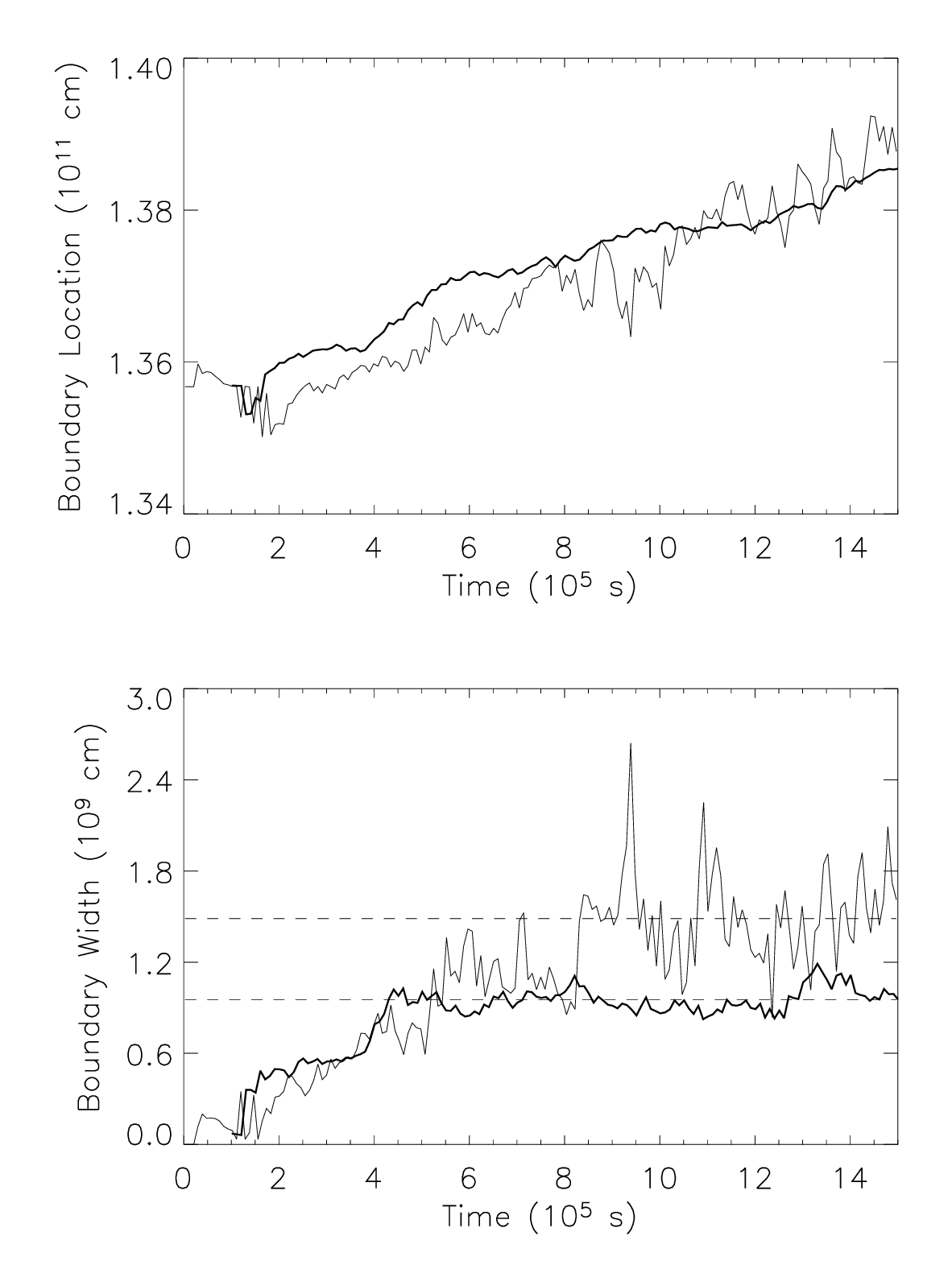}
  \caption{The time history of (top row) the convective boundary
  location, and (bottom row) the thickness of the convective
  interface for: (left) upper shell burning boundary;
  (middle) lower shell burning boundary; and
  (right) core convection boundary.
  The (thick line) identifies the 3D models, ob.3d.B and msc.3d.B; and the
  (thin line) identifies the 2D models, ob.2d.e and msc.2d.b.
  The (dashed lines) show the averaged interface thickness
  for $t>300$ s for oxygen burning, and $t>6.0\times10^5$ s for
  core convection. The letters (a-c) in the upper left panel mark
  times when the outward migration rate of the convective boundary
  rapidly adjusts to a new value in the 3D model.
    \label{interface-migration-fig}}
\end{figure*}

 In Figure \ref{interface-migration-fig} the
time histories of the averaged interface location $\havg{r_i}$ and
interfacial thickness $h$ are shown for the convective boundaries in
our simulations.  A 3D model and a representative 2D model are shown
for each boundary. The outer shell boundary layer adjusts rapidly in
the first 100s to a new position, due to the penetration event
discussed in \S\ref{transient-section}, after which a slow outward
migration ensues.  For the 3D model, the outward migration proceeds
in distinct stages, labeled (a - c) in Figure
\ref{interface-migration-fig}.  Each stage is well described by a
linear increase of radius with time, and ends with a rapid
adjustment to a new entrainment rate. This behavior can also be seen
in Figure \ref{ob3d-evol}, where the change in entrainment rate is
coincident with changes in the background composition gradient and
stability (compare to the initial buoyancy frequency profile in
Figure \ref{ob-profile}).

\par The downward migration of the lower shell boundary is more
uniform and proceeds at a significantly reduced rate compared to the
upper boundary. The core convection boundary evolution departs most
significantly from a linear trend, but Monotonic growth is clearly
established very soon after the simulation begins $t\ga2\times
10^5$s.

\par The interfacial thickness $h$ in the oxygen burning models are
initially large, reflecting the strong mixing event during the initial transient, but
settle down to relatively constant values for $t\ga 300$s.  In contrast, the boundary
thickness in the core convection model increases gradually with time until a steady
state value is achieved, reflecting the milder initial development. In all cases, the
time averaged values of $h$ during the quasi-steady states compare well to the
boundary displacement expected for eddies impacting the stable layer with the
characteristic convective velocities of the model, $h \sim v_c/N$.

\par The entrainment rate and the interfacial thickness is
larger in all of the 2D models as a consequence of the larger velocity scales in
those simulations. The interface migration rates and averaged interfacial layer
thicknesses  are tabulated for all of the models in Tables \ref{ob-entrain-table} and
\ref{msc-entrain-table}, and are broken down into various time intervals over which
linear growth of the boundary is a good approximation. Time averaged mass entrainment
rates are also included in Tables \ref{ob-table} and \ref{msc-table}.

\subsection{The Entrainment Energetics}

\par In order for entrainment to take place at a convective
boundary the buoyancy increment of the stable layer material over that of the mixed
layer material must be overcome.  This can happen in two distinct ways.  First,
non-adiabatic processes can change the relative stability of the stable layer. For
example, heating the convective region will cause an increase in its entropy, and the
buoyancy jump separating the overlying layer will decrease. The rate at which the
convective boundary will grow due to heating is $u_s = \dot{s}/(\partial_r s)$, where
$\partial_r s$ is the radial gradient of entropy at the boundary and $\dot{s}$ is the
time rate of change of entropy in the shell. This process will cause both the upper
and lower boundaries to migrate to larger radii -- the upper boundary will be
weakened, while the lower boundary will become stiffer. Non-adiabatic processes in
the boundary layers will affect their stability in the same way: cooling in the upper
and heating in the lower boundaries will weaken their stratification.

\par A related, but distinct process is "turbulent entrainment"
whereby turbulent kinetic energy does work against gravity to draw material into the
turbulent region.  In this process, the stratification is weakened at a convective
boundary by the turbulent velocity fluctuations. This is quantified in terms of the
buoyancy flux $q=g\rho' v'/\rho_0$. In the absence of heating and cooling sources the
buoyancy in the interfacial layers will evolve according to the buoyancy conservation
equation,

\begin{equation}\label{buoyancy-conserve-eq}
\partial_t b = -\hbox{div}(q)
\end{equation}

\noindent and a positive flux divergence at the boundary will lead
to a weakening of the stratification.  The relationship between
turbulent entrainment and the weakening of a boundary through
heating and cooling can be understood in terms of the enthalpy flux
which attends the buoyancy flux.  In fact, the buoyancy flux is
directly related to the enthalpy flux across the interface,

\begin{equation}
F_c = \rho_0 c_p \havg{T' v_r'} = \frac{c_p T_0}{\beta}\havg{\rho'
v_r'} = \rho_0 c_p \frac{T_0}{\beta g}\times q
\end{equation}

\noindent and is equivalent to heating and cooling processes
operating in the boundary layer (note the downward directed enthalpy
flux within the boundary layers in Figure \ref{fenth-fig}).

\par What drives the entrainment seen in the present simulations?  Can
the entrainment in the outer shell boundary be explained by the heating of the
convection zone by nuclear burning?  Comparing the entropy growth rate of the shell
to the entropy gradient at the boundary we find $u_s \sim 0.8\times 10^4$ cm/s, which
is at most 17\% of the measured growth rate for this boundary, and typically of order
a few percent.  Shell heating will reduce the growth rate of the lower boundary by
$u_s \sim 0.04\times 10^4$ cm/s, which is of order a few percent of the rate
measured. Therefore, the overall heating and cooling of the shell contributes very
modestly to the growth of the shell over the course of the simulation. The long
thermal timescale in the core convection model reduces this effect even more, where
it is lower by several orders of magnitude. Therefore, we turn to the turbulent
hydrodynamic processes operating in the boundary layer to understand the growth of
the convection zones.

\par
\begin{figure*}
\epsscale{0.8}
\plottwo{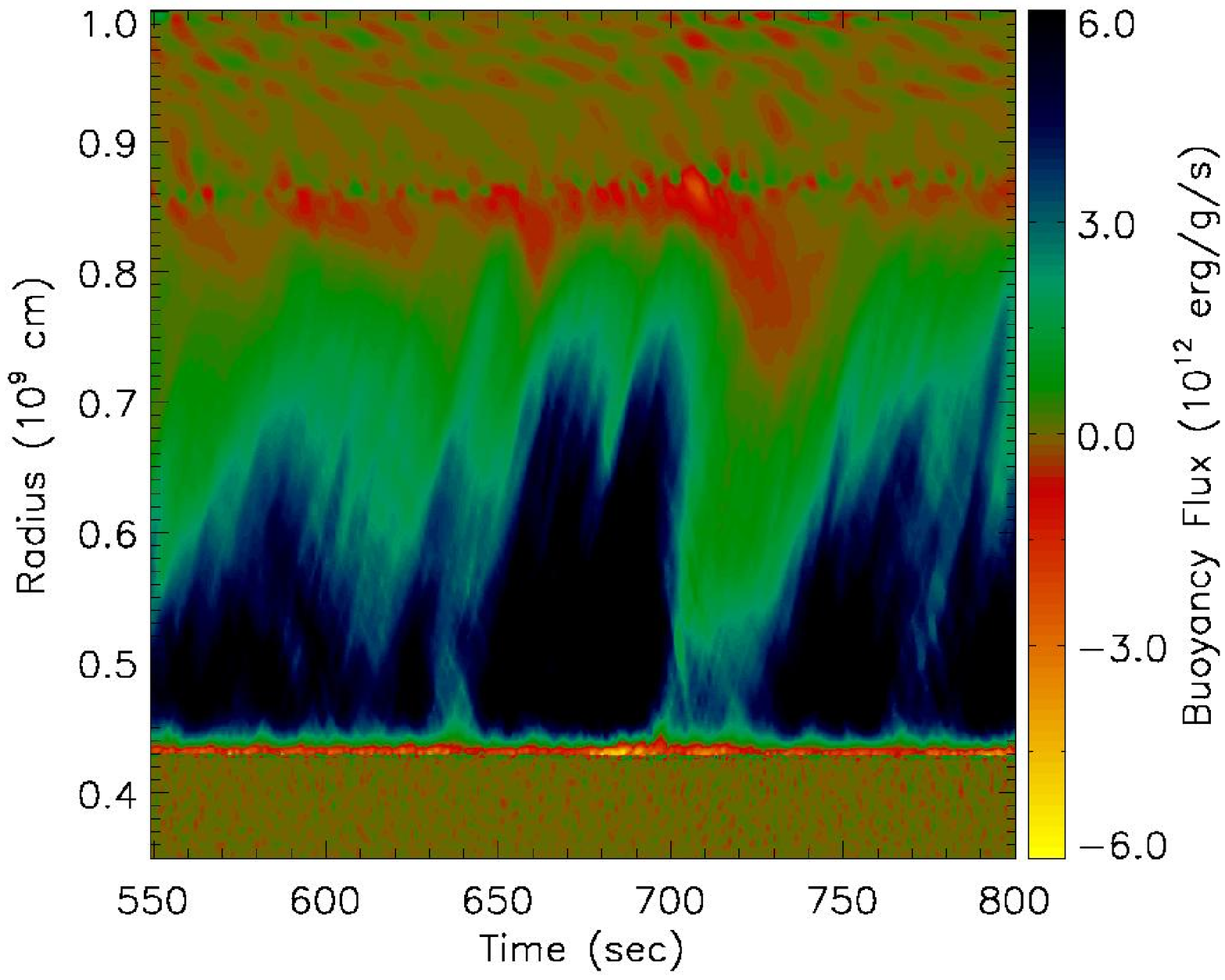}{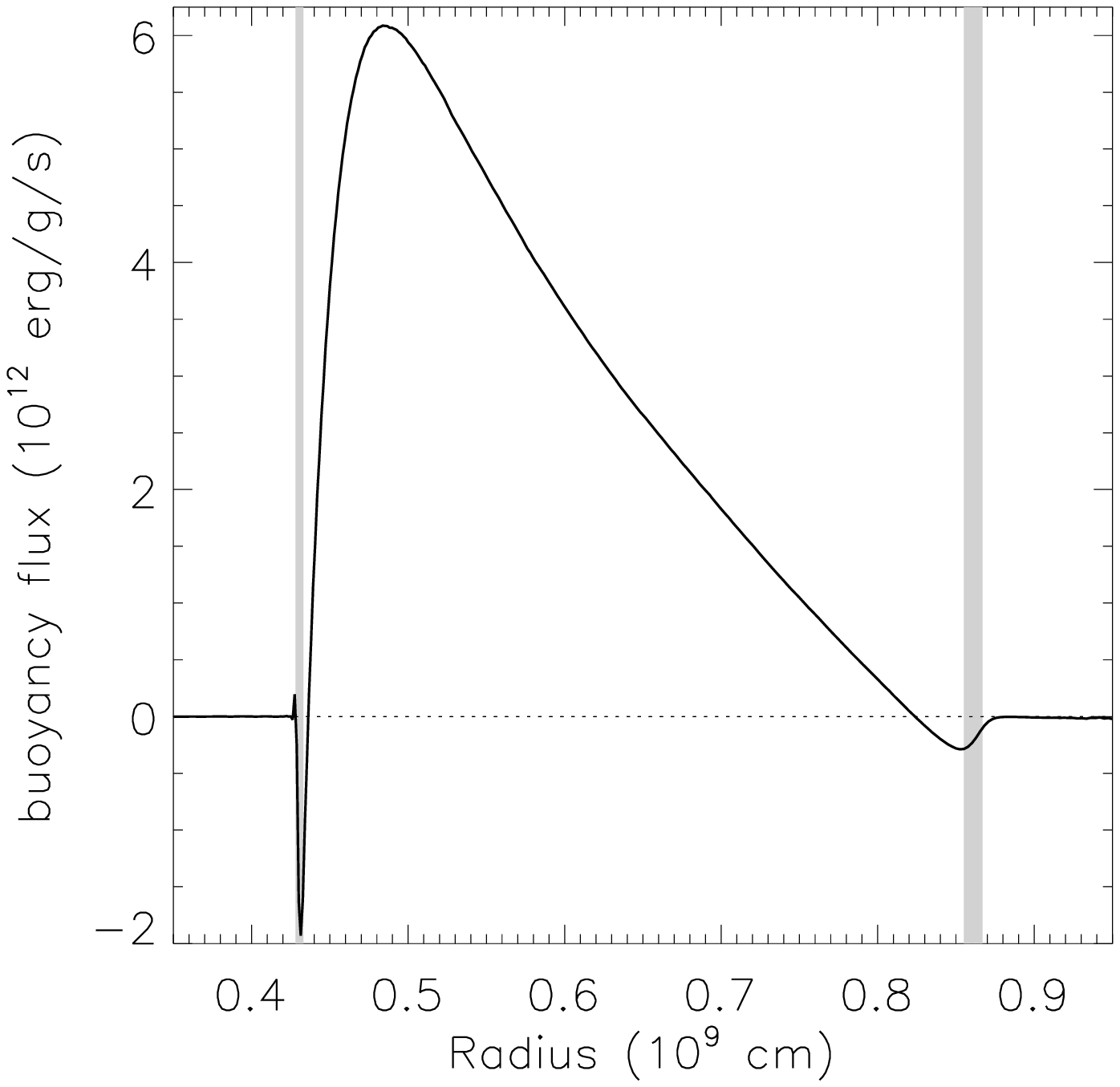}\plottwo{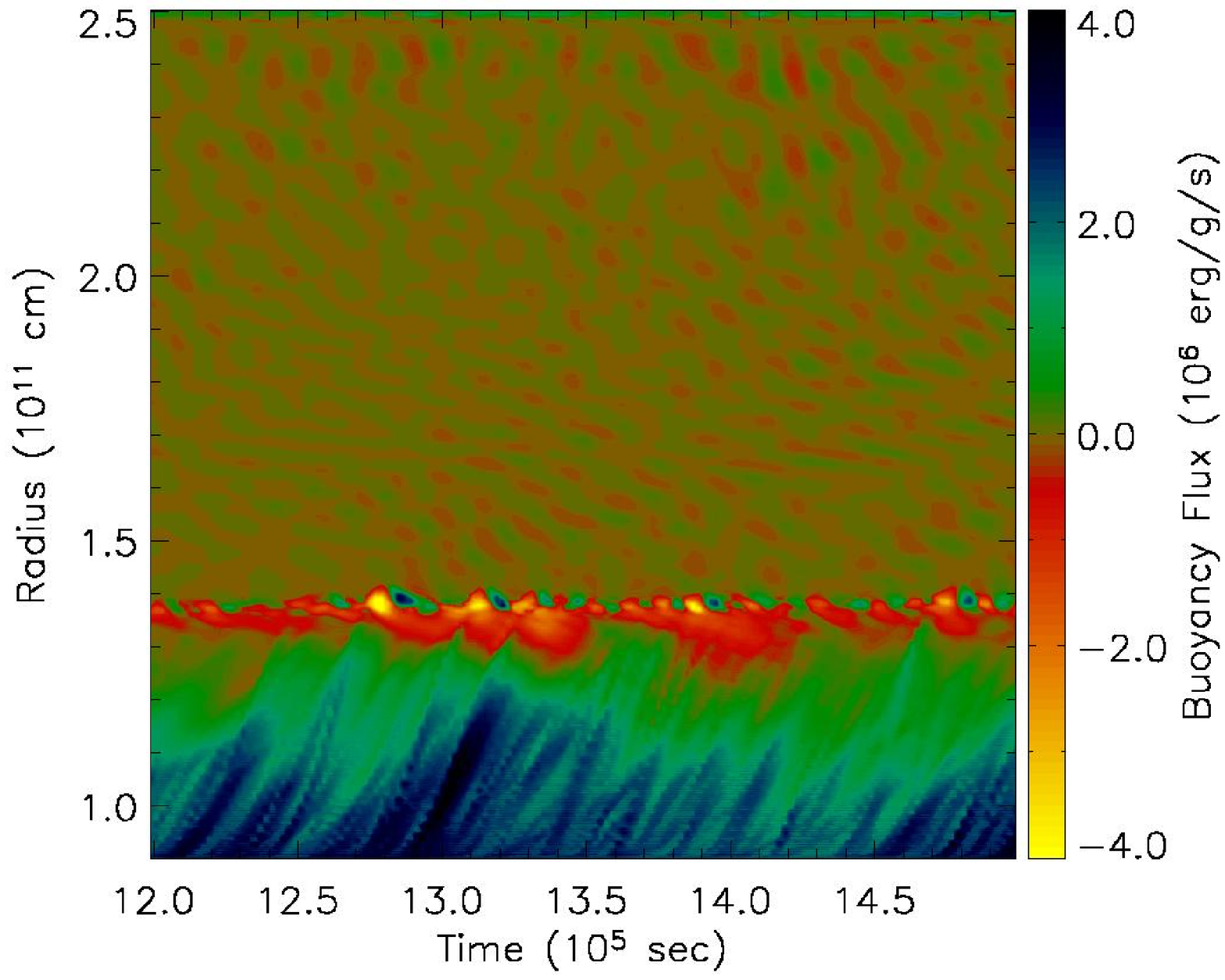}{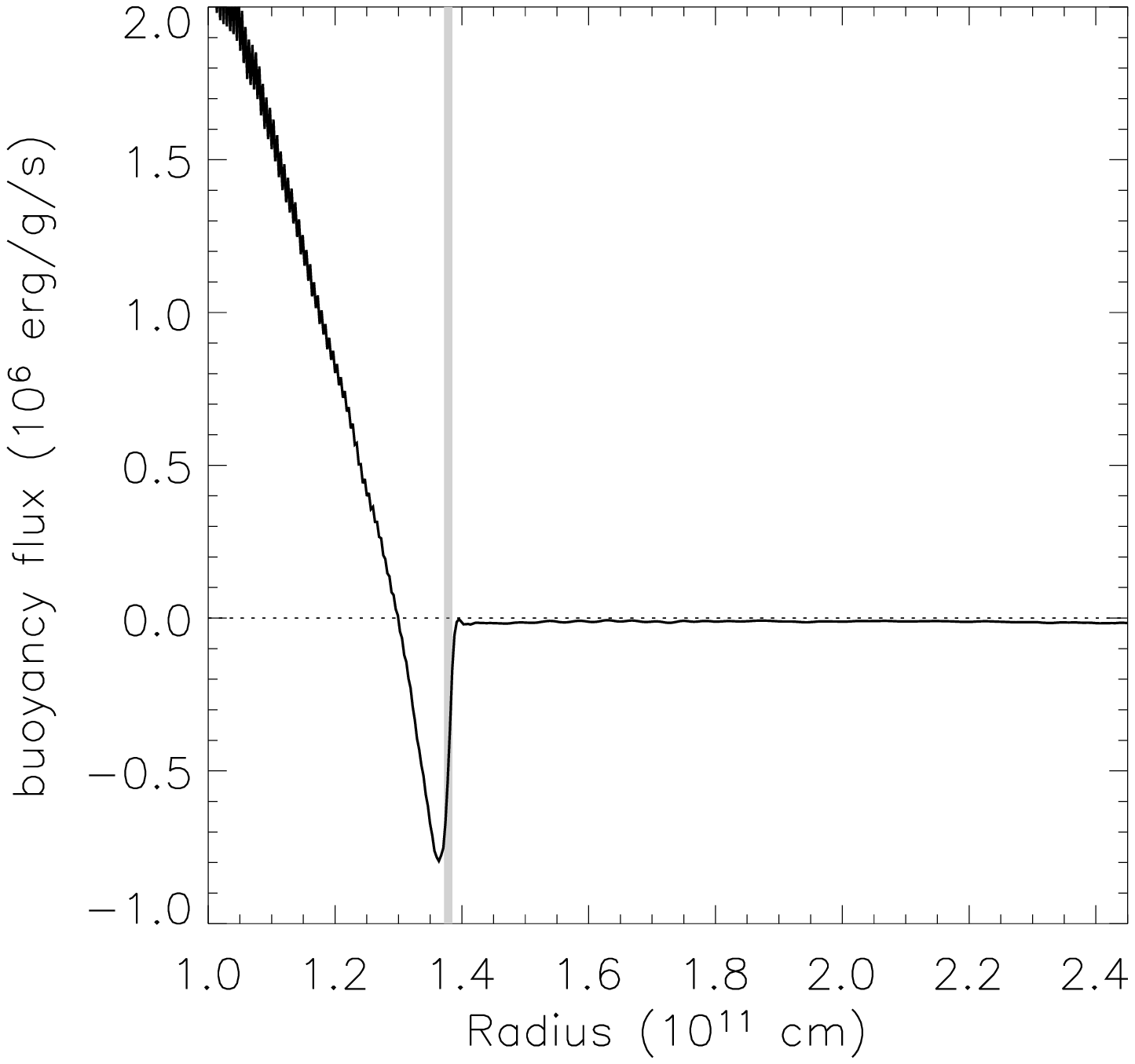} \caption{Buoyancy flux.
Time-series diagrams and time-averaged radial profiles are shown for: (top-row) the
3D oxygen shell burning model; and (bottom-row) the 3D core convection model.
\label{qflux-fig}}
\end{figure*}

 In Figure \ref{qflux-fig} we present the buoyancy flux profiles for our 3D
simulation models, including both time-series diagrams and time averaged radial
profiles. The properties of the buoyancy flux can be divided into three distinct flow
regimes: (1) the body of the buoyant convecting layer, which is dominated by positive
$q$; (2) the convective boundary layers, with negative $q$; (3) the stably stratified
layers, where $q$ is oscillatory, but has a nearly zero mean (in both a horizontal
and time average sense).

\par The buoyancy driving of the convective flow in regime (1) can
be appreciated by comparing the flow velocity to the commonly used buoyant convection
velocity scale $v_*^3 = 2.5\int\havg{q} dr$, where integration is taken over the
radial extent of the convection zone \citep[see e.g.][]{deardorff80}. $v_*$ for the
3D shell burning and core convection models are $v_* \sim 10^7$cm/s and $v_* \sim
3\times 10^5$ cm/s, which compare well to the radial r.m.s. turbulent velocities
measured in the simulation (Figures \ref{ob-2d-3d-vrms} and \ref{msc-2d-3d-vrms}).

In regime (2), which occurs in the convective boundaries, the buoyancy flux is
negative.  A negative value of $q$ signifies that the turbulent kinetic energy is
being converted into potential energy. The mixing that attends this negative buoyancy
flux underlies the entrainment that is taking place at the boundaries through
equation \ref{buoyancy-conserve-eq}. We demonstrate this by showing that the
entrainment speeds measured in the simulation are consistent with the measured
buoyancy fluxes. The interface migration speed is incorporated into the conservation
equation by writing the time derivative as an advective derivative,

\begin{equation}\label{buoyancy-time-deriv-eq}
\partial_t b \sim u_e\partial_r b = u_e N^2
\end{equation}

\noindent where we have used the relationship $\partial_r b= N^2$.
Using this time derivative in equation \ref{buoyancy-conserve-eq}
and solving for $u_e$ we find,

\begin{equation}
\tilde{u}_e = \frac{\Delta q}{h N^2}
\end{equation}

\noindent where we have approximated the divergence of the buoyancy flux with the
difference $\Delta q/h$.  We use the symbol $\tilde{u}_e$ to distinguish the
estimated rate from the values measured in the simulation.

\par If we adopt the buoyancy flux at the interface for $\Delta q$
(Figure \ref{qflux-fig}), the measured interface thickness $h$, and the buoyancy
frequency at the boundary, we find the following entrainment rates.  For the upper
shell boundary, lower shell boundary, and the core convection boundary we have:
$\tilde{u}_e\sim 5.1\times 10^4$ cm/s; $\tilde{u}_e\sim 1.1\times 10^4$ cm/s; and
$\tilde{u_e} \sim 2.2\times 10^3$ cm/s, respectively. These are to be compared with
$u_e = |\dot{r_i} - v_{exp}|$ measured in section
\S\ref{quantify-entrainment-section} and presented in Tables \ref{ob-entrain-table}
and \ref{msc-entrain-table}. The values corresponding to the same time period are:
$u_e = 4.6\times 10^4$ cm/s; $u_e = 1.2\times 10^4$ cm/s; and $u_e = 2\times 10^3$
cm/s. Although these estimates are only order of magnitude (e.g., using the crude
approximation for the time derivative in eq. [\ref{buoyancy-time-deriv-eq}]) they
compare well to the values measured in the simulations and the buoyancy flux due to
"turbulent entrainment" can account for the growth of the convective layers seen
here.

\subsection{Whence q?}

\par The buoyancy flux $q$ appears as a term in the turbulent kinetic
energy (TKE) equation, which we present in \S\ref{energy-appendix}
(eq. [\ref{tke-eq}]). In our notation, the buoyancy flux is related
to the buoyancy work term by $q=\havg{W_B}/\rho_0$.  The buoyancy
flux, therefore, is related to the rate at which turbulent kinetic
energy is advected into the stable layer $F_K$, the rate at which it
dissipates through viscous forces $\varepsilon_K$, and the rate at
which energy is transported through the boundary layer by
pressure-velocity correlations $F_p$. In essence, entrainment is the
process by which the turbulent kinetic energy in the boundary layer
does work against gravity to increase the potential energy of the
overall stratification.

\par Two theoretical approaches have been taken to study entrainment.  The first
approach ignores the TKE equation and instead posits an "entrainment law".
The entrainment law is merely a functional form for the rate at which
stable layer mass will flow into the turbulent region, and is therefore a
dynamic boundary condition. These laws are usually parameterized by the
stability properties of the interface and the strength of the turbulence
through $Ri_B$ \citep[see e.g.][]{fedorovich04}. Once an entrainment law is
adopted, the enthalpy flux can be calculated and the evolution of the
boundary can be self-consistently solved for. The advantage of such an
entrainment law is the simplicity with which it can be incorporated into
global circulation models of the atmosphere, for instance.

\par An alternative approach to adopting an entrainment law is an
explicit physical model for the terms in the TKE equation (eq. [\ref{tke-eq}]).  For
example, general forms for the buoyancy flux profile within the stable layer have
been applied with some success in reproducing the growth of the atmospheric boundary
layer and the deepening of the oceanic thermocline
\citep{stull76b,deardorff79,fedorovich95}.  In some respects, however, these models
are glorified entrainment laws since the buoyancy flux is prescribed in a simplified,
parameterized way. Moving beyond assumptions concerning the turbulence profiles
within the interfacial layer, are theoretical models which take into account the
interactions of waves and turbulence and incorporate non-linear models for the
evolution of instabilities \citep[e.g.][]{ch86,fernando97}. The approach adopted in
these theoretical studies is general enough that any adjustable parameters may turn
out to be universal and a predictive model can be developed.  In addition, the
framework employed is general enough that the production of turbulence by mean flows
(i.e., stellar rotation) can be incorporated, as well as long-range effects due to
internal waves. The internal waves are incorporated through the pressure-correlation
flux, $F_p$, and plays a central role in the evolution of the buoyancy flux when wave
breaking is important.

\subsection{An "Empirical" Entrainment Law}

\par The development of a sophisticated turbulence model to explain
entrainment is beyond the scope of the present work. Instead, we ask
to what extent do the entrainment laws used in geophysical models
apply to our simulations and stellar interiors? Guided by laboratory
study and geophysical large eddy simulation we study the dependance
of the entrainment rate on the bulk Richardson number.

\par
\begin{figure*}
\epsscale{0.8}
\plotone{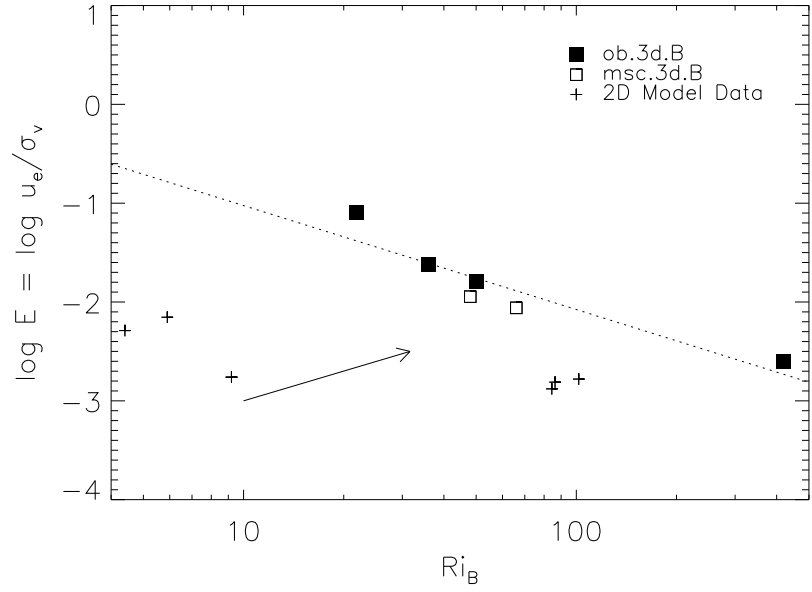}
 \caption{Normalized entrainment rate plotted against bulk
Richardson number $Ri_B$. The 3D models are marked with squares, and the 2D data by
plus signs.  The best fit power-law to the 3D model data is shown by the dashed line.
The 2D entrainment rates fall everywhere below the 3D trend.  The arrow indicates the
direction in the diagram that the 2D data points would move if the effective r.m.s.
turbulence velocity were lower. \label{entrain-law-fig}}
\end{figure*}

  $Ri_B$ is calculated according to equation
\ref{RiB-eq}, where we use the horizontal correlation length scale $L=\mathcal{L}_H$
defined in \S\ref{correlation-appendix}. The buoyancy jump is calculated by
performing the integration in equation \ref{buoyancy-equation} over the width of the
interface which we take to be the interval $r\in[\overline{r_i}-h,\overline{r_i}+h]$.
The normalized entrainment rates $E=u_e/\sigma$, the buoyancy jumps $\Delta b$, and
$Ri_B$ are presented in Tables \ref{ob-entrain-table} and \ref{msc-entrain-table}.
The dependance of the entrainment coefficient $E$ on  $Ri_B$ is presented in Figure
\ref{entrain-law-fig}.

\par The 2D and 3D data are found to obey similar trends (lower E for
higher $Ri_B$), but occupy signifanctly different regions of the diagram.  This can
be explained by the much higher r.m.s. velocities in the 2D simulation. The velocity
scale in 2D is apparently an artifact of the reduced dimensionality of the problem
which significantly influences the flow morphology. Although the velocity scale is
higher in the 2D models, it is much more laminar and accompanied by less turbulent
mixing. The arrow in Figure \ref{entrain-law-fig} indicates the direction that the 2D
data points would move if a lower effective r.m.s. velocity were assumed. In what
follows we focus our attention exclusively on the entrainment data found for the more
realistic 3D models.

\par What we find is that the entrainment coefficient $E$ is well described
by a power law dependance on $Ri_B$ of the form in equation
\ref{entrainment-law-equation}. Our best fit values for the parameters are $\log A =
0.027 \pm 0.38$ and $n = 1.05 \pm 0.21$. This entrainment law is shown by a dashed
line in Figure \ref{entrain-law-fig}.  Remarkably, the power law is of order unity,
in agreement with geophysical and laboratory studies. The fact that the entrainment
in our simulations are governed by the same, fairly universal dependance on $Ri_B$ as
these other studies may have been anticipated, considering the striking degree of
similarity between the buoyancy profiles and the character of the developed flow in
the vicinity of the boundary (Figure \ref{boundary-snapshots-fig}).

\subsection{A Dynamic Convection Zone Boundary Condition}

\par Mass entrainment is a fundamentally different phenomena from
diffusion, which is the typical route used to incorporate new mixing phenomena into a
stellar evolution code. Therefore, how might we incorporate this new process?
Schematically, the idea is very simple.  For each convective boundary, initially
found with the traditional stability criteria ($\partial s/\partial r = 0$,
$\partial^2 s/\partial r^2 \neq 0$), we can calculate the associated bulk Richardson
number based on the background stratification and an approximation of the turbulence
characteristics (e.g., from MLT). With $Ri_B$ in hand we can then input this into our
entrainment law, $E=E(Ri_B)$ which returns to us the entrainment rate.  The
entrainment rate, therefore, is the boundary growth rate as a function of $Ri_B$ and
possibly other parameters of the system.  The function $E(Ri_B)$ can be broken up
into at least three regimes for convenience.

\par {\it Low stability:} $Ri_B < Ri_B^{min}$. For low $Ri_B$ it is
observed that mass entrainment happens very quickly, on an advection timescale
(\S\ref{transient-section}).  Therefore, we can define a minimum $Ri_B^{min}$ at
which the expansion of the convection zone will proceed very quickly, eliminating
convective boundaries which are too weak to support the adjacent turbulence.

\par {\it Intermediate Stability: $Ri_B^{min}<Ri_B<Ri_B^{max}$}.
For an intermediate range of stability, we can use the fairly universal entrainment
law which matches our simulation data, defined by the two parameters $A$ and $n$.
Although scatter in mixing rates were found to be as large as a factor of a few
relative to the best fit law, the general monotonic, power-law dependance was found
to be robust. We can incorporate this physics into the stellar evolution code as a
mass entrainment rate,

\begin{equation}
  \dot{M_E} = \frac{\partial M}{\partial r} u_E = (4\pi
  r_i^2\rho_i)\sigma_H\times
  f_{A}\times 10^{(-n\log Ri_B)}
\end{equation}

\noindent where the normalization factor is written $f_{A} = 10^{(\log A)}$ and
represents the turbulent entrainment mixing efficiency.  More sophistication can
subsequently be incorporated as our understanding of the entrainment process
improves.

\par {\it High Stability: $Ri_B > Ri^{max}_B$}. The entrainment process
will cease to operate at some upper limit $Ri^{max}_B$, above which
the boundary evolution will be controlled by diffusive processes on
the molecular scale.  Following \citep{phillips66}, we have,

\begin{equation}
Ri_B^{max} \simeq \Big(\frac{u_E}{\sigma}\Big)\Big(\frac{\sigma L}{\kappa}\Big)
\end{equation}

\noindent which is based on the condition that the kinetic energy in the turbulence
is sufficient to lift the material from the interface, $\rho \sigma^2 \ga \rho N^2
\Delta^2$. Here, the interface thickness is taken to be that due to molecular
diffusion with $\Delta \ga \kappa/u_E$. The relatively small diffusion rates in
stellar interiors imply that turbulent entrainment will continue to operate to very
high Richardson numbers.  For comparison, the entrainment process in the ocean is
estimated to operate up to $Ri_B \sim 10^{5 - 6}$.

\par Additional details concerning the implementation of this type
of boundary condition into TYCHO will be presented in a subsequent
paper.

\subsection{Spatial Scales, Numerical Resolution, and Entrainment}

\par We conclude this section with a few comments on how well simulation can be
trusted in elucidating the process of entrainment, which is not very well understood.
The spatial scales which limit the entrainment rate at a convective boundary are not
also not well understood, and depend on the interplay between large eddy and small
scale turbulent transport processes. As discussed by \citet{lewellen98}, there is
feedback between the transport rate away from the turbulent boundary layer which is
controlled by large eddies, and the transport rate of material in the immediate
vicinity of the interface by small scale turbulence.  A full understanding of this
problem hinges on being able to resolve the entrainment zone in the presence of the
large scale eddies.

\par A code comparison and resolution study of the entrainment problem in the
planetary boundary layer context was conducted by \citet{bretherton99} and
\citet{stevens99}. In these studies, it was suggested that the appropriate criteria
for resolving boundary layer entrainment is that the grid zoning is smaller than the
fluctuations induced in the inversion layer by the large eddies so that shear
instabilities (e.g., Kelvin-Helmholtz) would not be suppressed. It was suggested that
a "nested grid" of refinement within the boundary layer was comparable to using fine
resolution throughout the simulation domain.  The suggested resolution criterion in
these papers, however, fail to account for the non-linear turbulent evolution which
proceeds the onset of instabilities. In addition, no simulations were conducted with
enough resolution that the boundary layers were turbulent, and the simulations
presented were even marginally resolved by the author's suggested criteria.

\par A related study by \citet{alexakis04} investigates the entrainment
and mixing at a boundary due to internal gravity wave breaking driven by a shear
flow. This process may be operating in the shear mixing layers that form when large
eddies impact the boundary layer. In this study, the mass entrainment rate was found
to depend on the numerical resolution in a non-monotonic way, first decreasing and
then increasing with finer resolution. The author concludes that low resolution
models are dominated by numerical diffusion until the resolution is fine enough to
resolve turbulence near the boundary, at which point the entrainment rate begins to
increase and is controlled by turbulent transport. Although the asymptotic mixing
rate was inconclusive and no resolution criteria was proposed, resolving the
turbulence ensuing from the instability was shown to be important.

\par Much more work needs to be done to address the role played
by both the small scale processes and the interplay with large eddies. Two
complimentary numerical approaches can be taken. First, the feedback between large
and small scale mixing and transport processes can be studied using large eddy
simulation with a range of subgrid scale mixing efficiencies. Such a study can help
develop insight into which scales control the mixing rate. Second, direct numerical
simulations (DNS) which resolve the turbulent processes operating at the interface
can be undertaken when sufficient computational resources are available. These
studies would provide more definitive conclusions concerning the interplay between
eddy scales and would provide guidance for a more general framework for future
theoretical analysis. Finally, it is important to keep in mind that laboratory
studies of high Reynolds number turbulent entrainment continue to provide useful
constraints, and improved flow visualization techniques are allowing a more direct
comparison to theory and simulation.

\par  The "empirical" entrainment law which we discuss in this paper is constrained by
only a few data points (the six 3D data points in Figure \ref{entrain-law-fig}).
Extending simulations to include an ever more diverse suite of stellar structures
would provide an even stronger mandate, and better constrained model for
incorporating this physics into stellar evolution codes.

\section{SUMMARY AND CONCLUSIONS}

\par In this paper, we have presented the results of three-dimensional,
reactive, compressible, hydrodynamic simulations of deep, efficient stellar
convection zones in massive stars.  Our models are unique in terms of the degree to
which non-idealized physics have been used, and the evolutionary stages simulated,
with fuel and ash clearly distinguished.

\par We find several general results regarding the basic properties of
the convective flow:

\begin{itemize}
\item the flow is highly intermittent, but has robust statistical
properties,
\item the 2D vs 3D velocity
scales differ by almost a factor of several, and the flow morphologies are completely
different,
\item stable layers interact with convection to decelerate plumes, and
consequently distort these layers, which then generate waves,
\item mixing is found to occur at convective boundaries in manner best described as
turbulent entrainment, rather than the traditional picture of convective overshooting
wherein turbulent eddies ballistically penetrate the stable layers.
\end{itemize}

\par We have compared our oxygen shell burning model to
mixing length theory assumptions.  We show that, while a reasonable representation of
the super-adiabatic temperature gradient and velocity scale can be fit with a single
mixing length, the values of the inferred mixing-length ``constants'' differ from
other simulations. This was already implied in \cite{pwj00}, who found difference for
slab and spherical geometries. There may be a dependence upon the depth of the
convection zone as well, and possibly upon the nature of the stable boundary regions
and/or the nature of the driving process (burning or radiative loss).

\par Why do we care about MLT in regions of efficient convection?
(1) temperature profile can affect the burning rates, which have a stiff temperature
dependence; (2) the velocity scale can affect the nucleosynthesis (such as s-process
branching ratios in double shell burning AGB stars) by dictating the exposure time of
the plasma to varying conditions throughout the burning region; (3) the velocity
scale and the kinetic energy flux is an important input needed for calculating the
mixing at convective boundaries.

\par We have found that the extent of mixing is better represented by an integrated
Richardson number rather than the convectional Schwarzschild or Ledoux criteria
alone. This incorporates the addition physics related to the resistance of stiff
boundaries to mixing. Related to the definition of boundary stiffness, we have
identified an important physical process which is missing from the standard theory of
stellar evolution: turbulent entrainment. This process is well known in the
meteorology and oceanographic communities, and has been extensively studied
experimentally. We show that the rate of entrainment is well represented as a simple
function of the buoyancy jump, in a manner similar to that determined experimentally.

\par The long term consequences of convective boundary inconsistencies such as the one
illustrated by the initial transient in our simulation, and for which the conditions
are common in 1D stellar models, can significantly alter the size of convective
cores, and thus the subsequent explosion and nucleosynthetic yields of the resultant
supernova. In a subsequent paper in this series, we will present case studies which
incorporate the physical insight gained through these simulations into the TYCHO
stellar evolution code. We expect to see effects in solar models, s-processing in AGB
stars, stellar core formation (white dwarfs, neutron stars, and black holes), stellar
nucleosynthesis yields, stellar ages, and HR diagrams.

\begin{acknowledgements}
  This work was supported in part by the ASCII FLASH center at the University of Chicago.
  CM would like to acknowledge the stimulating discussions at the 2006 Los Alamos
  {\em Summer Hydro Days} Workshop, made possible by Falk Herwig, which influenced the
  writing of this paper.  DA wishes to thank the Aspen Center for Physics for their
hospitality.
\end{acknowledgements}

\begin{appendix}

\section{THE ENERGY EQUATION} \label{energy-appendix}

\subsection{Total Energy}
\par The primitive energy equation solved by PROMPI is,

\begin{equation}
  \partial_t (\rho E) + \nabla\cdot\big[(\rho E + p)\mathbf{u} + \mathbf{F_r} \big]
  = \rho \mathbf{u}\cdot\mathbf{g} + \rho\epsilon_{net}
\end{equation}

\noindent where the total energy is composed of the internal and kinetic components,
$E = E_I + E_K$. We decompose the velocity, density, and pressure fields into mean
and fluctuating components according to,

\begin{equation}
\varphi = \varphi_0 + \varphi'
\end{equation}

\noindent where $\havg{\varphi} = \varphi_0$ and $\havg{\varphi'} = 0$. The
pressure-velocity correlation term is,

\begin{equation}
  \nabla\cdot\avg{p\mathbf{u}} =
  \nabla\cdot\avg{p_0 \mathbf{u_0}} +
  \nabla\cdot\avg{p_0 \mathbf{u'}} +
  \nabla\cdot\avg{p' \mathbf{u_0}} +
  \nabla\cdot\avg{p' \mathbf{u'}}.
\end{equation}

\noindent The gravity term is,

\begin{equation}
  \avg{\rho\mathbf{g}\cdot\mathbf{u}} =
  \avg{\rho_0\mathbf{u_0}\mathbf{g}} +
  \avg{\rho_0\mathbf{u'}\mathbf{g}} +
  \avg{\rho'\mathbf{u_0}\mathbf{g}} +
  \avg{\rho'\mathbf{u'}\mathbf{g}}.
\end{equation}

\noindent The averaging operator eliminates terms which are first order in
fluctuations (by definition) and we have,

\begin{equation}
  \partial_t\avg{\rho E} +
  \nabla\cdot\Big[
    \avg{\rho E\mathbf{u_0}} +
    \avg{\rho E\mathbf{u'}} +
    \avg{p_0\mathbf{u_0}} +
    \avg{p'\mathbf{u'}} +
    \mathbf{F_r}
    \Big] =
    \avg{\rho_0\mathbf{u_0}\mathbf{g}} +
    \avg{\rho'\mathbf{u'}\mathbf{g}} +
  \avg{\rho\epsilon_{net}}.
\end{equation}

\noindent We can further simplify this expression using the condition of hydrostatic
equilibrium, which holds to a high degree of accuracy in the simulation ($\nabla p_0
= \rho_0 \mathbf{g}$). The background velocity in this case, $\mathbf{u_0}$, is a
slow, highly subsonic, expansion or contraction that is driven on a thermal
timescale.  The background velocity field has only a radial component (i.e., there is
no rotation in the current model), $\mathbf{u_0} = (u_{0,(r)},0,0)$. The energy
equation can be then simplified to read,

\begin{equation}
  \partial_t\avg{\rho E} +
  \nabla\cdot\avg{\rho E\mathbf{u_0}} =
  -\nabla\cdot\avg{\mathbf{F_p} + \mathbf{F_I} + \mathbf{F_K} + \mathbf{F_r}}
  -\avg{p_0\nabla\cdot\mathbf{u_0}}
  +\avg{\mathbf{W_b}} + \avg{\rho\epsilon_{net}}.
\end{equation}

\noindent where we have used the following definitions,

\begin{eqnarray}
  \mathbf{F_I} = \rho E_I\mathbf{u'}\\
  \mathbf{F_K} = \rho E_K\mathbf{u'}\\
  \mathbf{F_p} = p'\mathbf{u'} \\
  \mathbf{W_b} = \rho'\mathbf{g}\cdot\mathbf{u'}.
\end{eqnarray}

\subsection{Kinetic Energy}

\par The kinetic energy equation is derived by forming the scalar
product of the velocity with the equation of motion (e.g., Shu, 1992, Ch.2). The
momentum equation can be written in vector form as,

\begin{equation}
\partial_t(\rho E_K) + \nabla\cdot(\rho E_K\mathbf{u})
    = -\mathbf{u}\cdot\nabla p + \rho\mathbf{u}\cdot\mathbf{g}
\end{equation}

\noindent Again, we decompose the fields into mean and fluctuating components, employ
the hydrostatic equilibrium condition, and perform averages.  The result is,

\begin{equation}\label{tke-eq}
\partial_t \avg{\rho E_K} + \nabla\cdot\avg{\rho E_K\mathbf{u_0}} =
-\nabla\cdot\avg{\mathbf{F_p} + \mathbf{F_K}} + \avg{p'\nabla\cdot\mathbf{u'}} +
\avg{\mathbf{W_b}} -\varepsilon_K.
\end{equation}

\par Here, $\varepsilon_K$ is the viscous dissipation of kinetic
energy.  In our simulations, this term is not modeled explicitly and arises due to
numerical dissipation. The term $p'\nabla\cdot\mathbf{u'}$ represents the
compressional work done by turbulent fluctuations, and the other terms are as defined
above.

\section{CORRELATION LENGTH SCALES}
\label{correlation-appendix}

\par The vertical correlation of the horizontal distribution of fluctuations
in a quantity $X' = X - \langle X\rangle$ at radial position $r$ and offset position
$r+\delta r$ is calculated according to,

\begin{equation}
    \label{vscale-equation}
  C^V(\delta r;r) =\frac{1}{\Delta\Omega}\frac{\int X'(r,\theta,\phi)
    X'(r+\delta r,\theta,\phi) d\Omega}{\sigma_X(r)\sigma_X(r+\delta r)}
\end{equation}

\noindent where the integral is taken over the angular direction with $d\Omega =
\sin(\theta)d\theta d\phi$. The correlation is normalized by the product of the
horizontal r.m.s. value of the quantity at the two levels being compared $\sigma_X$.

\par The horizontal correlation of fluctuations at radial
position $r$ is calculated using the autocorrelation function,

\begin{equation}
    \label{hscale-equation}
  C^H(\delta s;r) = \frac{\langle X'(r,s) X'(r,s+\delta s) \rangle}{\sigma_X(r)^2}
\end{equation}

\noindent where the brackets $\langle\cdot\rangle$ denote averaging over all
horizontal locations $s$ and fixed offset $\delta s$.  The horizontal correlation is
normalized by the variance of the quantity $\sigma_X^2$.

\par Characteristic length scales are defined as the offset position where the
correlation function drops to a value of 0.5.  For horizontal correlations, we define
this length as $\mathcal{L}_H$. We also define a value which is twice this length,
the full width at half maximum, which we denote by $L_H$.  (The value $\mathcal{L}_H$
provides a good approximation to the integral scale, $\int C^H (\delta s;r) d\delta
s$.)

\par In the vertical direction the sign of the offset $\delta r$ is retained and a
separate length scale is defined where the correlation function drops to 0.5 for
positive and negative offsets, which we denote by $L_V^+$ and $L_V^-$.  The full
width is denoted $L_V = L_V^{+} - L_V^{-}$.

\end{appendix}

\clearpage


 \clearpage

 \clearpage

\clearpage

\begin{deluxetable}{lll}
\tabletypesize{\scriptsize} \tablewidth{0pt} \tablecaption{Nuclei Included in Reduced
Nuclear Reaction Network\label{network-table}} \tablehead{
  Element &
  \mcol{1}{c}{Charge} &
  \mcol{1}{c}{Atomic} \\
  &  & \mcol{1}{c}{Weight}}

\startdata
Helium \dots \dotfill      & 2  & 4      \\
Carbon  \dots \dotfill     & 6  & 12     \\
Oxygen  \dots \dotfill     & 8  & 16     \\
Neon \dots \dotfill        & 10 & 20     \\
Sodium \dots \dotfill      & 11 & 23     \\
Magnesium \dots \dotfill   & 12 & 24     \\
Silicon \dots \dotfill     & 14 & 28     \\
Phosphorus  \dots \dotfill & 15 & 31     \\
Sulfur  \dots \dotfill     & 16 & 32, 34 \\
Chlorine \dots \dotfill    & 17 & 35     \\
Argon \dots \dotfill       & 18 & 36, 38 \\
Potassium  \dots \dotfill  & 19 & 39     \\
Calcium \dots \dotfill     & 20 & 40, 42 \\
Titanium \dots \dotfill    & 22 & 44, 46 \\
Chromium \dots \dotfill    & 24 & 48, 50 \\
Iron \dots \dotfill        & 26 & 52, 54 \\
Nickel \dots \dotfill      & 28 & 56     \\
\enddata

\tablecomments{Network also includes electrons, protons, and neutrons.}

\end{deluxetable}

\begin{deluxetable}{llllll}
  \tabletypesize{\scriptsize}
  \tablewidth{0pt}
  \tablecaption{Summary of Oxygen Shell Burning Models\label{ob-table}}
  \tablehead{
    \mcol{1}{c}{Parameter} &
    \mcol{1}{c}{Units} &
    \mcol{1}{c}{ob.3d.B} &
    \mcol{1}{c}{ob.2d.c} &
    \mcol{1}{c}{ob.2d.C} &
    \mcol{1}{c}{ob.2d.e}}
\startdata
r$_{in}$, r$_{out}$           & (10$^9$ cm)    & 0.3, 1.0 &  0.3, 1.0  & 0.3, 1.0 & 0.3, 5.0 \\
 $\Delta\theta$,$\Delta\phi$ & (deg.)          &  30, 30 & 90, 0 & 90, 0 & 90, 0 \\
Grid Zoning                  & -               & 400\mult (100)$^2$ & 400\mult 320&  800\mult 640  &  800\mult 320 \\
$t_{\hbox{max}}$             & (s)             & 800 & 574 & 450 & 2,400  \\
$v_{\hbox{conv}}$ & (10$^7$ cm/s)              & 0.8  & 2.0 & 1.8 & 1.8 \\
$t_{\hbox{conv}}$            & (s)             & 103 & 40 & 44 & 44\\
$\dot{M_i}|_u$\tnm{1}        & (10$^{-4}M_{\odot}$/s) & 1.1   &  1.33 & 1.25 & 1.3 \\
$\dot{M_i}|_l$               & (10$^{-4}M_{\odot}$/s) & -0.23 &  -0.52 & -0.5 & -0.5\\
\enddata
\tnt{1}{The subscripts $u$ and $l$ refer to the upper and lower convective shell
boundary.}

\end{deluxetable}

\begin{deluxetable}{llll}
  \tabletypesize{\scriptsize}
  \tablewidth{0pt}
  \tablecaption{Summary of ``Core Convection'' Models\label{msc-table}}
  \tablehead{
    \mcol{1}{c}{Parameter} &
    \mcol{1}{c}{Units} &
    \mcol{1}{c}{msc.3d.B}  &
    \mcol{1}{c}{msc.2d.b}}
\startdata
r$_{in}$, r$_{out}$              & (10$^{11}$ cm)  & 0.9, 2.5 & 0.9, 2.5\\
$\Delta\theta$,$\Delta\phi$      & (deg.)        &  30, 30 & 30,0 \\
Grid Zoning                      & -             & 400\mult (100)$^2$ & 400\mult 100\\
$t_{\hbox{max}}$                 & (s)           & 2.0\mult 10$^6$  & 2.0\mult 10$^6$ \\
$v_{\hbox{conv}}$                & (10$^5$ cm/s) & 2.5 & 13\\
$t_{\hbox{conv}}$                & (s)           & 3.6\mult10$^5$ & 6\mult10$^4$\\
$\dot{M_i}$                      & (10$^{-7}M_{\odot}$/s) & 2.72 & 4.73\\
\enddata
\end{deluxetable}

\begin{deluxetable}{llllllll}
\tabletypesize{\scriptsize} \tablewidth{0pt} \tablecaption{Assumed and Measured
Convection Parameters\tnm{a}\label{conv-table}} \tablehead{
  Study &
  P\'e\tablenotemark{b} &
  $\alpha_E$&
  $\alpha_{\Lambda,T}$&
  $\alpha_{\Lambda,v}$&
  $\alpha_{\Lambda}$&
  $L/H_p$\tablenotemark{c} &
  Grid Zoning \\
  }
\startdata

MLT               & $\gg$1    & 1                 & $\alpha$    & $\alpha$    & $\alpha$           & \dots                & \dots \\
This Study\tnm{1} & $\ga10^3$ & 0.70$\pm$0.03     & 0.8 - 1.4   & 1.35 - 1.73 & 0.87 - 1.31        & $\sim$2(3.7)\tnm{d}  & $100^2\times223(400)$\tnm{c} \\
Chan-Sofia\tnm{2} & \dots     & 0.81$\pm$0.03     & 1.32 - 3.75 & 3.39 - 6.4  & 1.90 - 4.4         &  7                   & $20^2\times(\la50)$ \\
Kim\tnm{3}        & \dots     & 0.80$\pm$0.01     & 2.96 - 4.2  & 1.5 - 3.4   &  1.4 - 3.2         &  6                   & $32^3$ \\
Robinson\tnm{4}   & \dots     & 0.65-0.85         & \dots       & \dots       &  \dots             & 8.5                  & $114^2\times170$  \\
Porter-Woodward\tnm{5a}&$10-8\times10^4$& 0.7-0.9 & 4.08        & 3.82        & 2.68(3.53)\tnm{5b} & 5                    & $512^2\times256$ \\
\enddata
\tnt{a}{See \S\ref{alphas-section} for parameter definitions: $\alpha_{\Lambda,T} =
2\times\alpha_T$ and $\alpha_{\Lambda,v} = \sqrt{2}\times\alpha_v$ where $\alpha_T$
and $\alpha_v$ are defined by equations \ref{alphat-eq} and \ref{alphav-eq}, and
$\alpha_{\Lambda} =
\sqrt{\alpha_E\times\alpha_{\Lambda,v}\times\alpha_{\Lambda,T}}$.} \tnt{b}{The
P\'eclet number is shown when provided by the author. In all cases the regions in the
simulations for which parameter values are quoted were efficient convection, with
$\Delta\nabla \la 10^{-2}$, and excluded the super-adiabatic layers in the surface
convection models where parameters deviate significantly from those quoted
here.}\tnt{c}{The number of pressure scale heights spanned by the convectively
unstable region.} \tnt{d}{The value in parentheses is for region spanning the entire
computational domain, including the stable bounding layers with the other value
referring to the convective region.}\tnt{1}{Model ob.3d.B: additional details in
Table \ref{ob-table}.} \tnt{2}{\citet{chan89}: The range in $\alpha_T$ and
$\alpha_{\Lambda,v}$ is calculated according to their Table 1 for the nearly
adiabatic portion of the simulation where $10^{-3}<\Delta\nabla<10^{-2}$.}
\tnt{3}{\citet{kim96}: The coefficient $\alpha_T$ is based on their Fig. 6. The
coefficients $\alpha_{\Lambda,v}$ and $\alpha_{\Lambda}$ are plotted in their Fig. 9
and the range quoted in the table above is for values at least one pressure scale
height from the boundaries.} \tnt{4}{\citet{rob04}: only the correlation between
radial velocity and temperature fluctuation is provided, which is a good surrogate
for $\alpha_E$. For the solar and subgiant cases see their Figs. 7 -
9.}\tnt{5}{(a)\citet{pw00}: In this paper the values for $\alpha_v$, $\alpha_T$, and
$\alpha_{\Lambda}$ are quoted using the same definitions as in this study. (b) The
lower value quoted by these authors for $\alpha_{\Lambda}$ is a results of
subtracting the kinetic energy flux from the enthalpy flux. The value in the
parentheses is the mixing length $\alpha_\Lambda$ according to the definition in note
\tnm{a} above.}
\end{deluxetable}

\begin{deluxetable}{llllrrrrrr}
\tabletypesize{\scriptsize} \tablewidth{0pt} \tablecaption{Convective Boundary Layer
Properties For Oxygen Shell Burning Models\label{ob-entrain-table}} \tablehead{
  \mcol{1}{c}{Model} &
   \mcol{1}{l}{Time Interval} &
  \mcol{1}{l}{$\overline{r_i}$}&
  \mcol{1}{l}{$\overline{h}$} &
  \mcol{1}{l}{$\overline{\dot{r_i}}$} &
  \mcol{1}{l}{$\overline{v_{exp}}$} &
  \mcol{1}{l}{$\sigma[v_{H}]$\tablenotemark{a}} &
  \mcol{1}{l}{$\overline{\Delta b}$\tablenotemark{b}} &
  \mcol{1}{l}{$\log \overline{E}$} &
  \mcol{1}{l}{$\overline{Ri_B}$}\\
  \mcol{1}{l}{} &
  \mcol{1}{l}{(10$^2$ s)} &
  \mcol{1}{l}{(10$^9$ cm)} &
  \mcol{1}{l}{(10$^7$ cm)} &
  \mcol{1}{l}{(10$^4$ cm/s)} &
  \mcol{1}{l}{(10$^4$ cm/s)} &
  \mcol{1}{l}{(10$^7$ cm/s)} &
  \mcol{1}{l}{(10$^7$ cm/s$^2$)} &
  \mcol{1}{l}{} &
  \mcol{1}{l}{}
  }
\startdata
ob.3d.B        & [1.5, 2.7] & 0.816 & 1.287 &25.766$\pm$0.869  & 0.6            & 0.313 & 0.574 & -1.095 & 21.8 \\
\mcol{1}{c}{"} & [2.7, 5.5] & 0.842 & 0.797 & 8.252$\pm$0.180  & \mcol{1}{r}{"} & 0.316 & 0.966 & -1.616 & 36.0 \\
\mcol{1}{c}{"} & [5.5, 8.0] & 0.861 & 0.586 & 5.171$\pm$0.179  & \mcol{1}{r}{"} & 0.281 & 1.062 & -1.789 & 50.0 \\
ob.2d.c        & [3.5, 5.7] & 0.857 & 0.191 &10.620$\pm$0.816  & 0.9            & 1.385 & 1.422 & -2.154 &  5.9 \\
ob.2d.C        & [2.0, 4.0] & 0.830 & 1.776 &19.117$\pm$0.988  & 0.5            & 1.436 & 1.010 & -1.887 &  3.2 \\
ob.2d.e        & [3.5, 8.0] & 0.868 & 1.900 &10.021$\pm$0.319  & 2.5            & 1.464 & 1.334 & -2.289 &  4.4 \\
\hline
ob.3d.B        & [3.0, 8.0] & 0.429 & 0.057 &-0.700$\pm$0.009  & 0.50  & 0.479 & 30.686  & -2.601 & 418.6 \\
ob.2d.c        & [3.5, 5.7] & 0.428 & 0.201 &-1.686$\pm$0.058  & 1.05  & 1.769  & 33.739 & -2.811 &  86.3 \\
ob.2d.C        & [2.0, 4.0] & 0.430 & 0.193 &-1.625$\pm$0.072  & 0.65  & 1.434  & 32.160 & -2.780 & 101.7 \\
ob.2d.e        & [3.5, 8.0] & 0.429 & 0.162 &-0.975$\pm$0.018  & 1.20  & 1.645  & 32.620 & -2.879 &  84.4\\
\enddata
\tablenotetext{a}{The rms fluctuations in the horizontal velocity at the interface
location.} \tablenotetext{b}{The buoyancy jump across the interface.}
\end{deluxetable}

\begin{deluxetable}{llllrlllll}
\tabletypesize{\scriptsize} \tablewidth{0pt} \tablecaption{Convective Boundary Layer
Properties For "Core Convection" Models\label{msc-entrain-table}} \tablehead{
  \mcol{1}{c}{Model} &
   \mcol{1}{l}{Time Interval} &
  \mcol{1}{l}{$\overline{r_i}$}&
  \mcol{1}{l}{$\overline{h}$} &
  \mcol{1}{l}{$\overline{\dot{r_i}}$} &
  \mcol{1}{l}{$\overline{v_{exp}}$\tnm{a}} &
  \mcol{1}{l}{$\sigma[v_{H}]$} &
  \mcol{1}{l}{$\overline{\Delta b}$} &
  \mcol{1}{l}{$\log \overline{E}$} &
  \mcol{1}{l}{$\overline{Ri_B}$}\\
  \mcol{1}{l}{} &
  \mcol{1}{l}{(10$^5$ s)} &
  \mcol{1}{l}{(10$^{11}$ cm)} &
  \mcol{1}{l}{(10$^9$ cm)} &
  \mcol{1}{l}{(10$^3$ cm/s)} &
  \mcol{1}{l}{(10$^2$ cm/s)} &
  \mcol{1}{l}{(10$^5$ cm/s)} &
  \mcol{1}{l}{(10$^2$ cm/s$^2$)} &
  \mcol{1}{l}{} &
  \mcol{1}{l}{}
  }
\startdata
msc.3d.B       & [ 6.0, 10.0] & 1.374 & 0.949 & 1.754$\pm$ 0.080  &\dots  & 2.011 & 6.07 & -2.0594 & 66 \\
\mcol{1}{c}{"} & [10.0, 12.0] & 1.378 & 0.897 &-0.020$\pm$ 0.140  &\dots  & 1.878 & 5.83 & \dots   & 72 \\
\mcol{1}{c}{"} & [12.0, 15.0] & 1.382 & 0.998 & 2.731$\pm$ 0.099  &\dots  & 2.411 & 5.70 & -1.9459 & 48 \\
msc.2d.b       & [ 6.0, 10.0] & 1.369 & 1.319 & 1.401$\pm$ 0.390  &\dots  & 8.070 & 6.43 & -2.7604 & 9.2\\
\enddata
\tnt{a}{The expansion velocity in these models remains very small with $v_{exp} < 10$
cm/s.}
\end{deluxetable}

\end{document}